\documentclass[12pt]{article}
\usepackage[textwidth=15.9cm,textheight=23.2cm]{geometry}
\usepackage{latexsym}
\usepackage{graphicx}
\usepackage{cite}
\usepackage{bbm}
\usepackage{mathrsfs}
\def\AdSs5{$AdS_5$}
\def\AdSS5{$AdS_5$}
\def\AdS5s5{$AdS_5 \times S^5$}
\def\al{{\alpha^{\prime}}}
\def\gs{g_{st}}
\def\gy{g_{_{\rm YM}}}
\def\er{{\rm e}}
\def\dr{{\rm d}}
\def\Tr{{\rm Tr}}
\def\tr{{\rm tr}}
\def\gs{g_{\rm s}}

\newcommand{\RR}{${\rm R}\!\otimes\!{\rm R}$\ }

\newcommand{\eg}{{\it e.g.~}}
\newcommand{\ie}{{\it i.e.~}}

\newcommand{\s}{\sigma}
\newcommand{\del}{\partial}
\newcommand{\aap}{{a^{\prime}}}
\newcommand{\be}{\begin{equation}}
\newcommand{\ee}{\end{equation}}
\newcommand{\ba}{\begin{eqnarray}}
\newcommand{\ea}{\end{eqnarray}}
\newcommand{\ra}{\rangle}
\newcommand{\la}{\langle}
\newcommand{\pp}{\prime}
\newcommand{\dpp}{{\prime\prime}}

\newcommand{\op}{\oplus}
\newcommand\fr[1]{\frac{1}{#1}}
\newbox\SlashedBox
\def\fs#1{\setbox\SlashedBox=\hbox{#1}
\hbox to
0pt{\hbox to 1\wd\SlashedBox{\hfil/\hfil}\hss}{#1}}
\def\hboxtosizeof#1#2{\setbox\SlashedBox=\hbox{#1}
\hbox to
1\wd\SlashedBox{#2}}

\def\ms#1{\setbox\SlashedBox=\hbox{$#1$}
\hbox to 0pt{\hbox to
1\wd\SlashedBox{\hfil/\hfil}\hss}#1}

\newcommand{\Dscrm}{\,{\raisebox{1pt}{$/$} \hspace{-9pt} {\scrD}}}

\newcommand{\one}{\mathbbm{1}}

\def\lam{\lambda}
\def\t2{\tau_2}
\def\IZ{\relax\ifmmode\mathchoice {\hbox{\cmss Z\kern-.4em Z}}
{\hbox{\cmss Z\kern-.4em Z}}
{\lower.9pt\hbox{\cmsss Z\kern-.4em Z}}
{\lower1.2pt\hbox{\cmsss Z\kern-.4em Z}}
\else{\cmss Z\kern-.4em Z}\fi}

\def\S{\Sigma}
\def\Sbar{{\bar\Sigma}}

\def\b{\beta}
\def\a{{\alpha}}
\def\g{\gamma}
\def\veps{\varepsilon}
\def\vt{\vartheta}
\def\bvt{{\bar\vartheta}}

\def\adot{{\dot\alpha}}
\def\bdot{{\dot\beta}}

\def\d{\delta}

\def\D{\Delta}
\def\Db{\bar \Delta}

\def\c1{{\chi^1}}

\def\v{\varphi}
\def\tc{{\tau^c}}

\def\N4{{\cal N}=4}
\def\half{\frac{1}{2}}
\def\quart{\frac{1}{4}}
\def\nn{\nonumber}

\def\nus{\begin{displaystyle}
    \bar\nu_{\rule{0pt}{2pt}}^{u[A}\nu^{B]}_u
    \end{displaystyle}}
\def\nut{\begin{displaystyle}
    \bar\nu_{\rule{0pt}{2pt}}^{u(A}\nu^{B)}_u
    \end{displaystyle}}
\def\nsix{(\bar\nu \nu)_{\bf 6}}
\def\nten{(\bar\nu \nu)_{\bf 10}}
\newcommand{\scrA}{{\mathscr A}}

\newcommand{\scrC}{{\mathscr C}}
\newcommand{\scrD}{{\mathscr D}}
\newcommand{\scrE}{{\mathscr E}}
\newcommand{\scrF}{{\mathscr F}}

\newcommand{\scrH}{{\mathscr H}}
\newcommand{\scrI}{{\mathscr I}}

\newcommand{\scrK}{{\mathscr K}}

\newcommand{\scrM}{{\mathscr M}}
\newcommand{\scrN}{{\mathscr N}}
\newcommand{\scrO}{{\mathscr O}}
\newcommand{\scrP}{{\mathscr P}}
\newcommand{\scrQ}{{\mathscr Q}}

\newcommand{\scrV}{{\mathscr V}}

\newcommand{\scrZ}{{\mathscr Z}}
\DeclareMathAlphabet{\mathpzc}{OT1}{pzc}{m}{it}

\newcommand{\scrd}{{\mathpzc d}}

\newcommand{\scrm}{{\mathpzc m}}

\newcommand{\scrmf}{{\mathpzc m}_{\rm \, f}}

\newcommand\hsp[1]{\hspace*{#1 cm}}
\newcommand\vsp[1]{\vspace*{#1 cm}}
\newcommand{\ndt}{\noindent}
\newcommand\atmp[3]{{\it Adv.\ Theor.\ Math.\ Phys.\ }{\bf #1} (#2) #3}

\newcommand\cqg[3]{{\it Class.\ and Quant.\ Grav.\ }{\bf #1} (#2) #3}

\newcommand\ijmpa[3]{{\it Int.\ J.\ Mod.\ Phys.\ }{\bf A#1} (#2) #3}

\newcommand\ijtp[3]{{\it Int.\ J.\ Theor.\ Phys.\ }{\bf #1} (#2) #3}

\newcommand\jhep[3]{{\it J. High Energy Phys.\ }{\bf #1} (#2) #3}

\newcommand\npb[3]{{\it Nucl.\ Phys.\ }{\bf B#1} (#2) #3}

\newcommand\pla[3]{{\it Phys.\ Lett.\ }{\bf A#1} (#2) #3}
\newcommand\plb[3]{{\it Phys.\ Lett.\ }{\bf B#1} (#2) #3}

\newcommand\prd[3]{{\it Phys.\ Rev.\ }{\bf D#1} (#2) #3}

\newcommand\prep[3]{{\it Phys.\ Rept.\ }{\bf #1} (#2) #3}

\newcommand\ibid[3]{{\it ibid.\ }{\bf #1} (#2) #3}
\newcommand{\hepth}[1]{{\tt hep-th/#1}}

\newcommand{\mb}[1]{\mathbf{#1}}
\newcommand{\mbb}[1]{\mathbf{\overline{#1}}}
\begin{document}

\thispagestyle{empty}

\begin{flushright}
AEI-2003-081 \\
{\tt hep-th/0310193}
\end{flushright}

\vsp{1}
\begin{center}
{\large \textbf{On instanton contributions to anomalous dimensions 
in $\scrN$=4 supersymmetric Yang--Mills theory\rule{0pt}{18pt}}}

\vsp{1}
\it{\textbf{Stefano Kovacs}} 

\vsp{0.5}
\it{Max-Planck Institut f\"ur Gravitationsphysik \\
Albert-Einstein-Institut \\
Am M\"uhlenberg, 1 -- 14476, Golm, Germany} 

\vsp{0.5}
{\tt stefano.kovacs@aei.mpg.de}
\end{center}

\vsp{0.5}
\begin{abstract}
\ndt 
Instanton contributions to the anomalous dimensions of gauge-invariant
composite operators in the $\scrN$=4 supersymmetric SU($N$)
Yang--Mills theory are studied in the one-instanton
sector. Independent sets of scalar operators of bare dimension
$\D_0\le5$ are constructed in all the allowed representations of the
SU(4) R-symmetry group and their two-point functions are computed in
the semiclassical approximation. Analysing the moduli space integrals
the sectors in which the scaling dimensions receive non-perturbative
contributions are identified. The requirement that the integrations
over the fermionic collective coordinates which arise in the instanton
background are saturated leads to non-renormalisation properties for a
large class of operators.  Instanton-induced corrections to the
scaling dimensions are found only for dimension 4 SU(4) singlets and
for dimension 5 operators in the $\mb6$ of SU(4). In many cases the
non-renormalisation results are argued to be specific to operators of
small dimension, but for some special sectors it is shown that they
are valid for arbitrary dimension. Comments are also made on the
implications of the results on the form of the instanton contributions
to the dilation operator of the theory and on the possibility of
realising its action on the instanton moduli space.

\end{abstract}

\newpage
\setcounter{page}{1}
{\footnotesize
\tableofcontents
}

\section{Introduction and summary of results}
\label{intro}

Recent progress in the study of the interconnections between string
and gauge theories has revived the interest for the computation of
anomalous dimensions in $\scrN$=4 supersymmetric Yang--Mills (SYM)
theory. In a conformal field the spectrum of scaling dimensions,
together with the set of OPE coefficients for a complete basis of
operators, constitutes the fundamental physical information. In the
original formulation of the AdS/CFT correspondence \cite{m,gkp,w},
which in its best understood form relates the $\scrN$=4 SYM theory
with SU($N$) gauge group to type IIB string theory in AdS$_5\times
S^5$, the possibility of comparing this physical information with the
dual string theory quantities is limited by the present inability of
quantising strings in the relevant background. In the correspondence
scaling dimensions of gauge-invariant composite operators are related
to the energy of the dual string/supergravity states. The parameters
of the string and gauge theories are related by
\be
4\pi\gs = \gy^2 \,, \qquad \left(\frac{L^2}{\al}\right)^2 = 
\gy^2N \equiv\lambda \, ,
\label{dict}
\ee 
which imply the strong/weak coupling nature of the duality: in the
limit in which classical supergravity is a good approximation, which
is the only limit in which string theory in AdS$_5\times S^5$ is under
control, the gauge theory  is strongly coupled.  This makes a direct
comparison of quantities calculated at weak coupling in field theory
with quantities computed in supergravity impossible. This is true in
particular for the comparison of field theory scaling dimensions and
energies of supergravity states. 

In the early days of the correspondence anomalous dimensions first
emerged in connection with logarithmic singularities in the short
distance limit of four-point functions \cite{bkrs1}. The singularities
in the so called conformal partial wave  expansion of four-point
amplitudes in supergravity \cite{adslog} are interpreted as due to
quantum corrections to the mass of states exchanged in intermediate
channels. A similar analysis of the operator product expansion (OPE)
of the dual four-point correlation functions in field theory displays
the same type of singularities \cite{cftlog}, which in turn are
attributed to anomalous dimensions of operators entering the OPE
\cite{bkrs1}. Although the physical mechanisms which are responsible
for the observed divergences are correctly identified, a quantitative
comparison is impossible because the two calculations have different
regimes of validity. This is a general feature and until recently, in
spite of what appear to be non-trivial tests of the duality,
relatively little insight into truly dynamical aspects had been
obtained.

An important step forward was made with the proposal of \cite{bmn}
where a limit of the AdS/CFT correspondence was identified in which
the comparison between string theory and gauge theory can be carried
on avoiding the strong/weak coupling problem. On the string side one
considers a maximally supersymmetric plane-wave background obtained as
Penrose limit of AdS$_5\times S^5$ \cite{bfhp}. Although there is a
nontrivial \RR condensate, the quantisation of string theory in this
background is feasible because the world sheet theory in the
light-cone gauge reduces to a free massive model \cite{mt}. The dual
of string theory in this background was identified in \cite{bmn} as a
particular sector of $\scrN$=4 SYM made of operators of large scaling
dimension, $\D$, and with large value of the charge, $J$, with respect
to one of the R-symmetry generators. The holographic nature of the
duality in this limit, referred to as BMN limit, is not well
understood, but a prescription for comparing the mass spectrum of the
string with the spectrum of dimensions of gauge theory operators has
been given. Using this prescription a remarkable agreement between
string and field theory results has been obtained in perturbation
theory \cite{sz,bkpss,cfhm}. What is particularly interesting in the
BMN correspondence is that the comparison can be extended beyond
supergravity to include massive string modes and moreover it can be
carried on in a quantitative way. This is because the duality arises
in a double scaling limit in which the rank of the gauge group is sent
to infinity and at the same time $J\to\infty$, in such a way that
$J^2/N$ is kept fixed. In this limit and in the BMN sector of the
gauge theory the relevant parameters are $\lambda^\prime=\gy^2N/J^2$
and $g_2=J^2/N$ \cite{kpss,cfhmmps}. The dual string theory can also
be formulated in terms of the same parameters and in the BMN limit on
both sides one can consider an expansion for small values of
$\lambda^\prime$ and $g_2$.

Another new idea, which was suggested in \cite{gkp2} extending
considerations in \cite{p}, has lead to further exciting
developments. What was noticed in these papers is that in certain
limits, in which some quantum numbers take large values, the
semiclassical description of the AdS$_5\times S^5$ string sigma model
can represent a good approximation. These concepts have been employed
in several recent papers \cite{ft0,r,bmsz,t1,ft1,ft2,ft3,afrt,bfst} in
which solitonic solutions of the sigma-model, corresponding to strings
rotating in AdS$_5\times S^5$, have been put in correspondence with
gauge theory operators. The comparison of the energy of such solitonic
string configurations with the anomalous dimensions of the dual
operators has opened the possibility of new quantitative tests of the
AdS/CFT correspondence, which appear to have a dynamical nature.  Very
accurate comparisons of the spectra on the two sides of the
correspondence have been performed. An important r\^ole in these
calculations in the gauge theory is played by the observation that the
problem of computing anomalous dimensions can be rephrased as an
eigenvalue problem for the dilation operator of the theory
\cite{j,mz,bkps,bks}. When the problem is recast in these terms the
structure of an integrable system emerges in the planar limit
\cite{mz,bs1}. This allows to apply the techniques of the Bethe ansatz
to the computation of anomalous dimensions. The integrability emerging
in the planar limit of the $\scrN$=4 theory appears to play a crucial
r\^ole in these recent tests of the AdS/CFT duality and is believed to
have a counterpart in the integrability of the free string theory in
AdS$_5\times S^5$ \cite{msw,bpr,dnw}.

All the recent developments mentioned here concern the perturbative
sector of the gauge theory and very little has been done in the study
of quantum corrections to anomalous dimensions beyond perturbation
theory. One of the purposes of this paper is to initiate a systematic
analysis of the non-perturbative instanton-induced corrections to the
spectrum of anomalous dimensions in $\scrN$=4 SYM. A fundamental
motivation for this comes from the observation that instantons are 
expected to play a special r\^ole in $\scrN$=4 SYM due the S-duality
of the theory. 

The study of instanton effects in the AdS/CFT correspondence provides
an example of a situation in which the agreement between string and
gauge theory results appears highly non-trivial already at the level
of supergravity. Here the duality relates effects of instantons in
$\scrN$=4 SYM and D-instantons in type IIB string theory
\cite{bgkr}. Instanton corrections to correlation functions have a
counterpart in contributions to string/supergravity amplitudes induced
by the presence of D-instantons.  In the large $N$ limit
multi-instanton effects in $\scrN$=4 SYM can be explicitly computed
and compared with AdS amplitudes involving certain D-instanton induced
vertices in the type IIB effective action. Such vertices appear in the
$\al$ expansion, \ie the low energy derivative expansion, of type IIB
supergravity and their structure is determined by supersymmetry and
S-duality constraints. For a large class of processes a striking
agreement between supergravity multi-particle amplitudes involving the
leading D-instanton vertices and multi-instanton contributions to the
dual Yang--Mills correlation functions was found in \cite{dhkmv},
generalising the results of \cite{bgkr,dkmv}. The correspondence has
been further extended to include higher order D-instanton terms in
\cite{gk}. What makes the result of the comparison remarkable is that
perfect agreement is found in the functional form of correlation
functions which depend non-trivially on the coupling.

The non-perturbative results just described refer to a class of four-
and higher-point correlation functions of protected operators. All the
cases analysed in the past involve 1/2 BPS operators, which are dual
to supergravity states and their Kaluza--Klein excited modes. In view
of the success in the comparison with D-instanton effects at the level
of supergravity on the one hand and of the recent progress in matching
the gauge and string theory spectra at the level of massive string
excitations on the other, it is interesting to study instanton
contributions to correlation functions of operators dual to massive
string modes. In the present paper we shall consider two-point
functions of gauge-invariant scalar operators of bare dimension
$\D_0\le5$. These include BPS operators dual to supergravity fields,
multi-trace operators dual to multi-particle bound states as well as
single-trace operators in long multiplets dual to massive excitations
of type IIB string theory.

As is well known, two-point functions in a conformal field theory like
$\scrN$=4 supersymmetric Yang--Mills encode the information about
scaling dimensions. For primary operators $\scrO(x)$ conformal
invariance determines the form of two-point functions to be
\be 
\la \bar\scrO(x)\scrO(y)\ra = \frac{c}{(x-y)^{2\D}} \, ,
\label{cft-2pt}
\ee
where $c$ is a constant and the scaling dimension is $\D=\D_0+\g$, with
$\g$ the anomalous part which has an expansion at weak
coupling. By computing (\ref{cft-2pt}) at the instanton level one can
extract the corresponding contribution to $\g$. In general the
anomalous dimensions have an expansion of the form
\be
\g(\gy,N) = \sum_{n=1}^\infty \g^{\rm pert}_n(N)\,\gy^{2n} + 
\sum_{K>0}\sum_{m=0}^\infty \left[\g^{(K)}_m(N)\,\gy^{2m}\,
\er^{2\pi i\tau K} +\;\mathrm{c.c.} \, \right] \, ,
\label{gamma-exp}
\ee
where $\tau = \frac{\theta}{2\pi}+i\frac{4\pi}{\gy^2}$ and
anti-instantons give the complex conjugate of the instanton
contributions, since the anomalous dimensions are real quantities.

The  $\scrN$=4 theory is believed to possess a SL(2,$\mathbbm{Z}$)
S-duality symmetry under which the complexified coupling $\tau$
transforms projectively, $\tau\to\frac{a\tau+b}{c\tau+d}$ (where
$ad-bc=1$ with integer $a$, $b$, $c$, $d$). In the dual type IIB
string theory D-instantons are instrumental in implementing the
constraints imposed by S-duality, for instance on the form of the low
energy effective action. On the field theory side S-duality requires
the invariance of the spectrum of scaling dimensions and instantons
are expected to play a crucial r\^ole in the implementation of this
symmetry. The invariance of the spectrum  under SL(2,$\mathbbm{Z}$)
transformations suggests that single anomalous dimensions should be
functions of $\tau$ and $\bar\tau$,
\be 
\D=\D(\tau,\bar\tau;N) \equiv \D_0 + \g(\tau,\bar\tau;N) \, ,
\label{dim-tau}
\ee
so that in general a contribution from the non-perturbative part in
(\ref{gamma-exp}) is expected. 

In order to compute instanton corrections to anomalous dimensions in
the following we shall study two-point functions of gauge-invariant
scalar operators of bare dimension $\D_0=2,3,4$ and $5$. The
calculations will be performed in the semiclassical approximation and
in the one-instanton sector of the SU($N$) $\scrN$=4 supersymmetric
Yang--Mills theory. This means that expectation values reduce to
finite dimensional integrals over the moduli space parametrised by the
bosonic and fermionic collective coordinates which arise in the
instanton background. The general aspects of these calculations are
reviewed in section \ref{sc-corrfunct} and the case of two-point
correlators is discussed in section \ref{k12ptfunct}. The methods
developed in this paper can be applied to non Lorentz-scalar operators
as well. In this case additional quantum numbers need to be taken into
account in the construction of independent sets of operators, but the
calculation of instanton corrections to two-point correlation
functions does not present any new complications.

The quantum numbers that characterise scalar operators are, besides
the dimension $\D_0$, the Dynkin labels, $[a,b,c]$, determining their
transformation under the SU(4) R-symmetry. The parameters
$(\D_0;[a,b,c])$ identify sectors in the theory and as discussed in
section \ref{genadim} in order to extract the anomalous dimensions one
has to compute all the two-point functions to resolve the mixing among
operators in any given sector. One of the outcomes of our analysis is
that the problem of operator mixing is in general more complicated at
the instanton level than in perturbation theory: in sectors where
instantons contribute we find that mixing occurs among all the
operators at the same leading order in the coupling.  We shall not
compute all the entries in the mixing matrices for the sectors we
shall examine and thus  it will not be possible to pin down the actual
values of the instanton induced anomalous dimensions. However it will
be possible, analysing the moduli space integrations, to unambiguously
identify the sectors in which the scaling dimensions receive instanton
corrections.

Rather surprisingly the  calculations of section \ref{scalaradim} show
that the majority of the scalar operators of dimension $\D_0\leq 5$ do
not get instanton  corrections. This is an unexpected result in view
of the above considerations on S-duality: many of the operators that
we examine are corrected in perturbation theory, but not at the
instanton level. A similar behaviour had been observed in
\cite{bkrs1,bkrs3} for operators in the Konishi multiplet. The results
presented here might suggest that this is a more general feature of
the theory, however in the concluding discussion we shall argue that
this is probably not the case and that the non-renormalisation
properties we find are likely to be specific to operators of small
dimension. At $\D_0=2$ there are two operators and all their two-point
functions vanish in the one-instanton sector. This was a known result
since one of the operators belongs to a protected 1/2 BPS multiplet
and the other is a component of the Konishi multiplet, whose anomalous
dimension was shown not to receive instanton  corrections in
\cite{bkrs1,bkrs3}. Among the operators of dimension 3 there is only
one whose structure allows instanton contributions, it is the unique
operator $\scrO^i_{3,\mb6}$ in the $\mb6$ of SU(4). However, although
this is hard to see from the two-point functions, we shall give
another argument, based on the OPE analysis of the four-point function
computed in appendix \ref{ope}, showing that $\scrO^i_{3,\mb6}$ has no
instanton induced  anomalous dimension. In all the other sectors at
$\D_0=3$, corresponding  to the representations $\mb{10}$, $\mbb{10}$
and $\mb{50}$, there are no instanton corrections. At $\D_0=4$ we find
the first non-vanishing instanton contributions. For operators in the
SU(4) singlet we find non-vanishing two-point functions. Even without
computing the actual values of the anomalous dimensions, the results
of section \ref{d4r1} show that operators in this sector have scaling
dimensions which are corrected by instantons. Operators in all the
remaining sectors at $\D_0=4$, \ie in the SU(4) representations
$\mb{15}$, $\mb{20^\pp}$, $\mb{45}$, $\mbb{45}$, $\mb{84}$ and
$\mb{105}$, are not renormalised by instantons. Similarly among the
operators of bare dimension $\D_0=5$ only those transforming in the
$\mb6$ have non-vanishing two-point functions in the one-instanton
background. These are more complicated to compute because their
evaluation at leading order in the coupling requires the inclusion of
the leading quantum fluctuations around the instanton
configuration. Explicit examples of calculations of such effects will
be given in section \ref{d5r6}. Again, in all the other SU(4) sectors
at $\D_0=5$ ($\mb{10}$, $\mbb{10}$, $\mb{50}$, $\mb{64}$, $\mb{70}$,
$\mbb{70}$, $\mb{126}$, $\mbb{126}$, $\mb{196}$ and $\mb{300}$) no
instanton corrections to two-point functions are found. For both the
singlet at $\D_0=4$ and the $\mb6$ at $\D_0=5$ the OPE analysis of the
four-point function computed in appendix \ref{ope} confirms the
presence of non-perturbative corrections to the scaling dimensions.

In the calculation of anomalous dimensions important simplifications
arise from the use of the constraints imposed by the PSU(2,2$|$4)
global symmetry. First of all supersymmetry implies that all the
operators in a multiplet must have the same anomalous
dimension. Therefore the above results for operators with $\D_0\le 5$
imply the absence of instanton corrections to a much larger set of
operators: all the components of the multiplets for which a
representative is found to be non-renormalised share the same
(vanishing) anomalous dimension. Taking into account this fact and the
structure of the PSU(2,2$|$4) multiplets, which can be determined
systematically using the method of \cite{bms}, the problem of
extracting the values of the anomalous dimension for operators that
receive corrections can be significantly simplified. When a direct
analysis is complicated it is usually possible to identify
superconformal descendants which can be more easily studied. Then
superconformal symmetry implies that the result for the anomalous
dimension extends to the original operators. An explicit example of
this is represented by the case of the Konishi multiplet: the absence
of instanton corrections to the superconformal primary (as well as to
the whole multiplet) is obtained as consequence of the
non-renormalisation of a suitably chosen descendant. The same idea can
be used to simplify the resolution of the operator-mixing when
instanton corrections are present.

As already observed the above results for $\D_0\le5$ probably do not
reflect a general  property of the theory, but rather are
characteristic of small dimension sectors. However some
non-renormalisation properties can be generalised to arbitrary
$\D_0$. This is in particular the case for the absence of instanton
corrections in the SU(2) subsector identified in \cite{bks}
consisting of scalar operators transforming in the representation
$[a,b,a]$ and with bare dimension $\D_0=2a+b$. In \cite{bks,b2} it
was shown that the action of the dilation operator on such operators
is particularly simple. We shall argue that operators of this type do
not receive instanton corrections. Superconformal symmetry implies
that this result is valid for all the components of multiplets
containing operators of this type.

The paper is organised as follows. In section \ref{sc-corrfunct} we
review general aspects of instanton calculus in $\scrN$=4 SYM. Bosonic
and fermionic collective coordinates in the one-instanton sector are
described together with the integration measure on the moduli space
used in the semiclassical calculations. The structure of the $\scrN$=4
instanton supermultiplet, \ie of the instanton solution for the
elementary fields, is described in section \ref{N4instmult}.
Generalities on the calculation of instanton induced two-point
functions and their relation with corrections to the anomalous
dimensions are discussed in section \ref{2ptfunct}.  Section
\ref{scalaradim} presents the calculation of one-instanton
contributions to two-point functions of operators with $\D_0\le 5$ and
contains the main results of the paper. The final section contains a
discussion of the results and some comments on the possibility of
realising the action of the dilation operator on the instanton moduli
space. Appendix \ref{convents} summarises our notation and in appendix
\ref{k1adhm} a brief summary of the ADHM description of the
one-instanton sector of $\scrN$=4 SYM is provided. In appendix
\ref{ope} we present the calculation of a four-point function, which
allows to resolve ambiguities left from the analysis of two-point
functions in sections \ref{d3r6} and \ref{d5r6}.

\section{One-instanton contributions to correlation functions in 
$\scrN$=4 SYM} 
\label{sc-corrfunct}

In this section we present general aspects of the computation of
instanton effects in semiclassical approximation in the $\scrN$=4
supersymmetric Yang--Mills theory.

The general instanton configuration can be described using the ADHM
formalism \cite{adhm}. The original construction of self-dual
configurations in pure Yang--Mills theory with arbitrary classical
gauge  group has been generalised to the case of theories with
different field content and in particular to supersymmetric
theories. Comprehensive reviews of instanton calculus in
supersymmetric gauge theories can be found in \cite{akmrv,dhkm}. In
particular \cite{dhkm} contains a detailed description of
multi-instantons in the $\scrN$=4 SYM theory. In the following we
shall focus on the one-instanton sector of the $\scrN$=4 theory with
SU($N$) gauge group. Many of the results of the present paper however
remain valid in sectors with arbitrary instanton number, $K$. The
extension to orthogonal and symplectic groups also does not present
particular problems from the point of view of the instanton
calculations. Appendix \ref{k1adhm} contains a brief summary of the
ADHM description of the one-instanton sector of the SU($N$) $\scrN$=4
SYM theory.

We begin by discussing instanton contributions to generic correlation
functions of gauge invariant composite operators in  semiclassical
approximation. The specific case of two-point functions of such
operators, which is relevant for the computation of instanton-induced
anomalous dimensions will be considered in section \ref{2ptfunct}.

At the semiclassical level expectation values are evaluated
using a saddle point approximation around the instanton
configuration. For a correlation function of generic local operators
$\scrO_i$ the path integral reduces to a finite dimensional
integration over the moduli space of the instanton, \ie an integration
over the unfixed parameters (moduli or collective coordinates) of the
generic instanton configuration as described by the ADHM construction
\be 
\la\scrO_1(x_1)\ldots\scrO_n(x_n)\ra  = \int \dr \mu_{\rm
inst}(\scrm_{\rm \,b},\scrm_{\rm \,f})  \, \er^{-S_{\rm inst}} \,
\hat\scrO_1(x_1;\scrm_{\rm \,b},\scrm_{\rm \,f}) \ldots
\hat\scrO_n(x_n;\scrm_{\rm \,b},\scrm_{\rm \,f})\, .
\label{semiclass}
\ee
In (\ref{semiclass}) we have denoted the bosonic and fermionic
collective coordinates by $\scrm_{\rm \,b}$ and $\scrm_{\rm \,f}$
respectively and by $\dr\mu_{\rm inst}(\scrm_{\rm \,b},\scrm_{\rm
\,f})$  the corresponding integration measure; $S_{\rm inst}$ is the
classical action evaluated on the instanton solution and
$\hat\scrO_r$, $r=1,\ldots,n$,  denotes the classical expression for
the operator $\scrO_r$ computed in the instanton background.

In the one-instanton sector and with SU($N$) gauge group there are
4$N$ bosonic collective coordinates entering into the integration
measure in (\ref{semiclass}). With a particular choice of
parametrisation these bosonic moduli can be identified with the size,
$\rho$, and position, $x_0$, of the instanton as well as its global
gauge orientation. The latter can in turn be described by three angles
identifying an SU(2) instanton and 4$N$ additional constrained
variables, $w_{u\adot}$ and $\bar w^{\adot u}$ (where $u$ is a colour
index), in the coset  SU($N$)/(SU($N-2$)$\times$U(1)) describing the
embedding of the SU(2)  configuration into SU($N$). In the $\scrN$=4
theory there are 8$N$ fermionic collective coordinates as well in the
one-instanton sector. These correspond to zero modes of the Dirac
operator in the background of an instanton. They comprise the 16 moduli
associated with Poincar\'e and special supersymmetries broken by the
instanton and denoted respectively by $\eta^A_\a$ and $\bar\xi^{\adot
A}$ (where $A$ is an index in the fundamental of the SU(4) R-symmetry
group) and 8$N$ additional parameters, $\nu^A_u$ and $\bar\nu^{Au}$,
which can be considered as the fermionic superpartners of the gauge
orientation parameters. The fermion modes $\nu^A_u$ and $\bar\nu^{Au}$
satisfy the constraints
\be
\bar w^{\adot u}\nu^A_u = 0 \, , \quad \bar\nu^{Au}w_{u\adot} = 0 \, ,
\label{nuconstraint}
\ee
which effectively reduce the number of independent variables of this
type to $8(N-2)$. 

In the computation of correlation functions of gauge-invariant
operators as the ones we shall be interested in the moduli space
integration measure in (\ref{semiclass}) simplifies. The only
non-trivial bosonic integrals are over $\rho$ and $x_0$. Moreover in
the case of the $\scrN$=4 theory of the total set of 8$N$ fermionic
moduli dictated by the index theorem, only the $\eta^A_\a$ and
$\bar\xi^{\adot A}$ moduli associated with broken supersymmetries
correspond to true zero-modes when the interactions are taken into
account. The instanton action acquires a non-trivial dependence on the
other modes, $\nu^A_u$ and $\bar\nu^{Au}$, which means that the
integration measure has an explicit dependence on these modes. The
moduli space integration measure for the $\scrN$=4 supersymmetric
Yang--Mills theory in the generic $K$ instanton sector was constructed
in \cite{dhkmv}. In the one-instanton sector the gauge-invariant
measure takes the form
\ba 
&& \int \dr\mu_{\rm phys} \, \er^{-S_{\rm inst}} 
\label{physmeasure} \\
&& = \frac{\pi^{-4N}\gy^{4N}\er^{2\pi i\tau}}{(N-1)!(N-2)!}  
\int \dr\rho\,\dr^4x_0 \,\prod_{A=1}^4
\dr^2\eta^A\dr^2\bar\xi^A \, \dr^{N-2}\nu^A \dr^{N-2}\bar\nu^A
\,\rho^{4N-13} \er^{-S_{4F}} \, , \nn
\ea
where the instanton action is \cite{dkm1}
\be
S_{\rm inst} = -2\pi i \tau + S_{4F} 
= -2\pi i \tau + \frac{\pi^2}{2\gy^2\rho^2}
\veps_{ABCD} \scrF^{AB}\scrF^{CD} 
\label{instaction1}
\ee
with
\be 
\tau=\frac{4\pi i}{\gy^2}+\frac{\theta}{2\pi} \, , \quad 
\scrF^{AB}=\frac{1}{2\sqrt{2}}(\bar\nu^{Au}\nu_u^B-\bar\nu^{Bu}\nu_u^A)
\,. 
\label{instaction2}
\ee
In (\ref{physmeasure}) we have omitted an overall numerical constant,
independent of $\gy$ and $N$, that will be reinstated in the final
expression. Following \cite{dkmv} the integration measure can be
written in the form
\ba
\frac{\pi^{-4N}\gy^{4N}\er^{2\pi i\tau}}{(N-1)!(N-2)!} 
&& \!\!\!\!\!\!\!
\int \dr\rho\,\dr^4x_0 \,\dr^6\chi \,\prod_{A=1}^4 
\dr^2\eta^A\dr^2\bar\xi^A \, \dr^{N-2}\nu^A \dr^{N-2}\bar\nu^A \nn \\
&& \rho^{4N-7} \exp\left[-2\rho^2\chi^i\chi^i
+\frac{4\pi i}{\gy}\chi_{AB}\scrF^{AB}\right] \, , 
\label{physmeaschi}
\ea
where auxiliary bosonic variables, $\chi^i$, $i=1,\ldots,6$, have been 
introduced to rewrite the exponential of the quartic fermionic action 
as a gaussian integral. In (\ref{physmeaschi}) 
\be
\chi_{AB} = \frac{1}{\sqrt{8}}\S^i_{AB}\chi^i \, ,
\label{chiAB}
\ee
where the symbols $\S^i_{AB}$ denote Clebsch-Gordan coefficients
projecting the product of two $\mbb4$'s of SU(4) onto the
$\mb6$ and are defined in appendix \ref{convents}. 

Once the measure on the instanton moduli space is written in the form
(\ref{physmeaschi}) the integration over $\nu^A_u$ and $\bar\nu^{Au}$
is gaussian and can be immediately performed. However  in general in
correlation functions of gauge-invariant operators there is also a
non-trivial dependence on these variables coming from the expressions
for  the operators in the instanton background. It is thus convenient
to construct a generating function as in \cite{gk}, which allows to
deal easily with the otherwise complicated combinatorics associated
with the fermionic integrations over $\nu^A_u$ and $\bar\nu^{Au}$. We
introduce sources, $\bar\vt^u_A$ and $\vt_{Au}$, coupled to $\nu^A_u$
and $\bar\nu^{Au}$ and define
\ba
Z[\vt,\bar\vt] &\!\!=\!\!&\frac{\pi^{-4N}\gy^{4N}\er^{2\pi i\tau}}
{(N-1)!(N-2)!} \int \dr\rho \, \dr^4x_0 \, \dr^6\chi
\prod_{A=1}^4 \dr^2\eta^A \, \dr^2 {\bar\xi}^A\,
\dr^{N-2}{\bar\nu}^A\, \dr^{N-2}\nu^A \nn \\
&& \rho^{4N-7} \exp\left[-2\rho^2 \chi^i\chi^i+
\frac{\sqrt{8}\pi i}{\gy}{\bar\nu}^{Au}\chi_{AB}\nu^B_u +
{\bar\vt}^u_{A}\nu^A_u + \vt_{Au}{\bar\nu}^{Au} \right] \, .
\label{genfunct}
\ea
Performing the gaussian  ${\bar\nu}$ and $\nu$ integrals and
introducing polar coordinates, 
\be
\chi^i \, \rightarrow \, (r,\Omega) \: , \quad 
\sum_{i=1}^6 (\chi^i)^2 = r^2 \, ,
\label{polaromega}
\ee
$Z[\vt,\bar\vt]$ can be put in the form
\ba
Z[\vt,\bar\vt] &\!\!=\!\!&
\frac{2^{-29}\pi^{-13}\,\gy^{8}\er^{2\pi i\tau}}{(N-1)!(N-2)!}
\int \dr\rho \, \dr^4x_0 \, \dr^5\Omega \prod_{A=1}^4 \dr^2\eta^A \,
\dr^2 {\bar\xi}^A \, \rho^{4N-7} \nn \\
&& \int_0^\infty \dr r \, r^{4N-3} \er^{-2\rho^2 r^2}
\scrZ(\vt,\bar\vt;\Omega,r) \, ,
\label{genfunctfin}
\ea
where all the numerical coefficients have been reinstated. In
(\ref{genfunctfin}) we have introduced the density
\be
\scrZ(\vt,\bar\vt;\Omega,r) = \exp\left[-\frac{i\gy}{\pi r}\,
{\bar\vt}_A^u \Omega^{AB} \vt_{Bu} \right] \, ,
\label{density}
\ee
where the symplectic form $\Omega^{AB}$ is given by 
\be
\Omega^{AB}= \frac{1}{\sqrt{8}}\bar\Sigma^{AB}_i\Omega^i \, ,
\quad \sum_{i=1}^6 \left(\Omega^i\right)^2 = 1 
\label{defOmega}
\ee
and the symbols $\bar\Sigma^{AB}_i$ are defined in appendix
\ref{convents}.  

In conclusion the semiclassical approximation to a correlation
function (\ref{semiclass}) is computed as 
\ba
&&\la\scrO_1(x_1)\ldots\scrO_n(x_n)\ra = 
\frac{2^{-29}\pi^{-13}\,\gy^{8}\er^{2\pi i\tau}}{(N-1)!(N-2)!}
\int \dr\rho \, \dr^4x_0 \, \dr^5\Omega \prod_{A=1}^4 \dr^2\eta^A \,
\dr^2 {\bar\xi}^A \, \rho^{4N-7} \nn \\
&& \hspace*{1cm}\int_0^\infty \dr r \, r^{4N-3} \er^{-2\rho^2 r^2}
\left[\scrZ(\vt,\bar\vt;\Omega,r) \, \hat\scrO_1\left(x_1;\rho,x_0;
\eta,\bar\xi,\nu(\Omega),\bar\nu(\Omega)\right)
\right. \nn \\
&& \hspace*{4.6cm} \left. \ldots \hat\scrO_n
\left(x_n;\rho,x_0;\eta,\bar\xi,\nu(\Omega),\bar\nu(\Omega)\right)
\right] \! \raisebox{-4pt}{$\Big|_{\vt=\bar\vt=0}$} \, , 
\label{smclasscorr}
\ea
where the $\nu^A_u$ and $\bar\nu^{Au}$ variables in each $\hat\scrO_r$
are understood to be rewritten in terms of derivatives with respect to
the sources, $\bar\vt^u_A$ and $\vt_{Au}$, before setting the latter
to zero. As discussed in \cite{gk} the use of the generating function
to express a correlator as in (\ref{smclasscorr}) allows in particular
to easily  determine the dependence on the parameters $\gy$ and $N$.

The variables $\nu^A_u$ and $\bar\nu^{Au}$ enter in the expressions of
gauge-invariant composite operators in the instanton background always
in colour singlet bilinears. They arise in pairs either in a
combination which is in the $\mb6$ of the SU(4) R-symmetry 
\be
(\bar\nu^A \nu^B)_{\mb6} \equiv
\nus = ({\bar\nu}^{Au}\nu^B_u - {\bar\nu}^{Bu}\nu^A_u) \, ,
\label{6bilinear}
\ee
or in the $\mb{10}$ of SU(4)
\be
(\bar\nu^A \nu^B)_{\mb{10}} \equiv
\nut = ({\bar\nu}^{Au}\nu^B_u + {\bar\nu}^{Bu}\nu^A_u) \, .
\label{10bilinear}
\ee
Using the generating function defined in (\ref{genfunctfin}) we  can
determine the dependence of a generic correlator on the parameters
$\gy$ and $N$. In the one-instanton sector the $N$-dependence can be
computed exactly. In particular in the large-$N$ limit from 
(\ref{smclasscorr}) we get 
\ba
\la\scrO_1(x_1)\ldots\scrO_n(x_n)\ra &\!\!\sim\!\!& 
\a(p,q;N)\,\gy^{8+p+q}\,\er^{2\pi i\tau}
\int \dr\rho \, \dr^4x_0 \, \dr^5\Omega \prod_{A=1}^4 \dr^2\eta^A \,
\dr^2 {\bar\xi}^A \, \rho^{p+q-5} \nn \\
&& \hat\scrO_1\left(x_1;\rho,x_0;
\eta,\bar\xi,\bar\nu\nu(\Omega)\right) \ldots \hat\scrO_n
\left(x_n;\rho,x_0;\eta,\bar\xi,\bar\nu\nu(\Omega)\right) \, , 
\label{largeN-dep}
\ea
with  
\ba
\a(p,q;N) &\!\!=\!\!&
\frac{2^{-2N+\half(p+q)}\,\pi^{-(p+q)}\,
\Gamma\left(2N-1-\half(p+q)\right)}
{(N-1)!(N-2)!} \left(N^{p+\frac{q}{2}}+
O(N^{p+\frac{q}{2}-1})\right) \nn \\
&\!\!\sim\!\!& N^{\half(p+1)}\left(1+O(1/N)\right) \, ,
\label{largeN-coeff}
\ea
where $p$ and $q$ denote respectively the number of $\nsix$ and
$\nten$ bilinears entering the integrand in (\ref{largeN-dep}). 

As will be discussed explicitly in the case of scalar operators of
bare dimension 5, in general the computation of two-point functions in
the instanton background at the first non-trivial order in the
coupling, $\gy$, requires to take into account the effect of the
leading quantum fluctuations around the classical configuration. To
compute leading order effects we need to include contributions in
which, instead of replacing all the fields with their background value
in the presence of an instanton, pairs of fields are contracted via a
propagator. The general construction of the scalar propagator in the
instanton background in terms of ADHM variables was given in
\cite{cgt}, an explicit expression for the scalar propagator in the
adjoint representation of the SU($N$) gauge group was presented in
\cite{gk}. In general the propagators for vectors and spinors are also
needed. These propagators can be deduced from the scalar propagator as
discussed in \cite{bccl}. The Green function for the adjoint scalars 
which will be relevant for some of the calculations in section
\ref{dim5scal} is given in appendix \ref{k1adhm}.

\section{The $\scrN$=4 instanton supermultiplet}
\label{N4instmult}

As discussed in the previous section, in order to evaluate instanton
induced correlation functions we need to integrate the classical
profiles~\footnote{In the following we shall use the word ``profile''
in a slightly loose sense to indicate the expression for a composite
operator computed in an instanton background including the dependence
on fermion zero-modes.} of the relevant composite operators over the
instanton moduli space. In preparation for such computations in this
section we shall discuss the structure of the instanton solution for
the elementary fields in the $\scrN$=4 supermultiplet.  We are
interested in the dependence on the collective coordinates and of
particular relevance will be the way the fermionic modes enter into
the expressions for the various fields.

The $\scrN$=4 multiplet consists of six real scalars, $\v^{AB}$, four
Weyl spinors, $\lambda^A_\a$, and a vector, $A_\mu$. Our notation is
summarised in appendix \ref{convents}. The field equations in the
$\scrN$=4 SYM theory take the form
\ba
&& \scrD_\mu F^{\mu\nu} +i\{\lambda^{\a A}
\s^\nu_{\a\adot},\bar\lambda_A^\adot \} + \half
[\bar\v_{AB},\scrD^\nu\v^{AB}] = 0 \nn \\
&& \scrD^2 \v^{AB} + \sqrt{2} \{\lambda^{\a A},\lambda^B_\a\} +
\frac{1}{\sqrt{2}}\veps^{ABCD}\{\bar\lambda_{\adot C},
\bar\lambda^\adot_D\} - \half [\bar\v_{CD},[\v^{AB},\v^{CD}]] = 0 
\nn \\
&& \bar{\Dscrm}_{\adot\a}\lambda^{\a A} +
i\sqrt{2}[\v^{AB},\bar\lambda_{\adot B}] = 0 \rule{0pt}{16pt}
\label{fieldeqs} \\
&& \Dscrm_{\a\adot}\bar\lambda_A^\adot - i\sqrt{2}
[\bar\v_{AB},\lambda^B_\a] = 0 \rule{0pt}{16pt} \nn \,. 
\ea
A solution to these equations is given by 
\be 
A_\mu = A_\mu^I \, , \quad \v^{AB} = \lambda^A_\a =
\bar\lambda^\adot_A = 0 \, ,
\label{bpstinst}
\ee
where $A_\mu^I$ is the standard instanton solution of SU($N$) pure
Yang--Mills theory. However the Dirac operator has zero modes in the
background of this solution, \ie the equation
$\bar{\Dscrm}_{\adot\a}\lambda^{\a A}=0$ has non-trivial solutions
when the covariant derivative is evaluated in the background of the
instanton. This is the origin of the fermionic zero-mode integrations
in the semiclassical expression for correlation functions
(\ref{smclasscorr}). Because of these fermionic integrations using the
solution (\ref{bpstinst}) to compute the classical profiles of the
operators $\hat\scrO_i$ in (\ref{smclasscorr}) is not sufficient and
we need to include the  zero-mode dependence in the operators. We must
thus solve the equations (\ref{fieldeqs}) iteratively in order to
determine the complete dependence on the fermion zero modes in the
multiplet of elementary fields \cite{bvv}. In the following we shall
use the notation $\Phi^{(n)}$ to denote a term in the solution for the
field $\Phi$ containing $n$ fermion zero modes.

Notice that in the simple case of SU(2) gauge group there are only 16
fermionic modes in a one-instanton background, the ones associated
with broken superconformal symmetries. In this case it is possible to
determine completely the zero mode dependence in the $\scrN$=4
supermultiplet acting with the broken supersymmetries on
(\ref{bpstinst}). Substituting $A_\mu^{(0)}\equiv A_\mu^I$ in the
supersymmetry transformation of $\lambda^A_\a$ gives a configuration,
$\lambda^{(1)A}_\a$, which is linear in the fermion modes $\eta^A_\a$
and $\bar\xi^{\adot A}$ and solves the corresponding field equation.
Then plugging $\lambda^{(1)A}$ into the variation of $\v^{AB}$
generates a solution for the scalar which is quadratic in the fermion
modes, $\v^{(2)AB}$. Iteration of this procedure gives rise to a
solution $\bar\lambda^{(3)}_{\adot A}$ and then to a contribution to
$A_\mu$ which corrects the original solution with the addition of a
term quartic in the fermion modes, $A^{(4)}_\mu$. In this way one can
construct the complete dependence on the fermion zero modes in the
$\scrN$=4 supermultiplet in the case of SU(2) gauge group. The
iteration continues until the number of zero modes in the fields
exceeds 16, at which point further variations do not produce new
independent terms in the solutions. This procedure can be implemented
in an efficient way using a superspace formalism. A general discussion
in the case of $\scrN$=1 supersymmetric Yang--Mills theory can be
found in \cite{sv}. 

In the case of SU($N$) gauge group the situation is more complicated
and the above procedure cannot be utilised. As already discussed in
this case there are the additional fermion zero modes $\nu^A_u$ and
$\bar\nu^{Au}$ which are not associated with symmetries broken by the
bosonic instanton solution. The dependence on these modes, which is
crucial in computing Green functions in semiclassical approximation,
cannot be obtained using symmetry arguments. Therefore we need to
explicitly construct the zero-mode dependence by solving the field
equations.

Starting with the solution $A_\mu^{(0)}$ in (\ref{bpstinst}), we solve
the equations (\ref{fieldeqs}) iteratively to generate solutions for
all the fields in the multiplet with an increasing number of fermionic
zero-modes. The first few steps in this construction have been carried
out in \cite{bvv}.

The term linear in the fermion modes in the solution for the spinor,
$\lambda^{(1)A}_\a$, is determined by the equation
\be
\bar{\Dscrm}^{(0)\adot\a} \lambda^{(1)A}_\a = 0 \, ,
\label{lambda1}
\ee
where the covariant derivative $\bar{\Dscrm}^{(0)\adot\a}$ contains
$A_\mu^{(0)}$. The subsequent steps give rise to $\v^{(2)AB}$ and
$\bar\lambda^{(3)}_{\adot A}$, which are obtained solving respectively  
\be
\Dscrm^{(0)2} \v^{(2)AB} + \sqrt{2} \{\lambda^{(1)\a A},
\lambda^{(1)B}_\a\} = 0 
\label{phi2}
\ee
and 
\be 
\Dscrm^{(0)}_{\a\adot}\bar\lambda_A^{(3)\adot} - i\sqrt{2}
[\bar\v^{(2)}_{AB},\lambda^{(1)B}_\a] = 0 \, .
\label{barlambda3}
\ee
Further iteration gives rise to corrections to the above lowest order
solution in which  each field has the minimal number of fermion
modes, $\{A^{(0)}_\mu,\,\lambda^{(1)A}_\a,\,\v^{(2)AB},\,
\bar\lambda^{(3)\adot}_A\}$. The following steps generate  the terms
$A^{(4)}_\mu$, $\lambda^{(5)A}_\a$ and $\v^{(6)AB}$ which solve
respectively
\be
\Dscrm^{(0)2}A^{(4)}_\mu - \Dscrm^{(0)}_\mu\Dscrm^{(0)}_\nu A^{(4)\nu}
+ 2 [F_{\mu\nu}^{(0)},A^{(4)\nu}] - i \{\bar\lambda^{(3)}_{\adot A}
\bar\sigma_\mu^{\adot\a},\lambda^{(1)A}_\a\} = 0 \, ,
\label{a4}
\ee
\be
\bar{\Dscrm}^{(0)}_{\adot\a}\lambda^{(5)\a A} + 
\bar{\Dscrm}^{(4)}_{\adot\a}\lambda^{(1)\a A} +
i\sqrt{2}[\v^{(2)AB},\bar\lambda^{(3)}_{\adot B}] = 0
\label{lambda5}
\ee 
and 
\ba
&& \scrD^{(0)2} \v^{(6)AB} + \scrD^{(4)2} \v^{(2)AB} + \sqrt{2} 
\{\lambda^{(1)\a A},\lambda^{(5)B}_\a\} 
+ \sqrt{2} \{\lambda^{(5)\a A},\lambda^{(1)B}_\a\} \nn \\ 
&& + \frac{1}{\sqrt{2}}\veps^{ABCD}\{\bar\lambda^{(3)}_{\adot C},
\bar\lambda^{(3)\adot}_D\} - \half [\bar\v^{(2)}_{CD},[\v^{(2)AB},
\v^{(2)CD}]] = 0 \, .
\label{phi6}
\ea

For generic operators in the $\scrN$=4 supersymmetric Yang--Mills
theory these new terms contribute and are needed to compute
correlation functions {\em at leading order in $\gy$}. In the special
case of 1/2 BPS operators, as those in the supercurrent multiplet, the
AdS/CFT correspondence suggests that only terms with the minimal
number of superconformal modes are allowed \cite{gk}. Operators of
this type are dual to the supergravity multiplet and its Kaluza--Klein
excitations in the type IIB string theory in AdS$_5\times S^5$ and the
AdS/CFT correspondence combined with knowledge of the structure of the
low energy effective action for the type IIB ``massless'' fields puts
restrictions on the set of correlation functions which can receive
instanton contributions. Such constraints restrict the maximal number
of superconformal modes that each operator can saturate. The same
constraints do not apply to the unprotected operators we shall
consider in the following and thus higher order terms will be needed
as well.

The procedure outlined here can in principle be employed to determine
the exact zero-mode structure of the solution. The actual
implementation of this construction becomes soon very involved as one
gets to higher order terms. The general solution takes the form
\ba
&& A_\mu = \hspace*{-0.3cm} \begin{array}[t]{c}
{\displaystyle \sum_{n=0}} \\
{\scriptstyle 4n \le 8N}
\end{array} \hspace*{-0.3cm}
A_\mu^{(4n)} \, , \hsp{1.4}  
\v^{AB} = \hspace*{-0.3cm} \begin{array}[t]{c}
{\displaystyle \sum_{n=0}} \\
{\scriptstyle 4n+2 \le 8N}
\end{array} \hspace*{-0.3cm}
\v^{(4n+2)AB} \nn \\
&& \lambda^A_\a = \hspace*{-0.3cm} \begin{array}[t]{c}
{\displaystyle \sum_{n=0}} \\
{\scriptstyle 4n+1 \le 8N}
\end{array} \hspace*{-0.3cm}
\lambda^{(4n+1)A}_\a \, , \hsp{0.5} 
\bar\lambda_{\adot A} = \hspace*{-0.3cm} \begin{array}[t]{c}
{\displaystyle \sum_{n=0}} \\
{\scriptstyle 4n+3 \le 8N}
\end{array} \hspace*{-0.3cm}
\bar\lambda^{(4n+3)}_{\adot A} \, , 
\label{N4multizeromod}
\ea
where it is also understood that in each field the number of
superconformal modes does not exceed 16 and the remaining modes are of
$\nu^A_u$ and $\bar\nu^{Au}$ type. 

In computing the expressions for gauge invariant composite operators
we shall make use of the ADHM description in which the elementary
fields are written as $[N+2]\times[N+2]$ matrices as discussed in
appendix \ref{k1adhm}. In the same appendix the leading order terms 
in the solution for $A_\mu$, $\lambda^A_\a$ and $\v^{AB}$, which
will be used in some of the examples presented in section
\ref{scalaradim}, are given explicitly.

Here we shall not discuss the details of the solution of the iterative
equations, but instead we shall only analyse the SU(4) structure,
which will suffice for the study of two-point functions to be carried
out in later sections. The scalar fields in the $\scrN$=4 multiplet,
$\v^i\sim\v^{AB}$, transform in the representation $\mb6$ (with
Dynkin labels $[0,1,0]$) of the  SU(4) R-symmetry group, the fermions,
$\lambda^A_\a$ and $\bar\lambda_A^\adot$, transform respectively in
the $\mb4$ ($[1,0,0]$) and $\mbb4$ ($[0,0,1]$) and the
vector, $A_\mu$, (as well as its field strength and the covariant
derivatives) is a singlet. The SU(4) structure of the combination of
fermion zero modes in the various terms in the iterative solution
discussed above can be determined without solving the equations
explicitly. All the fermion zero modes, both the superconformal ones,
$\eta^A_\a$ and $\bar\xi^{\adot A}$, and the modes of type $\nu^A_u$
and $\bar\nu^{Au}$, transform in the $\mb4$ of SU(4). We shall
denote a generic fermion mode by $\scrmf^A$. Inspecting the iterative
equations that determine the instanton multiplet we can deduce in
which SU(4) combinations the fermion modes enter each term. The
starting point is the classical instanton, $A_\mu^{(0)}$, which has no
fermions. The first term in $\lambda^A_\a$ is linear in the fermion
modes
\be
\lambda^{(1)A}_\a \sim \scrmf^A \, .
\label{lambda1su4}
\ee
For the term $\v^{(2)AB}$ in the scalar solution we find
\be
\v^{(2)AB} \sim \scrmf^{[A}\scrmf^{B]} \, ,
\label{phi2su4}
\ee
\ie the two fermion modes are antisymmetrised in order to form a
combination in the $\mb6$. The schematic notation of
(\ref{phi2su4}) indicates that the $[N+2]\times[N+2]$ matrix
$\v^{(2)AB}$ has entries which involve one mode of flavour $A$ and one
of flavour $B$ in all the possible combinations 
\ba
\eta^{[A}_\a\eta^{B]}_\b \,, \quad \eta^{[A}_\a\bar\xi^{\adot B]} \,, 
\quad \bar\xi^{\adot [A}\bar\xi^{\bdot B]} \,, \quad
\eta_\a^{[A}\bar\nu^{B]v} \,, \quad \nu^{[A}_u\eta^{B]}_\a \,, \quad 
\bar\xi^{\adot[A}\bar\nu^{B]v} \,, \quad \nu^{[A}_u\bar\xi^{\bdot B]} 
\,, \quad \nu_u^{[A}\bar\nu^{B]v} \, .
\label{phifermicomb}
\ea
The $\mbb4$ spinor $\bar\lambda^{(3)\adot}_A$ contains
fermion modes in the combination 
\be
\bar\lambda^{(3)\adot}_A \sim \veps_{ABCD} \,\scrmf^B\scrmf^C\scrmf^D
\,,  
\label{blambda3su4}
\ee
so that the component $\lambda^{(3)}_A$ has three fermion modes, one
of each of the flavours apart from $A$. Proceeding in the multiplet we
find the quartic term in the solution for the vector, $A_\mu^{(4)}$,
which contains one fermion mode of each flavour in a singlet
combination
\be
A_\mu^{(4)} \sim \veps_{ABCD}\,\scrmf^A\scrmf^B\scrmf^C\scrmf^D \, .
\label{a4su4}
\ee
The following term is $\lambda^{(5)A}_\a$, which has flavour structure
\be
\lambda^{(5)A}_\a \sim \veps_{A^\prime B^\prime C^\prime D^\prime}
\, \scrmf^A \scrmf^{A^\prime}\scrmf^{B^\prime}\scrmf^{C^\prime}
\scrmf^{D^\prime} \, ,
\label{lambda5su4}
\ee
\ie it involves a mode of flavour $A$ plus one of each
flavour. Then we find $\v^{(6)AB}$ that contains an antisymmetric
combination of a mode of flavour $A$ and one of flavour $B$ plus one
mode of each flavour
\be
\v^{(6)AB} \sim \veps_{A^\prime B^\prime C^\prime D^\prime}
\, \scrmf^{[A}\scrmf^{B]} \scrmf^{A^\prime}\scrmf^{B^\prime}
\scrmf^{C^\prime}\scrmf^{D^\prime} \, .
\label{phi6su4}
\ee
As observed after equation (\ref{phi2su4}) the previous expressions
are symbolic and the products of $\scrmf$'s in
(\ref{blambda3su4})-(\ref{phi6su4}) correspond to different
combinations of the modes $\eta^A_\a$, $\bar\xi^{\adot A}$, $\nu^A_u$
and $\bar\nu^{Au}$ in the various entries of the ADHM matrices for
each field. 

The SU(4) structure of the combinations of fermionic modes entering
into the higher order terms in the solution can be determined
analogously.  As already mentioned the knowledge of the flavour
structure of the solution described here will be enough to obtain some
interesting results concerning the anomalous dimensions of composite
operators without actually solving the field equations.

\section{Instanton induced two-point functions and anomalous
dimensions}
\label{2ptfunct}

Before analysing specific cases of scalar operators in section
\ref{scalaradim} we now describe the general strategy that will be
followed in such calculations. As usual in instanton calculus in the
evaluation of the moduli space integrals it is convenient to perform
the fermionic integrals first. In the case of two-point functions
these are particularly simple and their analysis will allow us to show
the absence of instanton corrections to a large class of operators.

\subsection{Two-point functions in the one-instanton sector}
\label{k12ptfunct}

In the special case of two-point functions the semiclassical
approximation takes the form~\footnote{In this general discussion we
omit overall numerical constants in the integration measure.}
\ba
&& \hsp{-0.5} \la \bar\scrO(x_1)\scrO(x_2) \ra = 
\frac{\pi^{-4N}\gy^{4N}\er^{2\pi i\tau}}{(N-1)!(N-2)!}  
\int \dr\rho\,\dr^4x_0 \,\prod_{A=1}^4
\dr^2\eta^A\dr^2\bar\xi^A \, \dr^{N-2}\nu^A \dr^{N-2}\bar\nu^A
\,\rho^{4N-13} \nn \\
&& \hsp{1}\er^{\frac{\pi^2}{16\gy^2\rho^2}\veps_{ABCD} 
(\bar\nu^{[A}\nu^{B]})(\bar\nu^{[C}\nu^{D]})} 
\,\,\,\hat{\!\!\bar{\scrO}}(x_1;x_0,\rho,\eta,\bar\xi,\nu,\bar\nu)
\hat{\scrO}(x_2;x_0,\rho,\eta,\bar\xi,\nu,\bar\nu) 
\, , \label{smclass2pt1}
\ea
where we have not yet rewritten the fermionic variables $\nu^A_u$ and
$\bar\nu^{Au}$ in terms of bosonic auxiliary variables. 
As already observed in the general discussion of section
\ref{sc-corrfunct} we have to distinguish between the superconformal
fermionic modes, $\eta^A_\a$ and $\bar\xi^{\adot A}$, and the
remaining ones, $\nu^A_u$ and $\bar\nu^{Au}$: the former do not appear
in the measure and must be soaked up by the operator insertions in
order for (\ref{smclass2pt1}) to yield a non vanishing result. 
The `non-exact' modes of type $\nu^A_u$ and $\bar\nu^{Au}$ appear in
the measure so that an explicit dependence on these variables in the
integrand is not required, although in general the classical profiles
of composite operators do depend on $\bar\nu^{A}\nu^B$ colour singlet 
bilinears. After translating the dependence on $\nu^A_u$ and
$\bar\nu^{Au}$ into a dependence on the angular variables
$\Omega^{AB}$ using the generating function as discussed in section
\ref{sc-corrfunct} the two-point function (\ref{smclass2pt1}) becomes 
\ba
\la \bar\scrO(x_1) \scrO(x_2) \ra = 
\frac{c(\gy,N)\gy^{8}\er^{2\pi i\tau}}{(N-1)!(N-2)!}
&& \!\!\!\!\!\!\!
\int \dr\rho \, \dr^4x_0 \, \dr^5\Omega \prod_{A=1}^4 \dr^2\eta^A \,
\dr^2 {\bar\xi}^A \rho^{4N-7} 
\label{smclass2pt2} \\ 
&& \,\,\,\hat{\!\!\bar{\scrO}}(x_1;x_0,\rho,\eta,\bar\xi,\Omega)
\hat\scrO(x_2;x_0,\rho,\eta,\bar\xi,\Omega) \, ,
\nn 
\ea
where the coefficient $c(\gy,N)$ contains additional dependence on
$\gy$ and $N$ arising from the integration over the radial variable $r$
introduced in (\ref{polaromega}). As mentioned previously $c(\gy,N)$
contains a power of $\gy$ for each $\bar\nu\nu$ bilinear in the integrand
and a factor of $\sqrt{N}$ plus $1/N$ corrections for each $\nsix$. 

As will be shown explicitly in the examples presented in section
\ref{scalaradim} the superconformal modes always appear in the
expressions for gauge-invariant composite operators in the combination 
\be
\zeta^A_\a(x) = \frac{1}{\sqrt{\rho}}\left[ \rho\,\eta^A_\a -
(x-x_0)_\mu \s^\mu_{\a\adot} \,\bar\xi^{\adot A} \right] \, .
\label{zetadef}
\ee
Since the two-component spinors $\zeta^A(x)$ satisfy
$\left(\zeta^A(x)\right)^3=0$, $\forall A,x$, it is clear that in a 
two-point function the only way of saturating the 16 integrations over
$\eta^A_\a$ and $\bar\xi^{\adot A}$ in (\ref{smclass2pt2}) is that
each of the two operators provides the combination 
\be
\left[\zeta^1(x)\right]^2 \left[\zeta^2(x)\right]^2 
\left[\zeta^3(x)\right]^2 \left[\zeta^4(x)\right]^2 \, ,
\label{saturmodes}
\ee
\ie each of the two operators must soak up two powers of the
superconformal combination (\ref{zetadef}) for of each of the four
flavours.  Unless this can be achieved the two-point function vanishes
and therefore the anomalous dimension of the operator does not get
instanton correction. This is a rather strong condition and it will
allow us, using the results of section \ref{N4instmult} to analyse
the dependence on the superconformal modes, to show the absence of
instanton corrections in many cases.

If the operators can indeed soak up the right combination of
superconformal modes the $\eta$ and $\bar\xi$ integrals are
non-vanishing and can be easily evaluated. Using a simple Fierz
rearrangement one finds
\be
\int \dr^2\eta^A \, \dr^2\bar\xi^A \, \left[\zeta^A(x_1)\right]^2 
\left[\zeta^A(x_2)\right]^2 = -(x_1-x_2)^2 \, ,
\quad \forall A=1,\ldots,4 \, .
\label{fermiint}
\ee

Once the integrations over the fermion superconformal modes have been
performed we are left with an integration over the five-sphere
parametrised by the angular variables $\Omega^{AB}$ and the bosonic
part of the moduli space integration over $x_0$ and $\rho$. The
five-sphere integration factorises and gives rise to further selection
rules. It gives a non-vanishing result only if the SU(4) indices
carried by the $\Omega$'s in the two operators can be combined to form
a SU(4) singlet.   This analysis of the fermionic moduli space
integrations can be repeated in the case of contributions with
contractions between pairs of fields in the operator insertions. The
same selection rules apply in these cases.

Before discussing specific examples we shall now briefly recall how
the information on anomalous dimensions is encoded in the two-point
functions and how it can be extracted from the final bosonic
integration over the moduli space.

\subsection{Anomalous dimensions}
\label{genadim}

Gauge invariant composite operators in the $\scrN$=4 theory are
classified in terms of their transformation under the PSU(2,2$|$4)
supergroup of global symmetries. Each operator is characterised by a
set of quantum numbers identifying the irreducible representation of
the maximal bosonic subgroup SO(2,4)$\times$SU(4) of PSU(2,2$|$4) it
transforms in. The quantum numbers defining such irreducible
representations are $(\D,J_1,J_2;[a,b,c])$, where the spins $J_1$ and
$J_2$ characterise the Lorentz group transformation and together with
the scaling dimension $\D$ determine the transformation under the
conformal group SO(2,4) and the remaining three numbers, $[a,b,c]$ are
Dynkin labels of SU(4).

Let us consider a sector in the $\scrN$=4 theory formed by operators
belonging to the same SU(4) and SO(1,3) representations (in the
following we shall restrict our attention to Lorentz scalars) and with
the same bare dimension, $\D_0$. In the quantum theory in general the
operators acquire an anomalous dimension which corrects the bare
value, $\D=\D_0+\g$, where the anomalous term is a function of the
parameters $\gy$ and $N$. Let us consider in such sector a complete
set of $n$ primary operators forming an orthonormal basis with respect
to the scalar product defined by the two-point functions. We shall
suppress Lorentz and SU(4) labels and denote the operators by
$\scrO_\D^r(x)$. We assume that the operators are the ones well
defined, \ie transforming properly under all the global symmetries, in
the full quantum theory. As is well known the spatial dependence in
two-point functions of primary operators is completely fixed by
conformal invariance. For the sector we are considering we have
\be
\la \bar\scrO^r_{\D_r}(x_1) \scrO^s_{\D_s}(x_2) \ra = 
\frac{\d^{rs}}{(x_1-x_2)^{2\D_r}} \, .
\label{gen2ptfunct}
\ee
In the full quantum theory $\D_r=\D_r(\gy,N)=\D_0+\g_r(\gy,N)$ and the
two-point function is non-zero only if $\bar\scrO^r_{\D_r}(x_1)$ and
$\scrO^s_{\D_s}(x_2)$ have the same dimension, $\D_r$. We assume that
in case there are operators with the same anomalous dimension they are
made orthogonal, so that the set $\left\{\scrO^r\right\}$ forms an
orthonormal  basis in the sector under investigation in this case as
well.

We want to compare (\ref{gen2ptfunct}) with the result of a small
coupling calculation in order to extract from the latter the
information about the anomalous dimensions. At small $\gy$ we assume
the anomalous dimension $\g(g,N)$ to be small and hence expand
(\ref{gen2ptfunct}) as 
\be
\la \bar\scrO^r_{\D_r}(x_1) \scrO^s_{\D_s}(x_2) \ra =
\frac{\d^{rs}}{(x_1-x_2)^{2\D_0}} \frac{1}{\mu^{2\g_r(\gy,N)}}
\left[ 1-\g_r(\gy,N) \log\left(\frac{x_1-x_2}{\mu}\right)^2 + \cdots
\right] \, ,
\label{smallgamma2pt}
\ee
where $\mu$ is an arbitrary length scale related to the
renormalisation scale in the small $\gy$ calculation. Clearly 
physical quantities, as the anomalous dimension $\g$, do not depend on
$\mu$. Equation (\ref{smallgamma2pt}) is a small-$\g$ expansion,
further expanding $\g(\gy,N)$ at small $\gy$ we get 
\ba
\la \bar\scrO^r_{\D_r}(x_1) \scrO^s_{\D_s}(x_2) \ra &\!\!=\!\!& 
\frac{c(\gy,N)\,\d^{rs}}{(x_1-x_2)^{2\D_0}} \left[ 
1-\gy^2\g^{(1)}_r(N) \log\left(\frac{x_1-x_2}{\mu}\right)^2 
\right. \nn \\
&& \left. + \cdots - \er^{2\pi i\tau}\g^{\rm (inst)}_r 
\log\left(\frac{x_1-x_2}{\mu}\right)^2 + \cdots \right] \, ,
\label{smallgym2pt}
\ea
where only the leading perturbative and instanton-induced terms have
been kept. At higher orders in the double expansion (small $\g$ and
small $\gy$) the situation is more complicated and terms with
different powers of logarithms appear at the same order in $\gy$. The
terms in (\ref{smallgym2pt}) are sufficient for the purposes of the
analysis to be carried on in the following sections.

Equation (\ref{smallgym2pt}) is the small coupling expansion of the
exact two-point function of primary operators in an orthonormal
basis. When performing explicit calculations, in perturbation theory
or semiclassically in an instanton background, one works with a set of
independent operators of bare dimension $\D_0$ ,
$\tilde\scrO^r_{\D_0}$, which are not in general orthonormal with
respect to the scalar product defined by the two-point function. In
extracting the physical information, \ie the anomalous dimensions, one
has thus to deal with the complications associated with operator
mixing.

The result for a  two-point function of operators
$\tilde\scrO^r_{\D_0}$ including the first order contributions in
perturbation theory and the leading semiclassical instanton term is of
the form
\ba
\la \bar{\tilde\scrO^r}_{\D_0}(x_1) \tilde\scrO^s_{\D_0}(x_2) \ra 
&\!\!=\!\!& \frac{1}{(x_1-x_2)^{2\D_0}} \left[ T^{rs} 
-\gy^2\,L^{rs} \log\left(\frac{x_1-x_2}{\mu}\right)^2 
\right. \nn \\
&& \left. + \cdots - \er^{2\pi i\tau}\,K^{rs} 
\log\left(\frac{x_1-x_2}{\mu}\right)^2 + \cdots \right] \, ,
\label{smallgym2pt-nd}
\ea
where we have denoted by $T$, $L$ and $K$ the matrices arising at
tree-level, one loop and in the one-instanton sector. To make contact
with the expansion (\ref{smallgym2pt}) of the exact function we
rewrite (\ref{smallgym2pt-nd}) as 
\ba
\la \bar{\tilde\scrO^r}_{\D_0}(x_1) \tilde\scrO^s_{\D_0}(x_2) \ra 
&\!\!=\!\!& \frac{1}{(x_1-x_2)^{2\D_0}} T^{rs^\prime}\left[
\one^{s^\prime s} - \er^{2\pi i\tau}\,\Gamma^{s^\prime s} 
\log\left(\frac{x_1-x_2}{\mu}\right)^2 \right] \, ,
\label{smallgym2pt-nd2}
\ea
where only the instanton part has been kept and we have defined 
$\Gamma^{rs} = \left( T^{-1} K\right)^{rs}$. In order to bring
(\ref{smallgym2pt-nd2}) into the form (\ref{smallgym2pt}) we need to
perform a change of basis which takes from $\tilde\scrO^r$ to
$\scrO^r$. This is achieved by acting with a matrix $M$
\be
\la\bar\scrO^r(x_1)\scrO^s(x_2)\ra = M^{rr^\prime}(M^*)^{s^\prime s}
\, \la \hsp{-0.2}\bar{\hsp{0.2}\tilde\scrO_{\D_0}^{r^\prime}}(x_1) 
\tilde\scrO^{s^\prime}_{\D_0}(x_2) \ra \, .
\label{diagonalise}
\ee
Substituting (\ref{smallgym2pt-nd2}) and (\ref{smallgym2pt}) into this
relation we find the matrix identity
\be
\mathrm{diag}(\{\gamma^{\rm (inst)}_r\}) = M\,\Gamma \,M^\dagger \, ,
\label{danommatrix}
\ee
which implies that the instanton contributions to the anomalous
dimensions in the sector under investigation can be identified with
the eigenvalues of the matrix $\Gamma^{rs}=\left(T^{-1}K\right)^{rs}$.
As discussed in \cite{j,bkps,bks} the above steps define the dilation
operator of the theory, $\hat D$. In this language $\Gamma$ is
therefore identified with the leading instanton correction to the
dilation operator.

\section{Non-perturbative contributions to anomalous dimensions of
scalar operators}
\label{scalaradim}

Having described the general strategy for the computation of anomalous
dimensions from two-point correlation functions we shall now consider
explicit examples of Lorentz scalar operators of bare dimension
$\D_0=2,3,4,5$.

We shall work with the normalisation of the fields which is standard
in instanton calculus, in which the action is written with an overall
factor of $1/\gy^2$. The usual normalisation of perturbative
calculations is recovered rescaling all the fields,
$\Phi\rightarrow\gy\Phi$, so that the kinetic terms become
$\gy$-independent, the cubic couplings have a factor of $\gy$ and the
quartic couplings are proportional to $\gy^2$. With our choice of
normalisation the propagators are proportional to $\gy^2$. We shall
define composite operators in such a way that their correlation
functions are independent of $\gy$ at tree-level. Therefore for an
operator made of $\ell$ elementary fields, which we denote generically
by $\Phi_a$, $a=1,\ldots,\ell$, we adopt the normalisation
\be 
\scrO^{(\ell)} = \frac{N}{(\gy^2 N)^{\ell/2}} \Tr\left(\Phi_1
\ldots \Phi_\ell\right) \, ,
\label{normlis}
\ee
which is easily verified to agree with the standard convention used in
the AdS/CFT correspondence in which two-point functions behave like
$N^2$ at large $N$ and are independent of the coupling at
tree-level. The same rescaling that leads to the ordinary perturbative
normalisation makes the composite operator (\ref{normlis}) independent
of $\gy$.

The operators we focus on are Lorentz scalars and for fixed bare
dimension, $\D_0$, they are characterised by their SU(4) Dynkin
labels. We use the notation $\scrO_{\D_0,\mb{r}}$ to denote a scalar
operator of bare dimension $\D_0$ transforming in the
$\mb{r}$-dimensional representation of SU(4). In each SU(4) sector we
shall construct  a complete set of independent operators and then
study the instanton contributions to their two-point functions.

\subsection{Dimension 2 scalar operators}
\label{dim2scal}

At bare dimension 2 the situation is particularly simple. All
operators are single trace and Lorentz scalars can only be obtained as
bilinears in the elementary scalar fields,
\be 
\scrO^{ij}_{2,\mb6\otimes\mb6} \sim \Tr \left(\v^i\v^j\right) \, .
\label{gendim2}
\ee
The scalars transform in the $\mb6$ of SU(4) with Dynkin labels
$[0,1,0]$ and thus the possible sectors for composite bilinears are
those in the decomposition
\be
[0,1,0]\otimes[0,1,0] = [0,0,0]_{\rm s}\oplus[1,0,1]_{\rm a}
\oplus[0,2,0]_{\rm s} \; \Leftrightarrow \;
\mb6\otimes\mb6 = \mb1_{\rm s}\oplus
\mb{15}_{\rm a}\oplus\mb{20^\prime}_{\rm s} \, , 
\label{6x6decomp}
\ee
where the subscripts s and a indicate the symmetric and antisymmetric
parts. There is actually no operator in the $\mb{15}$ because of
cyclicity of the trace and the only two sectors of scalars with
$\D_0=2$ are the singlet and $\mb{20^\prime}$ with one operator in
each of the two SU(4) representations
\ba
&& [0,0,0] \, : \qquad \scrO_{2,\mb1} = \frac{1}{\gy^2} 
\Tr \left(\v^i\v^i\right)  
\label{dim2-1} \\
&& [0,2,0] \, : \qquad \scrO^{\{ij\}}_{2,\mb{20^\prime}} =
\frac{1}{\gy^2} \Tr \left(\v^i\v^j - \frac{\d^{ij}}{6}
\v^k\v^k\right) \, ,
\label{dim2-20}
\ea
Here and in the following we indicate complete symmetrisation plus
removal of traces by curly brackets, $\{i_1i_2\ldots i_n\}$. We shall
denote complete antisymmetrisation by square brackets, $[i_1i_2\ldots
i_n]$ and symmetrisation without removal of traces by parenthesis
$(i_1i_2\ldots i_n)$.

The two operators in (\ref{dim2-1}) and (\ref{dim2-20}) are of course
well known. The singlet is the lowest component of the Konishi
multiplet, $\scrO_{2,\mb1} \equiv \scrK_{\mb1}$, and
$\scrO_{2,\mb{20^\prime}}^{ij}\equiv\scrQ^{ij}$ is the lowest
component of the supercurrent multiplet containing the energy-momentum
tensor and the supersymmetry and R-symmetry currents. The latter is a
1/2 BPS operator and thus its scaling dimension does not receive
quantum corrections. Operators in the Konishi multiplet do have
anomalous dimension in perturbation theory, but not at the instanton
level. The one-loop contribution was first computed in \cite{agj} and
re-derived in \cite{bkrs1} using OPE techniques in the context of the
AdS/CFT correspondence. The perturbative result has been extended to
two-loops in \cite{bkrs2,aeps} and a three loop result has been
obtained in \cite{bks} under certain assumptions studying the action
of the dilation operator.

The fact that the anomalous dimension of the Konishi multiplet does
not receive instanton corrections has been shown through the OPE
analysis of a four-point function of 1/2 BPS operators $\scrQ^{ij}$ in
\cite{bkrs1}. The derivation of this result directly from the study of
the two-point function is rather subtle. Rewriting the scalar fields
as $\v^{[AB]}$, $\scrK_{\mb1}$ takes the form
\be
\scrK_{\mb1} = \fr{\gy^2}\, \veps_{ABCD}\,\Tr\left( 
\v^{AB}\v^{CD} \right) \, . 
\label{dim2-1a}
\ee 
As discussed in section \ref{2ptfunct} in computing the instanton
induced two-point function $\la\scrK_\mb1(x_1)\scrK_\mb1(x_2)\ra$ we
integrate over the moduli space the classical expression of the
operators in the background of an instanton and the superconformal
fermionic integrations must be saturated. Each of the two operators
must soak up eight fermion modes in the combination (\ref{saturmodes}).
In order to get the correct number of fermion modes for each of the
two insertions we need to consider 
\be
\veps_{ABCD}\,\Tr\left( \v^{(2)AB}\v^{(6)CD} + \v^{(6)AB}\v^{(2)CD} 
\right) \, . 
\label{k1zeromod}
\ee
This combination contains the minimal required number of fermionic
modes. The other terms in the scalar solution, $\v^{(10)AB}$ and
$\v^{(14)AB}$, would give rise to contributions involving $\bar\nu\nu$
bilinears as well and so would be of higher order in $\gy$ and hence
not relevant in semiclassical approximation. Using (\ref{phi2su4}) and
(\ref{phi6su4}) the classical profile of $\scrK_\mb1$ is seen to be
proportional to 
\be
\veps_{ABCD}\veps_{A^\pp B^\pp C^\pp D^\pp} \scrmf^{[A}\scrmf^{B]}
\scrmf^{[C}\scrmf^{D]}\scrmf^{A^\pp}\scrmf^{B^\pp}\scrmf^{C^\pp}
\scrmf^{D^\pp} \, .
\label{k1zerosat}
\ee
This expression indeed contains the combination (\ref{saturmodes}).
This means that the vanishing of the instanton contribution to the
two-point function of the $\scrK_\mb1$ operator and thus of its
instanton induced anomalous dimension does not simply follow from
the impossibility of saturating the fermionic integrals. The direct
calculation of the (vanishing) anomalous dimension of $\scrK_\mb1$ from
the two-point function requires a cancellation among various terms for
which we need the exact form of $\v^{(6)AB}$, which we have not
determined. We shall find in section \ref{dim4scal} that the absence
of instanton correction for this operator can be shown analysing a
superconformal descendant of $\scrK_\mb1$ transforming in the
representation $\mb{84}$. 

For the 1/2 BPS operator $\scrQ^{ij}=\scrO^{\{ij\}}_{2,\mb{20^\pp}}$,
on the other hand, we can easily verify the absence of instanton
corrections to the two-point function. For this purpose it is
convenient to choose a specific component to evaluate the two-point
function, \eg 
\be
\la \scrQ^{12}(x_1)\scrQ^{12}(x_2) \ra \, .
\label{20-2pt-12}
\ee
Recalling that $\v^1=\sqrt{2}(\v^{14}+\v^{23})$ and
$\v^2=\sqrt{2}(\v^{24}+\v^{13})$ (see appendix \ref{convents}) we find
that in the instanton background the operator $\scrQ^{12}$ is
proportional to
\ba 
&& \hsp{-2}\veps_{A^\pp B^\pp C^\pp D^\pp} \, \scrmf^{A^\pp}
\scrmf^{B^\pp}\scrmf^{C^\pp}\scrmf^{D^\pp} \left(
\scrmf^{[1}\scrmf^{4]}\scrmf^{[2}\scrmf^{4]}
+\scrmf^{[2}\scrmf^{3]}\scrmf^{[2}\scrmf^{4]} \right. \nn \\ 
&& \rule{0pt}{16pt}\hsp{2.5}\left. +
\scrmf^{[1}\scrmf^{4]}\scrmf^{[1}\scrmf^{3]}+
\scrmf^{[2}\scrmf^{3]}\scrmf^{[1}\scrmf^{3]} \right) \, ,
\label{Q12zeromod}
\ea
and taking all the modes to be superconformal modes none of the four
terms in this expression contains the combination (\ref{saturmodes}). 

The above detailed discussion was clearly not needed for $\scrQ^{ij}$
which is known to be protected, as already remarked, but it is
included in order to illustrate the strategy utilised in more
complicated examples in the following sections. 

The operator $\scrO^{ij}_{2,\mb{20^\pp}}$ is the simplest example in a
class of operators characterised by the fact that their bare dimension
is $\D_0=\ell$ and they transform in the representation $[0,\ell,0]$
of SU(4). Operators of this type which are superconformal primaries
are 1/2 BPS \cite{dp,fz}. As shown in \cite{bks,b1} operators of this
type form a `sector' in the $\scrN$=4 SYM theory. They cannot mix
with operators involving fermions, field strengths or covariant
derivatives at any order in $\gy$. For operators of this type it is
always possible to choose a component which in $\scrN$=1 notation can
be written in terms of a single complex scalar. This makes proving the
absence of instanton corrections very simple. For $\scrQ^{ij}$ we can
consider the component
$\Tr\left(\phi^1\phi^1\right)\sim\Tr\left(\v^{14}\v^{14}\right)$ (see
appendix \ref{convents} for the definition of the complex combinations
$\phi^I$), so that it is immediately verified that it cannot saturate
the eight fermion modes as in (\ref{saturmodes}).

\subsection{Dimension 3 scalar operators}
\label{dim3scal}

Composite operators of bare dimension $\D_0=3$ are also necessarily
single trace. They can be obtained either as products of three scalars
or as fermion bilinears, which must involve spinors of the same
chirality in order for the composite to be a Lorentz scalar. So we
need to consider
\be
\scrO^{ijk}_{3,\mb6\otimes\mb6\otimes\mb6} \sim \Tr \left(
\v^i\v^j\v^k \right) \, , \quad
\scrO^{AB}_{3,\mb4\otimes\mb4} \sim \Tr\left( \lambda^A 
\lambda^B \right) \, , \quad 
\scrO_{3,\mbb4\otimes\mbb4;AB} \sim \Tr\left( \bar\lambda_A 
\bar\lambda_B \right) \, .
\label{gendim3}
\ee
Operators cubic in the elementary scalars belong to the sectors in the
decomposition
\ba
&&\hsp{-1} [0,1,0]\otimes[0,1,0]\otimes[0,1,0] = 3 [0,1,0]\oplus[2,0,0]
\oplus[0,0,2]\oplus [0,3,0]\oplus 2[1,1,1] \nn \\
&& \hsp{-1}\Leftrightarrow \; \mb6\otimes\mb6\otimes\mb6 = 3\cdot\mb6 
\oplus \mb{10} \oplus \mbb{10} \oplus \mb{50} \oplus 2\cdot\mb{64} \, , 
\label{decomp6x6x6}
\ea
whereas the fermionic bilinears contribute to the sectors 
\ba
&& [1,0,0]\otimes[1,0,0] = [0,1,0]_{\rm a}\oplus[2,0,0]_{\rm s} \; 
\Leftrightarrow \; \mb4\otimes\mb4 = \mb6_{\rm a} \oplus \mb{10}_{\rm s}
\label{decomp4x4}\\
&& [0,0,1]\otimes[0,0,1] = [0,1,0]_{\rm a}\oplus[0,0,2]_{\rm s} \;
\Leftrightarrow \; \mbb4\otimes\mbb4 = \mb6_{\rm a} \oplus 
\mbb{10}_{\rm s} \, .
\label{decomp4bx4b}
\ea

\subsubsection{$\D_0=3$, $[0,1,0]$}
\label{d3r6}

In this sector there is only one operator coming from
(\ref{decomp6x6x6}). The decomposition $\mb6\otimes\mb6\otimes\mb6$
contains the $\mb6$ with multiplicity 3, but taking into account the
cyclicity of the trace there is only one independent operator,
\be
\scrO^i_{3,\mb6} = \frac{1}{\gy^3N^{1/2}} 
\Tr \left( \v^i\v^j\v^j \right) \, .
\label{dim3-6}
\ee
The $\mb6$ is also contained in $\mb4\otimes\mb4$ and
$\mbb4\otimes\mbb4$, but the corresponding fermionic bilinears
vanish for the cyclicity of the trace, 
\be 
\S^i_{AB}\,\Tr\left(\lambda^{\a A}\lambda^B_\a\right) = 
\Sbar^{AB}_i\,\Tr\left(\bar\lambda_{\adot A}\bar\lambda^\adot_B
\right) = 0 \, , 
\label{zero-f-3-6}
\ee
since $\S^i_{AB}$ and $\Sbar^{AB}_i$ are antisymmetric in $A,B$. 

Let us then consider the two-point function of operators
$\scrO^i_{3,\mb6}$. For definiteness we take the component $i=1$, the
correlator $\la\scrO^i_{3,\mb6}(x_1) \scrO^j_{3,\mb6}(x_2)\ra$ is
clearly proportional to $\d^{ij}$. The first step is to check whether
each of the two operators can saturate the superconformal fermion
integrations, \ie whether their instanton profile contains the
combination (\ref{saturmodes}). We need to consider  
\ba
&&\Tr\left(\v^{(2)1}\v^{(2)j}\v^{(6)j} + \v^{(2)1}\v^{(6)j}\v^{(2)j} +
\v^{(6)1}\v^{(2)j}\v^{(2)j} \right) 
\label{O-3-6-zerom} \\
&& \sim \veps_{ABCD} \veps_{A^\pp B^\pp C^\pp D^\pp} 
\left[ \left(\scrmf^{[1}\scrmf^{4]} +
\scrmf^{[2}\scrmf^{3]}\right) \left(\scrmf^{[A}\scrmf^{B]}\right)
\left(\scrmf^{[C}\scrmf^{D]}\right)\left(\scrmf^{A^\pp}\scrmf^{B^\pp}
\scrmf^{C^\pp}\scrmf^{D^\pp}\right) \right] \, , \nn
\ea
where we have used (\ref{phi2su4}) and (\ref{phi6su4}) and written the
contraction $\v^j\v^j$ as $\veps_{ABCD}\v^{AB}\v^{CD}$. Inspecting
(\ref{O-3-6-zerom}) it is  easily verified that indeed the
superconformal modes can be saturated in the two-point function. A
potentially non-vanishing contribution is obtained for instance from
the first term in (\ref{O-3-6-zerom}) replacing $\v^1$ by $\bar\nu\nu$
bilinears and using the remaining scalars to soak up the 16 $\zeta$'s in
each of the two operators. The other terms give identical contributions
up the overall coefficient. The resulting contribution to the two-point
function is schematically of the form
\ba 
\la\scrO^1_{3,\mb6}(x_1)\scrO^1_{3,\mb6}(x_2) \ra &\!\!\sim\!\!&  
\int \dr\mu_{\rm phys} \, \er^{-S_{\rm inst}}
\label{2pt-O-3-6a} \\ 
&& \hsp{-0.5} f(x_1;x_0,\rho) \left(\bar\nu^{[1}\nu^{4]}\right) 
\prod_{A=1}^4\left[\zeta^A(x_1)\right]^2  
f(x_2;x_0,\rho) \left(\bar\nu^{[2}\nu^{3]}\right) 
\prod_{A^\pp=1}^4\left[\zeta^{A^\pp}(x_2)\right]^2 \nn \, ,
\ea
where the function $f(x;x_0,\rho)$ depends on the details of the
solutions $\v^{(2)}$ and $\v^{(6)}$ for the scalars. Writing the
measure explicitly as in section \ref{sc-corrfunct} to  we get
\ba
&& \la\scrO^1_{3,\mb6}(x_1)\scrO^1_{3,\mb6}(x_2) \ra \sim 
\frac{\gy^{2}\er^{2\pi i\tau}}{N(N-1)!(N-2)!}
\int \dr\rho \, \dr^4x_0 \, \dr^5\Omega \prod_{A=1}^4 \dr^2\eta^A \,
\dr^2 {\bar\xi}^A \, \rho^{4N-7} \nn \\
&& \hsp{1} \int_0^\infty \dr r \, r^{4N-3} \er^{-2\rho^2 r^2}
f(x_1;x_0,\rho) \prod_{A=1}^4\left[\zeta^A(x_1)\right]^2  
f(x_2;x_0,\rho) \prod_{A^\pp=1}^4\left[\zeta^{A^\pp}(x_2)\right]^2 
\nn \\
&& \hsp{1}\frac{\d^4}{\d\bvt^u_{[1}\d\vt_{u4]}\d\bvt^v_{[2}\d\vt_{v3]}}
\scrZ(\vt,\bar\vt;\Omega,r) 
\raisebox{-4pt}{$\Big|_{\vt=\bar\vt=0}$} \nn \\
&& \rule{0pt}{22pt} \hsp{1} 
\sim \frac{\gy^{2}\Gamma(2N-2) 2^{-2N}\er^{2\pi i\tau}}
{(N-1)!(N-2)!} (x_1-x_2)^8 \int \frac{\dr\rho \, \dr^4x_0}{\rho^3} \, 
f(x_1;x_0,\rho) f(x_2;x_0,\rho) \nn \\
&& \hsp{1} \int \dr^5\Omega \, \left[(N-2)^2 \,\Omega^{14}\Omega^{23}
-(N-2)\,\Omega^{13}\Omega^{24} \right] \, .
\label{2pt-O-3-6b}
\ea
The five-sphere integral is evaluated using 
\be
\int \dr^5\Omega \, \Omega^{AB}\Omega^{CD} = \quart \veps^{ABCD} \, 
\label{s5int1}
\ee
and we finally find
\be
\la\scrO^1_{3,\mb6}(x_1)\scrO^1_{3,\mb6}(x_2) \ra \sim 
\gy^{2} c(N) \, \er^{2\pi i\tau} (x_1-x_2)^8 
\int \frac{\dr\rho}{\rho^3} \, \dr^4x_0 
\, f(x_1;x_0,\rho) f(x_1;x_0,\rho) \, ,
\label{2pt-O-3-6}
\ee
where for large $N$ $c(N)=N\sqrt{N}\left[1+O(1/N)\right]$.  The
complete result for the two-point function is a sum of terms of this
form. A logarithmic divergence in the remaining bosonic integrals
would signal an instanton contribution to the anomalous dimension of
$\scrO^i_{3,\mb6}$. Since we have not determined the complete
expression for $\v^{(6)}$ we cannot compute the exact coefficient in
(\ref{2pt-O-3-6}) and verify the presence of the divergence.

Notice however that (\ref{2pt-O-3-6}) contains, apart from the
standard instanton weight, a factor of $\lambda=\gy^2N$. Since in this
sector there is only one operator there is no problem of mixing to
take into account and the instanton induced anomalous dimension can be
read directly from the two-point function (\ref{2pt-O-3-6}) (after
dividing by the tree-level coefficient). The result is that if
$\scrO^i_{3,\mb6}$ does acquire an anomalous dimension, it is of order
\be
\g_{3,\mb6}^{\rm inst} \sim \lambda \, \er^{2\pi i\tau} \, .
\label{gam-3-6}
\ee
This is to be contrasted with the case of other operators that we
shall examine in the following in which the leading instanton
contribution to two-point functions behaves as $\er^{2\pi i\tau}$
with no further factors of $\gy$.

We shall come back to the case of the operator $\scrO^i_{3,\mb6}$ in
section \ref{d3r6rev}.

\subsubsection{$\D_0=3$, $[2,0,0]$ and $[0,0,2]$}
\label{d3r10}

There are two scalar operators with $\D_0=3$ in the $[2,0,0]$, one is
cubic in the scalars and the other is bilinear in the $[1,0,0]$
fermions,
\ba
&& \scrO^{(1)\,(AB)}_{3,\mb{10}} = \frac{1}{\gy^3N^{1/2}}\,
t^{(AB)}_{[ijk]} \, \Tr \left(\v^i\v^j\v^k\right) 
\label{O-3-10a} \\
&& \scrO^{(2)\,(AB)}_{3,\mb{10}} = \frac{1}{\gy^2} \, \Tr \left(
\lambda^{\a A}\lambda^B_\a \right) \, ,
\label{O-3-10b}
\ea
where the projector $t^{(AB)}_{[ijk]}$ of the product
$\mb6\otimes\mb6\otimes\mb6$ onto the $\mb{10}$ is defined as
\be
t^{(AB)}_{\mb{10}\,[ijk]} = \Sbar^{AC}_{[i} \S_{j\,CD}
\Sbar^{DB}_{k]} \,,
\label{10proj}
\ee
with the brackets indicating complete antisymmetrisation in $i$, $j$
and $k$. A similar projector onto the $\mbb{10}$ can be constructed as 
\be
\bar{t}^{[ijk]}_{\mbb{10}\,(AB)} = \S_{AC}^{[i} \Sbar^{j\,CD}
\S_{DB}^{k]} \,,
\label{10bproj}
\ee
so that in the $[0,0,2]$ we find
\ba
&& \bar\scrO^{(1)}_{3,\mbb{10}\,(AB)} = \frac{1}{\gy^3N^{1/2}} \,
\bar{t}_{(AB)}^{[ijk]} \, \Tr \left(\v^i\v^j\v^k\right) 
\label{O-3-b10a} \\
&& \bar\scrO^{(2)}_{3,\mb{10}\,(AB)} = \frac{1}{\gy^2} \, \Tr \left(
\bar\lambda_{\adot A}\bar\lambda^\adot_B \right) \, .
\label{O-3-b10b}
\ea

We now consider the two-point functions in this sector. 
In order to saturate the fermionic integrals over the superconformal
modes we need to consider the following terms
\ba
&& \scrO^{(1)\,(AB)}_{3,\mb{10}} \sim t^{(AB)}_{[ijk]} \, 
\Tr \left(\v^{(2)i}\v^{(2)j}\v^{(6)k} + (\v^{(2)i}\v^{(6)j}\v^{(2)k} +
(\v^{(6)i}\v^{(2)j}\v^{(2)k} \right) 
\label{O-3-10a-exp} \\
&& \scrO^{(2)\,(AB)}_{3,\mb{10}} \sim \Tr \left(
\lambda^{(5)\a A}\lambda^{(5)B}_\a + 
\lambda^{(1)\a A}\lambda^{(9)B}_\a
+ \lambda^{(9)\a A}\lambda^{(1)B}_\a \right) \, 
\label{O-3-10b-exp}
\ea
and similarly for the conjugate operators we need
\ba
&& \bar\scrO^{(1)}_{3,\mb{10}\,(AB)} \sim \bar t_{(AB)}^{[ijk]} \, 
\Tr \left(\v^{(2)i}\v^{(2)j}\v^{(6)k} + (\v^{(2)i}\v^{(6)j}\v^{(2)k} +
(\v^{(6)i}\v^{(2)j}\v^{(2)k} \right) 
\label{O-3-b10a-exp} \\
&& \bar\scrO^{(2)}_{3,\mb{10}\,(AB)} \sim \Tr \left(
\bar\lambda^{(3)}_{\adot A}\bar\lambda^{(7)\adot}_B + 
\bar\lambda^{(7)}_{\adot A}\bar\lambda^{(3)\adot}_B \right) \,. 
\label{O-3-b10b-exp}
\ea
As usual it is convenient to pick a specific component, so \eg we
consider $\la\scrO^{(r)\,(11)}_{3,\mb{10}}(x_1)
\bar\scrO^{(s)}_{3,\mbb{10}\,(11)}(x_2)\ra$, with $r,s=1,2$. The two
operators $\scrO^{(r)\,(11)}_{3,\mb{10}}$ are
\be
\scrO^{(1)\,(11)}_{3,\mb{10}} = \frac{9\sqrt{8}}{\gy^3N^{1/2}}\,\Tr
\left( \v^{12}[\v^{13},\v^{14}] \right) \, , \qquad
\scrO^{(2)\,(11)}_{3,\mb{10}} = \frac{1}{\gy^2}\,\Tr
\left( \lambda^{\a 1}\lambda_\a^1 \right)
\label{O-3-10-11comp}
\ee
and their conjugates are
\be
\bar\scrO^{(1)}_{3,\mb{10}\,(11)} = \frac{9\sqrt{8}}{\gy^3N^{1/2}}\,\Tr
\left( \v^{23}[\v^{24},\v^{34}] \right) \, , \qquad
\bar\scrO^{(2)}_{3,\mb{10}\,(11)} = \frac{1}{\gy^2}\,\Tr
\left( \bar\lambda_{\adot 1}\bar\lambda_1^\adot \right) \,.
\label{O-3-b10-11comp}
\ee
Recalling the analysis of section \ref{N4instmult} we find that the
two operators  $\scrO^{(r)\,(11)}_{3,\mb{10}}$ can soak up the
fermionic modes in the combination required to get a non-zero
contribution to the two-point function. Expanding
(\ref{O-3-10-11comp}) as in (\ref{O-3-10a-exp})-(\ref{O-3-10b-exp}) we
obtain terms which contain the combination (\ref{saturmodes}) and  are
proportional to
\be
\left(\bar\nu^1\nu^1\right) \left[\zeta^1(x)\right]^2 
\left[\zeta^2(x)\right]^2 \left[\zeta^3(x)\right]^2 
\left[\zeta^4(x)\right]^2 \, .
\label{satur-O3-10}
\ee
None of the conjugate operators however can provide
(\ref{saturmodes}), they can at most contain one mode $\zeta^1$. 
Hence the two-point functions are all zero in the instanton
background and the scaling dimensions do not receive instanton
corrections in this sector.  

Notice that even before analysing the conjugate operators we could
conclude that the two-point functions are not corrected by
instantons. From (\ref{satur-O3-10}) it is clear that the five-sphere
integration would vanish as a consequence of (\ref{s5int1}). 

These results were to be expected. Two orthogonal operators of
dimension 3 in the $\mb{10}$ can be defined as
\ba
&& \scrE^{(AB)} = - \frac{1}{\gy^2}\,\Tr\left(\lambda^{\a A}
\lambda^B_\a\right) + \frac{\sqrt{2}}{\gy^2} \, t^{(AB)}_{[ijk]} 
\,\Tr \left( \v^i\v^j\v^k \right)
\label{E-10} \\
&& \scrK^{(AB)} = \frac{3\sqrt{2}}{\gy^2} \, t^{(AB)}_{[ijk]}
\,\Tr \left( \v^i\v^j\v^k \right) + \frac{6N}{32\pi^2} \, 
\Tr\left(\lambda^{\a A}\lambda^B_\a \right) \, .
\label{K-10} 
\ea
$\scrE^{(AB)}$ is a component of the 1/2 BPS multiplet of the
$\scrN$=4 supercurrents, it is found at level $\d^2$ starting from the
lowest component $\scrO^{\{ij\}}_{2,\mb{20^\pp}}$, and $\scrK^{(AB)}$
is a component at the same level of the Konishi multiplet.  Therefore
the former is protected and the latter does not receive instanton
corrections. Notice that the relative coefficients in (\ref{E-10}) can
be fixed for instance by the requiring that the three-point function
$\la\scrE^{(AB)}\scrE^{(CD)}\scrQ^{ij}\ra$ does not receive
perturbative corrections \cite{dfs}, whereas the coefficients in
(\ref{K-10}) are determined by the Konishi anomaly \cite{bkrs3}.

\subsubsection{$\D_0=3$, $[0,3,0]$}
\label{d3r50}

In this sector there is only one operator which is obtained from the
fully symmetrised product of three scalars
\ba
\scrO^{\{ijk\}}_{3,\mb{50}} &\!\!=\!\!& 
\frac{1}{\gy^3N^{1/2}} \, \Tr \left(\v^{\{i}\v^j\v^{k\}} \right) 
\label{O-3-50} \\
&\!\!=\!\!& \frac{1}{\gy^3N^{1/2}} \left[ \Tr \left(
\v^i\v^j\v^k + \v^i\v^k\v^j \right) -\quart \Tr \left( 
\d^{ij}\v^k\v^l\v^l + \d^{ik}\v^j\v^l\v^l + 
\d^{jk}\v^i\v^l\v^l \right) \right] \nn \, .
\ea
This sector is the next example of the type discussed at the end of
section \ref{dim2scal}. The operator in (\ref{O-3-50}) is a protected
1/2 BPS operator and using the notation of section \ref{dim2scal} we
define $\scrQ^{ijk}\equiv\scrO^{\{ijk\}}_{3,\mb{50}}$. The 1/2 BPS
operator in the $\mb{20^\pp}$ is the lowest component of the
supercurrent multiplet. It is dual in the AdS/CFT correspondence to a
scalar in the type IIB supergravity multiplet which is a linear
combination of the trace part of the metric and the \RR 4-form with
indices in internal directions. The operator $\scrQ^{ijk}$ is dual to
the first Kaluza--Klein excited mode of the same field~\footnote{In
this sense the operator $\scrO^i_{3,\mb6}$ of equation (\ref{dim3-6})
can be thought of as the ``first Kaluza--Klein excitation'' of the
Konishi operator, $\scrK_\mb1$, considered in section
\ref{dim2scal}.}.  It is straightforward to verify the absence of
instanton corrections to two point-functions of $\scrQ^{ijk}$ by
choosing a suitable component to analyse the zero-mode structure. The
simplest choice is a component written in terms of a single complex
scalar, $\Tr\left(\phi^1\phi^1\phi^1\right)$. The terms in this
operator with at least eight fermion modes in the instanton background
are
\be
\Tr\left(\phi^{(6)1}\phi^{(2)1}\phi^{(2)1}\right) \sim 
\veps_{ABCD} \left(\scrmf^{[1}\scrmf^{4]}\scrmf^{[1}\scrmf^{4]}
\scrmf^{[1}\scrmf^{4]}\scrmf^A\scrmf^B\scrmf^C\scrmf^D \right) \, ,
\label{O-3-50-zerom}
\ee
from which one immediately verifies that only one mode of flavours 2
and 3 can be saturated, so that the two-point functions of this
operator cannot get instanton correction.

\subsubsection{$\D_0=3$, $[1,1,1]$}
\label{d3r64}

Operators with $\D_0=3$ in the $\mb{64}$, corresponding to Dynkin
labels $[1,1,1]$, also involve only elementary scalars. The $\mb{64}$
occurs twice in the product $\mb6\otimes\mb6\otimes\mb6$, respectively
in $\mb6\otimes\mb{15}$ and $\mb6\otimes\mb{20^\pp}$. To obtain the
operators in the $\mb{64}$ we should suitably project the two
combinations 
\ba
&& \Tr\left(\v^i\v^{[j}\v^{k]}\right) 
\label{6x15} \\ 
&& \Tr\left(\v^i\v^{\{j}\v^{k\}}\right) \, .
\label{6x20}
\ea 
Taking into account the cyclicity of the trace one finds that
(\ref{6x15}) actually only contributes to the already examined
$\mb{10}$ and $\mbb{10}$, since the operator is automatically fully
antisymmetric in $i$, $j$ and $k$. Similarly (\ref{6x20}) is found to
only contribute to the $\mb{50}$ and $\mb6$ representations: it is not
possible to make (\ref{6x20}) orthogonal to both (\ref{dim3-6}) and
(\ref{O-3-50}). Hence the $\mb{64}$, although allowed by the group
theoretical analysis, is not realised in terms of gauge invariant
operators.

\subsubsection{$\D_0=3$, $[0,1,0]$ revisited}
\label{d3r6rev}

The non-renormalisation results for operators in the representations
$\mb{10}$, $\mbb{10}$ and $\mb{50}$ (and the absence of
gauge-invariant operators in the $\mb{64}$) simplify  the computation
of the anomalous dimension of $\scrO^i_{3,\mb6}$ using an alternative
approach, based on the OPE analysis of a four-point function. This
method is more complicated, because it involves the computation of an
instanton induced four-point function, but does not require to solve
the field equation for $\v^{(6)}$. 

A four-point function of protected operators $\scrQ$ can be expanded
in a double OPE as 
\be
\la\scrQ^{r}(x_1)\scrQ^{s}(x_1)\scrQ^{u}(x_3)\scrQ^{v}(x_4)\ra = 
\sum_k \frac{C^{rs}_k(x_{12},\del_1)\,C^{uv}_k(x_{34},\del_3)}
{(x_{12})^{\D_r+\D_s-\D_k}(x_{34})^{\D_u+\D_v-\D_k}} \, 
\la\scrO^k(x_1)\scrO^k(x_3)\ra \, ,
\label{OPE}
\ee
where $x_{ij}=x_i-x_j$, $\del_i=\del/\del x_i$ and the sum is over a
generally infinite set of primary operators. The dependence on
derivatives in the Wilson coefficients $C^{rs}_k$ implicitly
represents the inclusion of descendants. The operators $\scrQ$ are
chosen to be  protected, but in general some of the $\scrO_k$'s may
have anomalous dimension, so that $\D_k=\D_k^{(0)}+\g_k$. Similarly
for the $C^{rs}_k$ coefficients we have
$C^{rs}_k=C^{(0)rs}_{\:k}+\omega^{rs}_k$. Expanding (\ref{OPE}) for
small $\g_k$ and $\omega^{rs}_k$ we get
\ba
&&\la\scrQ^{r}(x_1)\scrQ^{s}(x_1)\scrQ^{u}(x_3)\scrQ^{v}(x_4)\ra = 
\sum_k \frac{\la\scrO^k(x_1)\scrO^k(x_3)\ra_{(0)}}
{(x_{12})^{\D_r+\D_s-\D{(0)}_k}(x_{34})^{\D_u+\D_v-\D^{(0)}_k}} \nn \\ 
&& \left[C^{(0)rs}_{\:k}C^{(0)uv}_{\:k} + 
C^{(0)rs}_{\:k}\omega^{uv}_{\:k} + \omega^{rs}_{\:k}C^{(0)uv}_{\:k} 
+ \frac{\g_k}{2} C^{(0)rs}_{\:k}C^{(0)uv}_{\:k} 
\log\left(\frac{x_{12}^2x_{34}^2}{x_{13}^4}\right)\right] \, , 
\label{OPE-exp}
\ea
which shows how anomalous dimensions and corrections to the OPE
coefficients can be extracted from the analysis of four-point
functions. 

In order to have the exchange of the a scalar in the $\mb6$ we can
consider the instanton contribution to the correlation function
\be  
G(x_1,x_2,x_3,x_4) = \la \scrO_{3,\mb{50}}(x_1)
\scrO_{2,\mb{20^\pp}}(x_2) \scrO_{3,\mb{50}}(x_3)
\scrO_{2,\mb{20^\pp}}(x_4) \ra \,,
\label{4ptope}
\ee
involving two $\D_0=2$ and two $\D_0=3$ 1/2 BPS operators. In
computing (\ref{4ptope}) at leading order we only need the scalar
solution $\v^{(2)}$ which is given in appendix \ref{k1adhm}.

The operators exchanged in the $s$-channel, corresponding to
$x_{12}\to 0$, $x_{34}\to 0$, are in the decomposition
\be
\mb{20^\pp}\otimes\mb{50} = \mb6 \oplus \mb{50} \oplus \mb{64} \oplus
\mb{196} \oplus \mb{300} \oplus \mb{384} \, .
\label{20x50}
\ee
According to the general formula (\ref{OPE-exp}) the presence of an
instanton induced anomalous dimension for an operator of dimension 3
would give rise to a singularity of the type 
\be
\frac{1}{(x_{12})^2(x_{34})^2} \log\left( 
\frac{(x_{12})^2(x_{34})^2}{(x_{13})^4}\right) \,. 
\label{pole-log}
\ee
Since we have verified that $\scrO^i_{3,\mb6}$ is the only operator of
bare dimension 3 which can have contribution at the instanton level,
the OPE analysis of a single four-point function unambiguously
determines its anomalous dimension. It can be read off from the
coefficient of the singular term (\ref{pole-log}) with no problem of
mixing to take into account and no necessity to project onto the
representation we are interested in. This is the same type of argument
used in \cite{bkrs1} to show that the anomalous dimension of the
Konishi operator, $\scrK_\mb1$, is not corrected
non-perturbatively. In that case the argument took advantage of the
fact that $\scrK_\mb1$ is the only operator of bare dimension 2 which
could possibly be corrected.

The computation of the correlator (\ref{4ptope}) in the one-instanton
sector is presented in appendix \ref{ope}. As shown there in the limit
$x_{12}\to 0$, $x_{34}\to 0$ this four-point function does not have a
singularity of the type (\ref{pole-log}), indicating that the operator
$\scrO^i_{3,\mb6}$ does not have anomalous dimension at the instanton
level. This means that the coefficient of the logarithmically
divergent term in the two-point function (\ref{2pt-O-3-6}) actually
vanishes.

\subsection{Dimension 4 scalar operators}
\label{dim4scal}

The analysis operators of bare dimension $\D_0=4$ is more involved, at
this level derivatives and field strengths also contribute to scalar
operators and moreover we need to consider single- as well as
double-trace operators. 

The combinations of elementary fields we need to consider here are
\ba
&& \scrO^{(a)}_{4,\mb1} \sim \Tr\left(F_{\mu\nu}F^{\mu\nu}\right)
\, , \quad \scrO^{(b)}_{4,\mb1} \sim \Tr\left(F_{\mu\nu}
\tilde{F}^{\mu\nu}\right) 
\label{d4-fieldstr} \\
&& \scrO^i_{4,\mb6\otimes\mb6} \sim
\Tr\left(\scrD_\mu\v^i\scrD^\mu\v^j\right)  \label{d4-deriv} \\
&& \scrO^{iAB}_{4,\mb6\otimes\mb4\otimes\mb4} \sim \Tr \left(
\v^i\lambda^{\a A}\lambda^B_\a \right) \, , \quad
\scrO^i_{AB;4,\mb6\otimes\mbb4\otimes\mbb4} \sim \Tr\left(
\v^i\bar\lambda_{\adot A}\lambda^\adot_B\right) 
\label{d4fermi} \\
&& \scrO^{{\rm(s)}\,ijkl}_{4,\mb6\otimes\mb6\otimes\mb6\otimes\mb6} 
\sim \Tr \left( \v^i\v^j\v^k\v^l \right) \, , \quad 
\scrO^{{\rm (d)}\,ijkl}_{4,\mb6\otimes\mb6\otimes\mb6\otimes\mb6} 
\sim \Tr \left(\v^i\v^j\right) \Tr\left(\v^k\v^l \right) \, ,
\label{d4scal}
\ea
where the notation is the same used in previous sections and
superscripts ${\rm (s)}$ and ${\rm (d)}$ denote single- and double-trace
operators. This shows that operators involving field strengths can
only contribute to the SU(4) singlet sector. Operators with covariant
derivatives can be found in the sectors in the decomposition
\be
[0,1,0]\otimes[0,1,0] = [0,0,0]\oplus[1,0,1]\oplus[0,2,0] 
\Leftrightarrow \mb6\otimes\mb6\ = \mb1\oplus\mb{15}\oplus\mb{20^\pp}
\, . \label{6x6d4}
\ee
The operators of the two types in (\ref{d4fermi}), made out of a scalar
and a fermion bilinear, contribute respectively to 
\ba
&& [0,1,0]\otimes[1,0,0]\otimes[1,0,0] = [0,0,0]\oplus 2[1,0,1]\oplus
[0,2,0] \oplus [2,1,0] \nn \\
&& \Leftrightarrow \mb6\otimes\mb4\otimes\mb4 = \mb1\oplus
2\cdot\mb{15}\oplus\mb{20^\pp} \oplus\mb{45} 
\label{6x4x4d4}
\ea
and
\ba
&& [0,1,0]\otimes[0,0,1]\otimes[0,0,1] = [0,0,0]\oplus 2[1,0,1]\oplus
[0,2,0] \oplus [0,1,2] \nn \\
&& \Leftrightarrow \mb6\otimes\mbb4\otimes\mbb4 = \mb1\oplus
2\cdot\mb{15}\oplus\mb{20^\pp} \oplus\mbb{45} \, . 
\label{6xb4xb4d4}
\ea
The single- and double-trace operators in (\ref{d4scal}) can
contribute to the representations in 
\ba
&&[0,1,0]\otimes[0,1,0]\otimes [0,1,0]\otimes[0,1,0] =
3[0,0,0]\oplus 7[1,0,1] \oplus 6[0,2,0]\nn \\
&&\hsp{1.3}{}\! \oplus 3([2,1,0]\oplus[0,1,2]) \oplus 2[2,0,2]\oplus [0,4,0]
\oplus[1,2,1] \label{6x6x6x6d4} \\
&& \Leftrightarrow \mb6\otimes\mb6\otimes\mb6\otimes\mb6  =
3\cdot\mb1\op 7\cdot\mb{15}\op 6\cdot\mb{20^\pp}\op 3\cdot(\mb{45}
\op\mbb{45})\op 2\cdot\mb{84}\op \mb{105}\op \mb{175} \, .
\nn
\ea

\subsubsection{$\D_0=4$, $[0,0,0]$}
\label{d4r1}

In the singlet sector at $\D_0=4$ one can consider the following basis
of operators.

There are the two operators (\ref{d4-fieldstr}) involving field
strengths, which including the normalisation read 
\be 
\scrO^{(1)}_{4,\mb1} = \frac{1}{\gy^2}\,\Tr\left( 
F_{\mu\nu}F^{\mu\nu}\right) \quad {\rm and} \quad 
\scrO^{(2)}_{4,\mb1} = \frac{1}{\gy^2}\,\Tr\left( 
F_{\mu\nu}\tilde{F}^{\mu\nu}\right) \, .
\label{O-1-fieldstr}
\ee
With two derivatives there is the operator 
\be
\scrO^{(3)}_{4,\mb1} = \frac{1}{\gy^2} \, \Tr\left(\scrD_\mu
\v^i\scrD^\mu\v^i \right) \, .
\label{O-1-deriv}
\ee
The two singlets in (\ref{d4fermi}) are respectively 
\be
\scrO^{(4)}_{4,\mb1} = \frac{1}{\gy^3N^{1/2}}\,\S^i_{AB} \,
\Tr\left(\v^i\lambda^{\a A}\lambda^B_\a \right) \quad {\rm and}
\quad \scrO^{(5)}_{4,\mb1} = \frac{1}{\gy^3N^{1/2}}\,\Sbar_i^{AB} 
\,\Tr\left(\v^i\bar\lambda_{\adot A}\bar\lambda^\adot_B \right) \,.
\label{O-1-fermi}
\ee
Finally using cyclicity of the trace one finds four (out of the
possible six) operators made out of four scalars, two single-trace and
two double-trace,  
\ba
&& \scrO^{(6)}_{4,\mb1} = \frac{1}{\gy^4N} \,\Tr\left(
\v^i\v^j\v^i\v^j\right) \, , \quad \scrO^{(7)}_{4,\mb1} = 
\frac{1}{\gy^4N} \,\Tr\left(\v^i\v^i\v^j\v^j\right) \, ,
\label{O-1-scal-s} \\
&& \scrO^{(8)}_{4,\mb1} = \frac{1}{\gy^4N} \,\Tr\left(
\v^i\v^j\right)\Tr\left(\v^i\v^j\right) \, , \quad 
\scrO^{(9)}_{4,\mb1} = \frac{1}{\gy^4N} \,
\Tr
\left(\v^i\v^i\right)\Tr\left(\v^j\v^j\right) \, .
\label{O-1-scal-d}
\ea
Among the $\D_0=4$ operators in the SU(4) singlet two are known to be
protected. They are the $\theta^4$ and $\bar\theta^4$ components of
the $\scrN$=4 supercurrent multiplet, $\scrC^-$ and $\scrC^+$. These
are linear combinations of (\ref{O-1-fieldstr}), (\ref{O-1-deriv}),
(\ref{O-1-fermi}) and (\ref{O-1-scal-s}), 
\be 
\scrC^- = \frac{1}{\gy^2}\,\Tr\left(F^-_{\mu\nu}F^{-\,\mu\nu} \right) 
+\cdots \, , \quad
\scrC^+ = \frac{1}{\gy^2}\, \Tr\left( F^+_{\mu\nu}F^{+\,\mu\nu}
\right)+\cdots \,.
\label{O-4-1-protec}
\ee 
These two operators are 1/2 BPS and dual to complex combinations of the
dilaton and \RR scalar in the AdS/CFT correspondence, $\tau$ and
$\bar\tau$, where $\tau = \tau_1 + i\tau_2 = C^{(0)} + i \er^{-\phi}$.
The sum $\scrC^-+\scrC^+$ is proportional to the $\scrN$=4 on-shell
lagrangian.

Another known operator in this sector is a component of the Konishi
multiplet, $\scrK_\mb1^\pp$, which is a different linear combination of
the same operators, orthogonal to $\scrC^-$ and $\scrC^+$. Since it
belongs to the Konishi multiplet this operators is not corrected at
the instanton level. 

The operators (\ref{O-1-scal-s}) and (\ref{O-1-scal-d}) have been
studied in perturbation theory in \cite{appss,bks} and their anomalous
dimensions were computed at one loop, where there is no mixing with
the remaining operators in the sector. At higher loops mixing is
expected to occur. At the non-perturbative level there is mixing among
all the operators (\ref{O-1-fieldstr})-(\ref{O-1-scal-d}) at leading
order in $\gy$. As will be now shown all the operators
$\scrO^{(r)}_{4,\mb1}$, $r=1,\ldots,9$, can soak up the correct
combination of fermionic modes and hence all their two-point functions
can get instanton corrections.

For the two operators (\ref{O-1-fieldstr}) the terms contributing to
two-point functions in the one-instanton sector are
\ba
&& \scrO^{(1)}_{4,\mb1} \rightarrow \fr{\gy^2} \Tr\left(
F_{\mu\nu}^{(4)}F^{(4)\mu\nu} \right) \label{O1-ff-zm} \\
&& \scrO^{(2)}_{4,\mb1} \rightarrow \fr{\gy^2} \Tr\left(
F_{\mu\nu}^{(4)}\tilde F^{(4)\mu\nu} \right) \, , 
\label{O1-fft-zm}
\ea
which using (\ref{a4su4}) can be shown to be proportional to 
\be
\veps_{ABCD}\veps_{A^\pp B^\pp C^\pp D^\pp} \left(
\scrmf^A\scrmf^B\scrmf^C\scrmf^D \right) 
\left(\scrmf^{A^\pp}\scrmf^{B^\pp}\scrmf^{C^\pp}\scrmf^{D^\pp} \right)
\, , 
\label{d4-fstr-zm}
\ee
so that they contain the combination (\ref{saturmodes}). 

Similarly for the operator (\ref{O-1-deriv}) one gets a potentially
non vanishing contribution considering
\be
\scrO^{(3)}_{4,\mb1} \rightarrow \fr{\gy^2} \Tr\left(\scrD^{(0)}_\mu
\v^{(6)i}\scrD^{(0)\mu}\v^{(2)i} + \scrD^{(4)}_\mu
\v^{(2)i}\scrD^{(0)\mu}\v^{(2)i} \right) \, ,
\label{d4-deriv-zm}
\ee
which yields the combination
\be
\veps_{ABCD}\veps_{A^\pp B^\pp C^\pp D^\pp} \left( \scrmf^{A^\pp}
\scrmf^{B^\pp}\scrmf^{C^\pp}\scrmf^{D^\pp} \scrmf^{[A}\scrmf^{B]}
\scrmf^{[C}\scrmf^{D]} \right) \, ,
\label{d4-deriv-zm-m}
\ee
which can saturate the superconformal modes. 

In the expansion of the two operators (\ref{O-1-fermi}) the relevant
terms are
\ba
&& \hsp{-1}
\scrO^{(4)}_{4,\mb1} \rightarrow \frac{1}{\gy^3N^{1/2}}\,\S^i_{AB} \,
\Tr\left(\v^{(6)i}\lambda^{(1)\a A}\lambda^{(1)B}_\a + 
\v^{(2)i}\lambda^{(5)\a A}\lambda^{(1)B}_\a + 
\v^{(2)i}\lambda^{(1)\a A}\lambda^{(5)B}_\a \right) 
\label{d4-ferm-zm} \\
&& \hsp{-1}
\scrO^{(5)}_{4,\mb1} \rightarrow \frac{1}{\gy^3N^{1/2}}\,\Sbar_i^{AB} 
\,\Tr\left(\v^{(2)i}\bar\lambda^{(3)}_{\adot A}
\bar\lambda^{(3)\adot}_B \right) \,,
\label{d4-bferm-zm} 
\ea
which again can saturate the fermion integrations in a two-point
function since these expressions contain respectively
\be
\veps_{ABCD}\veps_{A^\pp B^\pp C^\pp D^\pp}\left(\scrmf^{[C}
\scrmf^{D]}\scrmf^A\scrmf^B \scrmf^{A^\pp} \scrmf^{B^\pp}
\scrmf^{C^\pp}\scrmf^{D^\pp} \right)
\label{d4-ferm-zm-m}
\ee
and
\be
\veps_{AC^\pp D^\pp E^\pp}\veps_{BC^{\dpp} D^{\dpp} E^{\dpp}}
\left(\scrmf^{[A}\scrmf^{B]}\scrmf^{C^\pp}\scrmf^{D^\pp}
\scrmf^{E^\pp} \scrmf^{C^\dpp}\scrmf^{D^{\dpp}}\scrmf^{E^{\dpp}} 
\right) \, .
\label{d4-bferm-zm-m}
\ee

Finally the operators (\ref{O-1-scal-s}) and (\ref{O-1-scal-d}),
quartic in the scalars, can contribute to two-point correlators
through the terms
\ba
&& \hsp{-1} \scrO^{(6)}_{4,\mb1} \rightarrow \frac{1}{\gy^4N} \,
\Tr\left(\v^{(2)i}\v^{(2)j}\v^{(2)i}\v^{(2)j}\right) 
\label{d4-scal-s1-zm} \\ 
&& \hsp{-1} \scrO^{(7)}_{4,\mb1} \rightarrow \frac{1}{\gy^4N} \,
\Tr\left(\v^{(2)i}\v^{(2)i}\v^{(2)j}\v^{(2)j}\right) \, 
\label{d4-scal-s2-zm} \\
&& \hsp{-1} \scrO^{(8)}_{4,\mb1} \rightarrow \frac{1}{\gy^4N} \,
\Tr\left(\v^{(2)i}\v^{(2)j}\right)\Tr\left(\v^{(2)i}\v^{(2)j}\right) 
\label{d4-scal-d1-zm} \\
&& \hsp{-1} \scrO^{(9)}_{4,\mb1} \rightarrow \frac{1}{\gy^4N} \,
\Tr\left(\v^{(2)i}\v^{(2)i}\right)\Tr\left(\v^{(2)j}\v^{(2)j}\right) 
\label{d4-scal-d2-zm}
\ea
whose zero-mode structure is
\be
\veps_{ABCD}\veps_{A^\pp B^\pp C^\pp D^\pp}\left(\scrmf^{[A}\scrmf^{B]}
\scrmf^{[C}\scrmf^{D]} \scrmf^{[A^\pp} \scrmf^{B^\pp]}
\scrmf^{[C^\pp}\scrmf^{D^\pp]} \right) \, .
\label{d4-scal-zm-m}
\ee

From the above analysis it is clear that the classical expressions of
all the fields in the singlet sector at $\D_0=4$ can contain the correct
combination of fermion zero-modes to produce non-vanishing two-point
functions. The expansion of all the combinations (\ref{d4-fstr-zm}),
(\ref{d4-deriv-zm-m}), (\ref{d4-bferm-zm-m}) and (\ref{d4-scal-zm-m}),
in which all the modes are taken to be superconformal, involves exactly
two $\zeta$'s for each flavour. Moreover at leading order in the
coupling there is no dependence on the $\nu$ and $\bar\nu$ other than
what comes from the measure. So the integration over the fermionic
part of the moduli space is expected to be non-zero for all two-point 
correlation functions $\la\scrO^{(r)}_{4,\mb1}(x_1)
\bar\scrO^{(s)}_{4,\mb1}(x_2)\ra$, $r,s=1,\ldots,9$. This is a general
result, in sectors in which instanton contributions are present the
non-perturbative mixing should be expected to be more complicated
than at the first few orders in perturbation theory. In this
particular case one has to diagonalise the whole 9$\times$9
problem to extract the instanton induced anomalous dimensions. As
already remarked at least three of the eigenvalues of the anomalous
dimensions matrix should vanish, since they correspond to the
operators $\scrC^-$, $\scrC^+$ and $\scrK^\pp_\mb1$ discussed above. 
As  discussed in the introduction further simplifications arise from
taking into account the constraints imposed by the PSU(2,2$|$4), which
imply that all the operators in a multiplet have the same anomalous
dimension. Therefore other operators can be eliminated from the mixing
problem if they are identified as superconformal descendants of
operators whose anomalous dimension is known. 

We shall now compute explicitly the two-point functions involving the
quartic scalar operators in (\ref{O-1-scal-s}) and (\ref{O-1-scal-d})
in the one-instanton sector~\footnote{In the following discussion we
omit the subscripts indicating the dimension and SU(4)
representation.}.  This will prove that the situation in this sector
is different from what was found in sections \ref{dim2scal} and
\ref{d3r6} and indeed singlet operators at $\D_0=4$ do receive
instanton corrections.

In order to evaluate the semiclassical two-point functions of these
operators we need their expression in the instanton background, \ie
the explicit form of (\ref{d4-scal-s1-zm})-(\ref{d4-scal-d2-zm}). As
usual it is convenient to work with the scalar fields written as 
$\v^{AB}$. The composite operators become
\ba
\scrO^{(6)} &\!\!=\!\!& \fr{\gy^4N} \veps_{A_1B_1A_3B_3}
\veps_{A_2B_2A_4B_4} \Tr \left( \v^{A_1B_1}\v^{A_2B_2}\v^{A_3B_3}
\v^{A_4B_4} \right) 
\label{O-1-s1-AB} \\
\scrO^{(7)} &\!\!=\!\!& \fr{\gy^4N} \veps_{A_1B_1A_2B_2}
\veps_{A_3B_3A_4B_4} \Tr \left( \v^{A_1B_1}\v^{A_2B_2}\v^{A_3B_3}
\v^{A_4B_4} \right) 
\label{O-1-s2-AB} \\
\scrO^{(8)} &\!\!=\!\!& \fr{\gy^4N} \veps_{A_1B_1A_3B_3}
\veps_{A_2B_2A_4B_4} \Tr \left( \v^{A_1B_1}\v^{A_2B_2} \right)
\Tr\left(\v^{A_3B_3} \v^{A_4B_4} \right) 
\label{O-1-d1-AB} \\
\scrO^{(9)} &\!\!=\!\!& \fr{\gy^4N} \veps_{A_1B_1A_2B_2}
\veps_{A_3B_3A_4B_4} \Tr \left( \v^{A_1B_1}\v^{A_2B_2} \right)
\Tr\left(\v^{A_3B_3} \v^{A_4B_4} \right) \,.
\label{O-1-d2-AB}
\ea
The two combinations that are needed are thus
\ba
&& \hsp{-1.5}\Tr\left[\left( \v^{A_1B_1}\v^{A_2B_2}\v^{A_3B_3}
\v^{A_4B_4} \right)(x)\right] = \frac{2^8\rho^8}{[(x-x_0)^2+\rho^2]^8} 
\left[ \left( \zeta^{B_1}_\a \zeta^{\b\,A_1} - \zeta^{A_1}_\a 
\zeta^{\b\,B_1} \right) \right. 
\label{tr4sc-inst} \\
&& \left.\left(\zeta^{B_2}_\b \zeta^{\g\,A_2} - \zeta^{A_2}_\b 
\zeta^{\g\,B_2}\right)\left(\zeta^{B_3}_\g \zeta^{\d\,A_3} - 
\zeta^{A_3}_\g \zeta^{\d\,B_3}\right)\left(\zeta^{B_4}_\d 
\zeta^{\a\,A_4} - \zeta^{A_4}_\d \zeta^{\a\,B_4}\right) \right](x)
+ \cdots \nn
\ea
and 
\ba
\Tr\left[\left( \v^{A_1B_1}\v^{A_2B_2} \right)(x)\right] &\!\!=\!\!&
\frac{2^4\rho^4}{[(x-x_0)^2+\rho^2]^4} 
\label{tr2sc-inst} \\
&& \left[ \left(\zeta^{B_1}_\a \zeta^{\b\,A_1} - \zeta^{A_1}_\a 
\zeta^{\b\,B_1} \right) \left(\zeta^{B_2}_\b \zeta^{\a\,A_2} - 
\zeta^{A_2}_\b \zeta^{\a\,B_2} \right) \right](x) + \cdots \, . \nn
\ea
In both the previous expressions the ellipsis stands for terms
involving the modes of type $\nu$ and $\bar\nu$ as well as terms with
more than eight fermion zero modes which are not relevant in
semiclassical approximation. 

Substituting into the expressions for the composite fields we find
that only the operators $\scrO^{(6)}$ and
$\scrO^{(8)}$ can soak up the superconformal modes in the
combination (\ref{saturmodes}). $\scrO^{(9)}$ vanishes identically,
whereas $\scrO^{(7)}$ contains terms with eight superconformal modes,
but not in the required combination. We thus find that the
non-vanishing correlation functions in the one-instanton sector in
semiclassical approximation are 
\ba
G^{(\rm a)}(x_1,x_2) &\!\!=\!\!& \la \scrO^{(6)}(x_1) 
\scrO^{(6)} (x_2) \ra \nn \\ 
G^{(\rm b)}(x_1,x_2) &\!\!=\!\!& \la \scrO^{(8)}(x_1) 
\scrO^{(8)}(x_2) \ra \label{nonzero-O1-2pt} \\
G^{(\rm c)}(x_1,x_2) &\!\!=\!\!& \la \scrO^{(6)}(x_1) 
\scrO^{(8)} (x_2) \ra \, . \nn
\ea
In computing these correlation functions at leading order in the
coupling the $\nu$ and $\bar\nu$ modes appear only in the moduli space
integration measure and we find
\ba
&& \la \scrO^{(r)}(x_1) \scrO^{(s)}(x_2) \ra = 
\frac{\er^{2\pi i\tau}}{N^2(N-1)!(N-2)!}
\int \dr\rho \, \dr^4x_0 \, \dr^5\Omega \prod_{A=1}^4 \dr^2\eta^A \,
\dr^2 {\bar\xi}^A \, \rho^{4N-7} \nn \\
&& \int_0^\infty \dr r \, r^{4N-3} \er^{-2\rho^2 r^2} \,
\hat\scrO^{(r)}(x_1;\rho,x_0;\zeta) 
\hat\scrO^{(s)}(x_2;\rho,x_0;\zeta) \, ,
\label{O-1-2pt-a}
\ea
where as usual the classical expressions for the operators in the
presence of an instanton are denoted by a hat. The exact numerical
coefficients will be reinstated in the final formulae. 

Since there is no dependence on $\nu$ and $\bar\nu$ in the integrand
the integrations over the five-sphere and the radial variable $r$ can
be immediately performed and one obtains 
\ba
\la \scrO^{(r)}(x_1) \scrO^{(s)}(x_2) \ra &\!\!=\!\!&  
\frac{2^{-2N}\Gamma(2N-1)\,\er^{2\pi i\tau}}{N^2(N-1)!(N-2)!} 
\label{O-1-2pt-b} \\
&& \int \frac{\dr\rho \, \dr^4x_0}{\rho^5} \prod_{A=1}^4 
\dr^2\eta^A \, \dr^2 {\bar\xi}^A \,   
\hat\scrO^{(r)}(x_1;\rho,x_0;\zeta) 
\hat\scrO^{(s)}(x_2;\rho,x_0;\zeta) \, .
\nn 
\ea
Using (\ref{tr4sc-inst}) and (\ref{tr2sc-inst}) to compute
(\ref{O-1-s1-AB}) and (\ref{O-1-d1-AB}) and integrating over the
superconformal modes we then get 
\ba
\la \scrO^{(r)}(x_1) \scrO^{(s)}(x_2) \ra &\!\!=\!\!&  c^{rs}\,
\frac{3^4\,5^2\,\pi^{-13}\,2^{-2N-15}\Gamma(2N-1)\,\er^{2\pi i\tau}}
{N^2(N-1)!(N-2)!}(x_1-x_2)^8 \nn \\
&&\int \dr\rho \, \dr^4x_0
\frac{\rho^{11}}{[(x_1-x_0)^2+\rho^2]^8[(x_2-x_0)^2+\rho^2]^8} \, , 
\label{O-1-2pt-c}
\ea
where the coefficients $c^{rs}$ take into account the different
prefactors in the term (\ref{saturmodes}) in the expansion of the
operators. By explicitly computing the contractions with the
Levi-Civita tensors in the definitions of the operators we find for
these coefficients
\ba
c^{66} &\!\!=\!\!& 1 \nn \\
c^{88} &\!\!=\!\!& 2^4 \label{crs-coeff} \\
c^{68} &\!\!=\!\!& c^{86} = -2^2 \, . \nn
\ea
The bosonic integrals which remain to be evaluated in
(\ref{O-1-2pt-c}) are logarithmically divergent as can be seen
introducing Feynman parameters to rewrite
\ba
&& \scrI = \int\dr\rho\,\dr^4x_0
\frac{\rho^{11}}{[(x_1-x_0)^2+\rho^2]^8[(x_2-x_0)^2+\rho^2]^8} \nn \\
&& = \frac{\Gamma(16)}{\left[\Gamma(8)\right]^2} \int_0^1 \dr\a_1\,
\dr\a_2 \, \d(\a_1+\a_2-1)\,\a_1^7\a_2^7 \nn \\
&& \times \int \dr\rho\,\dr^4x_0
\frac{\rho^{11}}{\left[(x_0-\sum_i\a_i x_i)^2 + \rho^2 + 
\a_1\a_2 x_{12}^2 \right]^{16}} \, .
\label{feynpar}
\ea
After the standard manipulations the $\rho$ integral can be performed
and using dimensional regularisation for the $x_0$ integration we get 
\be
\scrI = \pi^{2+\epsilon} c(4-\epsilon) 
\frac{\Gamma(14+\epsilon)}{\left[\Gamma(8)\right]^2} 
\frac{1}{(x_1-x_2)^{16-\epsilon}} \int_0^1\frac{\dr\a}
{\left[\a(1-\a)\right]^{1-\epsilon}} \, , 
\label{eps-pole}
\ee
where $c(d) = 3840/[(d-20)(d-22)(d-24)(d-26)(d-28)(d-30)]$. The final
integration over $\a$ produces a $1/\epsilon$ pole which is the signal
of a logarithmic divergence in dimensional regularisation.  In
conclusion we find non-zero entries in the matrix $K^{rs}$ defined in
section \ref{genadim} for to the two-point functions (\ref
{nonzero-O1-2pt}), which can be read from (\ref{O-1-2pt-c}),
(\ref{crs-coeff}) and (\ref{eps-pole}). As anticipated after equation
(\ref{gam-3-6}), these matrix elements behave as $\er^{2\pi i\tau}$
with no additional powers of $\gy$. 

The above calculation shows that $\D_0=4$ operators in the singlet do
acquire an instanton induced anomalous dimension. This can be
confirmed analysing the OPE of a four-point function as in section
\ref{d3r6rev}. In fact the $t$-channel, $x_{13}\to0$, $x_{24}\to0$, of
the same four-point function (\ref{4ptope}), which is computed in
appendix \ref{ope}, displays a singularity corresponding to the
exchange of operators of dimension 4. The results of the following
subsections show that this contribution to the OPE must come from
operators in the singlet.

Although we have not computed all the entries in the matrix $K^{rs}$
and diagonalised $\Gamma^{rs}=(T^{-1}K)^{rs}$, our results appear to
disagree with those of \cite{afp}. There, on the basis of an OPE
analysis, it was argued that only operators whose dimension remains
finite in the large $N$ limit get an instanton correction. These are
multi-trace operators dual to supergravity multi-particle bound states
in the AdS/CFT correspondence. We have instead shown that single trace
operators, which are dual to massive string modes and whose dimension
diverges in the $N\to\infty$ limit, have non-zero two-point functions
in the one-instanton sector.

\subsubsection{$\D_0=4$, $[1,0,1]$}
\label{d4r15}

The representation $[1,0,1]\equiv\mb{15}$ occurs in (\ref{6x6d4}), in
(\ref{6x4x4d4}) and (\ref{6xb4xb4d4}) with multiplicity 2 and in
(\ref{6x6x6x6d4}) with multiplicity 7, so there are in principle many
operators to be considered. The actual number of independent
gauge-invariant operators however reduces when the cyclicity of the
trace is taken into account. In particular this implies that there are
only single trace operators. 

No operators can be constructed from (\ref{6x6d4}), \ie with two
scalars and two covariant derivatives, because of antisymmetry. 

A basis of independent operators can be obtained as follows. 
To realise the two operators in (\ref{6x4x4d4}) we need to consider
\be
\scrO^{(1)[ij]}_{4,\mb{15}} = \fr{\gy^3N^{1/2}} \,\Tr\left(
\S^{[i}_{AB}\v^{j]}\lambda^{\a A}\lambda^B_\a \right) 
\label{O-15-fermi-1}
\ee 
and 
\be
\scrO^{(2)[ijkl]}_{4,\mb{15}} = \fr{\gy^3N^{1/2}} \,\Tr\left(
t^{[ijk}_{AB}\v^{l]}\lambda^{\a A}\lambda^B_\a \right) \, ,
\label{O-15-fermi-2}
\ee
where the symbols $\S^i_{AB}$ are defined in (\ref{defsig6}) and
$t^{[ijk]}_{AB}$ is the projector defined in (\ref{10proj}). 

Similarly from (\ref{6xb4xb4d4}) we get 
\be 
\scrO^{(3)}_{4,\mb{15}\,[ij]} = \fr{\gy^3N^{1/2}} \, \Tr\left(
\Sbar^{AB}_{[i} \v_{j]} \bar\lambda_{\adot A}\bar\lambda^\adot_B
\right) 
\label{O-15-fermi-3}
\ee 
and 
\be
\scrO^{(4)}_{4,\mb{15}\,[ijkl]} = \fr{\gy^3N^{1/2}} \, \Tr\left(
{\bar t}^{AB}_{[ijk} \v_{l]} \bar\lambda_{\adot A}
\bar\lambda^\adot_B \right) \, .
\label{O-15-fermi-4}
\ee

Operators in the $\mb{15}$ made of four scalars can be constructed
antisymmetrising a pair of indices and contracting the remaining two
or completely antisymmetrising the four indices.  We obtain however
only one independent gauge-invariant operator,
\be
\scrO^{(5)[ij]}_{4,\mb{15}} = \fr{g^4N}\, 
\Tr\left(\v^{[i}\v^{j]}\v_k\v^k \right) \, ,
\label{O-15-scal-1} 
\ee
The other combinations of four $\mb6$'s forming a $\mb{15}$ are
identically zero because of cyclicity of the trace.
Antisymmetrisation in (\ref{O-15-scal-1}) implies that no double trace
operators can be realised.

We can now analyse the instanton contributions to two-point functions
in this sector. An analysis on the lines of that in the previous
subsection will show that none of the above operators can soak up the
correct combination of superconformal modes to give rise to
non-vanishing two-point correlators. As usual the strategy is to
choose a specific component for the operators and study the
zero-mode structure, the result is the same for all the components.
For the operator (\ref{O-15-fermi-1}) we consider for instance the
component $\scrO^{(1)[12]}_{4,\mb{15}}$ which is proportional to 
\ba
&& \hsp{-0.6}\Tr\left( \v^{13}\lambda^{[1}\lambda^{4]} + 
\v^{13} \lambda^{[2}\lambda^{3]} + \v^{24}\lambda^{[1}\lambda^{4]} 
+ \v^{24}\lambda^{[2}\lambda^{3]} \right. \nn \\
&& \left. - \v^{14}\lambda^{[1}\lambda^{3]} - 
\v^{14}\lambda^{[2}\lambda^{4]} - \v^{23}\lambda^{[1}\lambda^{3]} 
- \v^{23}\lambda^{[2}\lambda^{4]} \right) \, .
\label{O-15-fer1-12}
\ea
In order to saturate the eight superconformal modes required to give
rise to a non-zero two-point functions in all these terms the
contributions one needs to consider are
\be
\Tr\left( \v^{(2)AB}\lambda^{(1)[C}\lambda^{(5)D]} + 
\v^{(2)AB}\lambda^{(5)[C}\lambda^{(1)D]} + 
\v^{(6)AB}\lambda^{(1)[C}\lambda^{(1)D]} \right) \, .
\label{O-15-fer1-12satur}
\ee
Expanding in this way the first term in (\ref{O-15-fer1-12}) yields a
contribution proportional to 
\be
\veps_{ABCD}\zeta^1\zeta^3\zeta^1\zeta^4 \zeta^A\zeta^B\zeta^C\zeta^D
\, , \label{O-15-1-fer12-zm}
\ee
which does not contain the required combination (\ref{saturmodes}),
having only one mode of flavour 2. Similar considerations apply to the
other terms in (\ref{O-15-fer1-12}) as well as to the other components
of the same operator. For the operator (\ref{O-15-fermi-3}) the
analysis is completely analogous and gives the same result. 

For the operator (\ref{O-15-fermi-2}) we can consider the component
$\scrO^{(2)[1234]}_{4,\mb{15}}$. After contracting the indices
the result one obtains is a sum of terms of the form
\be
\Tr\left(\v^{AB}\lambda^C\lambda^C + \v^{AC}\lambda^B\lambda^C
\right) \, ,
\label{O-15-fer2-1234}
\ee
which when expanded as in (\ref{O-15-fer1-12satur}) in order to select
the terms with eight fermion modes again give contributions of the
type (\ref{O-15-1-fer12-zm}), where one flavour occurs thrice and one
flavour only once. The result is analogous for other components
and also for the operator (\ref{O-15-fermi-4}), which can be analysed
much in the same way.

Similar results hold for the operator (\ref{O-15-scal-1}) constructed
from only scalar fields. The $[12]$ component of (\ref{O-15-scal-1}) is
\ba
\scrO^{(5)[12]}_{4,\mb{15}} &\!\!\sim\!\!& \Tr\left\{
\left(\v^{14}+\v^{23}\right)\left(-\v^{13}+\v^{24}\right)\left[
\left(\v^{14}+\v^{23}\right)^2+\left(-\v^{13}+\v^{24}\right)^2 
\right. \right. \label{O-15-sc1-12} \\ 
&& \left.\left. +\left(\v^{12}+\v^{34}\right)^2+
\left(-\v^{14}+\v^{23}\right)^2+\left(-\v^{13}-\v^{24}\right)^2+
\left(\v^{12}-\v^{34}\right)^2 \right] \right\} \, . \nn
\ea
and a combination of eight fermion modes is obtained selecting in each
scalar $\v$ the term with two fermion modes, $\v^{(2)AB}$. It is then
easy to verify that none of the terms in the resulting expansion can
soak up two modes of each flavour. As in the case of the operators
(\ref{O-15-fermi-1})-(\ref{O-15-fermi-4}) in each term there is always
a flavour occurring thrice and one occurring only once.  Analogous
considerations can be repeated for all the components.

In conclusion we find that none of the two-point functions
$\la\scrO^{(r)}_{4,\mb{15}}(x_1)\scrO^{(s)}_{4,\mb{15}}(x_2)\ra$,
$r,s=1,\ldots,7$, in this sector receives instanton corrections, so
that the corresponding non-perturbative contributions to all the
anomalous dimensions vanish.

\subsubsection{$\D_0=4$, $[0,2,0]$}
\label{d4r20}

Scalar operators in the $[0,2,0]\equiv\mb{20^\pp}$ were studied in
\cite{bers}, where the mixing at the perturbative level was resolved
at order $\gy^2$. In the same paper it was also argued that anomalous
dimensions in this sector do not receive instanton corrections. In
this section we shall re-derive this result by directly analysing the
instanton contributions to two-point functions.

The basis of operators we consider is different from the starting
point of \cite{bers}. Moreover as in previously considered sectors at
the non-perturbative level there is mixing among all the operators at
leading order in $\gy$. From (\ref{d4-deriv}) which admits the
decomposition (\ref{6x6d4}) there is one operator that can
be constructed as 
\be
\scrO^{(1)\{ij\}}_{4,\mb{20^\pp}} = \fr{\gy^2}\, \Tr\left(\scrD_\mu 
\v^{\{i}\scrD^\mu\v^{j\}} \right) \equiv  \fr{\gy^2}\, 
\Tr\left(\scrD_\mu \v^i\scrD^\mu\v^j \right) - \fr{6\gy^2}\d^{ij}
\,\Tr\left(\scrD_\mu \v^k\scrD^\mu\v_k \right) \, ,
\label{O-20-deriv}
\ee
where as usual $\{ij\}$ indicates symmetrisation and removal of traces
in flavour space, as explicitly indicated in (\ref{O-20-deriv}). 

We then find two operators involving fermionic bilinears respectively
from (\ref{6x4x4d4}) and (\ref{6xb4xb4d4}). They read
\be
\scrO^{(2)\{ij\}}_{4,\mb{20^\pp}} = \fr{\gy^3N^{1/2}} \, \Tr\left(
\S^{\{i}_{AB} \v^{j\}}\lambda^{\a A}\lambda_\a^B \right) 
\label{O-20-fermi-1}
\ee 
and 
\be
\scrO^{(3)\{ij\}}_{4,\mb{20^\pp}} = \fr{\gy^3N^{1/2}} \, \Tr\left(
\Sbar^{AB\{i} \v^{j\}}\bar\lambda_{\adot A}\bar\lambda^\adot_B 
\right) \, . 
\label{O-20-fermi-2}
\ee

Operators made out of four scalars in the $\mb{20^\pp}$ can be single
or double trace. There is a total of four operators of this type and
as a basis we consider the single trace 
\ba
\scrO^{(4)\{ij\}}_{4,\mb{20^\pp}} &\!\!=\!\!& \fr{\gy^4N} \, \Tr\left(
\v^{\{i}\v^{j\}}\v_k\v^k \right) \, , \label{O-20-sc-s1} \\
\scrO^{(5)\{ij\}}_{4,\mb{20^\pp}} &\!\!=\!\!& \fr{\gy^4N} \, \Tr\left(
\v^{\{i}\v_k\v^{j\}}\v^k \right)  \label{O-20-sc-s2}
\ea
and the double trace combinations
\ba
\scrO^{(6)\{ij\}}_{4,\mb{20^\pp}} &\!\!=\!\!& \fr{\gy^4N} \, \Tr\left(
\v^{\{i}\v^{j\}}\right) \Tr\left(\v_k\v^k \right) \, , 
\label{O-20-sc-d1} \\
\scrO^{(7)\{ij\}}_{4,\mb{20^\pp}} &\!\!=\!\!& \fr{\gy^4N} \, \Tr\left(
\v^{\{i}\v_k\right) \Tr\left(\v^{j\}}\v^k \right) \, . 
\label{O-20-sc-d2}
\ea

The analysis of the zero-mode structure of these operators follows
closely what was done for the $[1,0,1]$ sector. Notice that the
antisymmetrisation was irrelevant in deriving the results of the
previous subsection and only played a r\^ole in constructing the basis
of operators. Therefore considering the $\{12\}$ component of
(\ref{O-20-fermi-1})-(\ref{O-20-sc-d2}) we can immediately conclude
that all the two-point functions $\la\scrO^{(r)\{12\}}_{4,\mb{20^\pp}}
(x_1)\scrO^{(s)\{12\}}_{4,\mb{20^\pp}}(x_2)\ra$, for $r,s=2,\ldots,7$,
vanish in the one-instanton sector and in semiclassical
approximation. The zero-mode content of
(\ref{O-20-fermi-1})-(\ref{O-20-sc-d2}) is in fact exactly the same
found in the analogous antisymmetric combinations belonging to the
$\mb{15}$. If one chooses a component which is diagonal in flavour
space, $\scrO^{\{ii\}}_{4,\mb{20^\pp}}$, the analysis is slightly more
involved since the subtraction of the trace is then crucial in
cancelling terms which could saturate the superconformal modes in a
two-point function. A careful analysis of the instanton profiles of
the operators shows that indeed for these components the two-point
correlation functions vanish as well. 

The operator (\ref{O-20-deriv}) must be analysed separately since it
has no analogue in the $[1,0,1]$ sector. Considering again the
$\{12\}$ component the terms in the expansion which soak up eight
fermion modes are
\ba
\scrO^{(1)\{12\}}_{4,\mb{20^\pp}} &\!\!\to\!\!& \fr{\gy^2} \, \Tr\left(
\scrD^{(0)}_\mu\v^{(2)\{1}\scrD^{(4)\mu}\v^{(2)2\}} + 
\scrD^{(4)}_\mu\v^{(2)\{1}\scrD^{(0)\mu}\v^{(2)2\}} \right) \, .
\nn \\
&\!\!\sim\!\!& \fr{\gy^2} \, \Tr\left[
\scrD^{(0)}_\mu\left(\v^{(2)14}+\v^{(2)23}\right)\scrD^{(4)\mu}
\left(-\v^{(2)13} + \v^{(2)24}\right) + \cdots \right] \, ,
\label{O-20-der-satur}
\ea
where the ellipsis stands for symmetrisation. Recalling that
$\scrD^{(4)}$ involves one fermion mode of each flavour we find that
(\ref{O-20-der-satur}) contains the terms 
\be
\veps_{ABCD}\zeta^A\zeta^B\zeta^C\zeta^D \left(
\zeta^1\zeta^4\zeta^1\zeta^3 + \zeta^1\zeta^4\zeta^2\zeta^4 + 
\zeta^2\zeta^3\zeta^1\zeta^3 + \zeta^2\zeta^3\zeta^2\zeta^4 \right)
\label{O-20-der-zm}
\ee
none of which equals the required combination (\ref{saturmodes}). 

Hence we find that in the scalar $[0,2,0]$ sector there are no
operators which acquire an anomalous dimension at the instanton
level.

\subsubsection{$\D_0=4$, $[2,1,0]$ and $[0,1,2]$}
\label{d4r45} 

Operators in the representations $[2,1,0]\equiv\mb{45}$ and
$[0,1,2]\equiv\mbb{45}$ arise respectively from (\ref{6x4x4d4}) and
(\ref{6xb4xb4d4}), involving a scalar and a fermionic bilinear, and
from (\ref{6x6x6x6d4}), quartic in the scalars, which contains both
$\mb{45}$ and $\mbb{45}$ with multiplicity 3. Operators of the latter
type can in principle be single- or double-trace. 

To project (\ref{d4fermi}) onto the $\mb{45}$ we consider 
\ba
\scrO^{(1)(AB)}_{4,\mb{45}[CD]} &\!\!=\!\!& \fr{g^3N^{1/2}}\, 
\Tr\left[\bar\v_{CD}\lambda^{\a(A}\lambda_\a^{B)} - \fr6
\left(\d^A_C\,\bar\v_{ED}\lambda^{\a(E}\lambda_\a^{B)} 
\right. \right. \nn \\
&& \left.\left. - \d^A_D\,\bar\v_{CE}\lambda^{\a(E}\lambda_\a^{B)} + 
\d^B_C\,\bar\v_{ED}\lambda^{\a(A}\lambda_\a^{E)} -
\d^B_D\,\bar\v_{CE}\lambda^{\a(A}\lambda_\a^{E)}\right) 
\rule{0pt}{16pt}\!\right] \, .
\label{O-45-fermi} 
\ea
The operator in the $\mbb{45}$ made of two fermions a scalar is
obtained analogously as 
\ba 
\scrO^{(1)[CD]}_{4,\mbb{45}(AB)} &\!\!=\!\!& \fr{g^3N^{1/2}}\, 
\Tr\left[\v^{CD} \bar\lambda_{\adot(A}\bar\lambda^\adot_{B)} - \fr6
\left(\d_A^C \v^{ED}\bar\lambda_{\adot(E}\bar\lambda^\adot_{B)} 
\right. \right. \nn \\
&& \left. \left. 
- \d_A^D \v^{CE}\bar\lambda_{\adot(E}\bar\lambda^\adot_{B)}
+ \d_B^C \v^{ED}\bar\lambda_{\adot(A}\bar\lambda^\adot_{E)}
- \d_B^D \v^{CE}\bar\lambda_{\adot(A}\bar\lambda^\adot_{E)}
\right) \right] \, .
\label{O-45b-fermi}
\ea

Because of cyclicity of the trace from the product of four scalars we
obtain only one operator in the $\mb{45}$ and one in the $\mbb{45}$.
In order to project (\ref{d4scal}) we  can construct a projector onto
the $\mb{45}$ as
\ba
\scrP^{\mb{45}\; (AB)}_{[ijk]l\,[CD]} &\!\!=\!\!& 
\S^{AE}_{[i}\Sbar_{jEF}\S^{FB}_{k]}\Sbar_{lCD} - 
\fr6 \left(\d^A_C\,\S^{GE}_{[i}\Sbar_{jEF}
\S^{FB}_{k]}\Sbar_{lGD} - \d^A_D\,\S^{GE}_{[i}\Sbar_{jEF}
\S^{FB}_{k]}\Sbar_{lCG} \right. \nn \\ 
&& \left. + \d^B_C\,\S^{AE}_{[i}\Sbar_{jEF}
\S^{FG}_{k]}\Sbar_{lGD} - \d^B_D\,\S^{AE}_{[i}\Sbar_{jEF}
\S^{FG}_{k]}\Sbar_{lCG} \right)
\label{proj45}
\ea
and similarly for the $\mbb{45}$ we consider 
\ba
\scrP^{\mbb{45}\; [CD]}_{[ijk]l\,(AB)} &\!\!\!=\!\!\!& 
\Sbar_{AE[i}\S^{EF}_j\Sbar_{k]FB}\S^{CD}_l - 
\fr6 \left(\d_A^C\,\Sbar_{GE[i}\S^{EF}_j
\Sbar_{k]FB}\S^{GD}_l - \d_A^D\,\Sbar_{GE[i}\S^{EF}_j
\Sbar_{k]FB}\S^{CG}_l \right. \nn \\ 
&& \left. + \d_B^C\,\Sbar_{AE[i}\S^{EF}_j
\Sbar_{k]FG}\S^{GD}_l - \d_B^D\,\Sbar_{AE[i}\S^{EF}_j
\Sbar_{k]FG}\S^{CG}_l \right)
\label{proj45b}
\ea
Then the two operators are respectively 
\be
\scrO^{(2)(AB)}_{4,\mb{45}[CD]} = 
\scrP^{\mb{45}\;(AB)}_{[ijk]l\,[CD]} \, \fr{\gy^4N}\, \Tr\left(
\v^i\v^j\v^k\v^l \right) 
\label{O-45-scal}
\ee
and 
\be
\scrO^{(2)[CD]}_{4,\mbb{45}(AB)} = 
\scrP^{\mbb{45}\;[CD]}_{[ijk]l\,(AB)} \, \fr{\gy^4N}\, \Tr\left(
\v^i\v^j\v^k\v^l \right) \, .
\label{O-45b-scal}
\ee
Notice that because of the symmetry properties of the projectors
(\ref{proj45}) and (\ref{proj45b}) there is no double trace operator
in these sectors. 

To compute possible instanton corrections to the anomalous dimensions
of the above operators we need to consider two-point functions
$\la\scrO_{4,\mb{45}}(x_1)\scrO_{4,\mbb{45}}(x_2)\ra$. As usual it is
convenient to pick a component and analyse the dependence on the
superconformal zero-modes. A simple choice is to consider the set
correlation functions  
\be
G^{(r,s)}(x_1,x_2) = \la \scrO^{(r)\;(12)}_{4,\mb{45}[34]}(x_1) 
\scrO^{(s)\;[34]}_{4,\mbb{45}(12)}(x_2) \ra \, , \qquad r,s=1,2 \, ,
\label{2pt-45}
\ee
since for these components there is no trace to subtract in flavour
space and (\ref{O-45-fermi}), (\ref{O-45b-fermi}), (\ref{O-45-scal})
and (\ref{O-45b-scal}) simplify. For the operators involving fermionic
bilinears the possible non-vanishing contributions to (\ref{2pt-45})
in semiclassical approximation come from the following terms
\ba
\scrO^{(1)(12)}_{4,\mb{45}[34]} &\!\!\to\!\!& \fr{g^3N^{1/2}} \, 
\Tr\left\{\bar\v^{(2)}_{[34]}\left[\left(\lambda^{(1)}\right)^{\a(1}
\left(\lambda^{(5)}\right)_\a^{2)} + \left(\lambda^{(5)}\right)^{\a(1}
\left(\lambda^{(1)}\right)_\a^{2)} \right] \right\} 
\label{O-45-1-satur} \\
\scrO^{(1)[34]}_{4,\mbb{45}(12)} &\!\!\to\!\!& \fr{g^3N^{1/2}} \, 
\Tr\left[\v^{(2)[34]}\left(\bar\lambda^{(3)}\right)_{\adot(1}
\left(\bar\lambda^{(3)}\right)^\adot_{2)} \right] \, .
\label{O-45b-1-satur}
\ea
The combination of fermionic modes contained in (\ref{O-45-1-satur})
is 
\be
\scrO^{(1)(12)}_{4,\mb{45}[34]} \sim \veps_{ABCD} \scrmf^1 \scrmf^2
\scrmf^1 \scrmf^2 \scrmf^A \scrmf^B \scrmf^C \scrmf^D 
\label{O-45-1-zm}
\ee
and the operator cannot provide two modes of each flavour as
required. From (\ref{O-45b-1-satur}) we find a contribution
proportional to 
\be
\scrO^{(1)[34]}_{4,\mbb{45}(12)} \sim \veps_{1ABC}\veps_{2A^\pp B^\pp
C^\pp} \scrmf^3 \scrmf^4 \scrmf^A \scrmf^B \scrmf^C \scrmf^{A^\pp}
\scrmf^{B^\pp} \scrmf^{C^\pp} \, ,
\label{O-45b -1-zm}
\ee
so that it cannot give rise to a non-vanishing contribution when
inserted in a two-point point function since it contains only one mode
of flavours 1 and 2. 

For the two operators quartic in the scalars the terms in the solution
which need to be considered for the semiclassical analysis are 
\ba
\scrO^{(2)(12)}_{4,\mb{45}[34]} &\!\!\to\!\!& 
\scrP^{\mb{45}\;(12)}_{[ijk]l\,[34]} \, \fr{\gy^4N}\, \Tr\left(
\v^{(2)i}\v^{(2)j}\v^{(2)k}\v^{(2)l} \right) 
\label{O-45-2-satur} \\
\scrO^{(2)[34]}_{4,\mbb{45}(12)} &\!\!\to\!\!& 
\scrP^{\mbb{45}\;[34]}_{[ijk]l\,(12)} \, \fr{\gy^4N}\, \Tr\left(
\v^{(2)i}\v^{(2)j}\v^{(2)k}\v^{(2)l} \right) \, .
\label{O-45b-2-satur}
\ea
Using the solution (\ref{phi2-inst-sol}) and performing the
contraction to project onto the $\mb{45}$ and $\mbb{45}$ one verifies
that none of these two operators can saturate the superconformal modes
as required for producing a non-zero two-point correlation function. 

The same results can be shown for other components, in which case a
cancellation involving the additional terms in the operators
subtracting flavour traces is required. We conclude that the scaling
dimensions of operators in these sectors do not receive instanton
corrections.

\subsubsection{$\D_0=4$, $[2,0,2]$}
\label{d4r84}

The representation $[2,0,2]\equiv\mb{84}$ at $\D_0=4$ can only be
obtained from (\ref{6x6x6x6d4}), \ie the operators in this sector are
all quartic in the scalars. This sector is the simplest example of the
type described in \cite{bks,b1}, which comprises operators
transforming in representations with Dynkin labels $[a,b,a]$ with bare
dimension $\D_0=2a+b$. Operators of this type cannot involve covariant
derivatives, field strengths or fermions and are only made out of
elementary scalars. Among these operators those which are
superconformal primaries belong to 1/4 BPS multiplets
\cite{dp,dhhr}. Moreover, as shown in \cite{bks}, for fields in this
class it is always possible to choose components that can be written
in terms of only two complex scalars. The action of the dilation
operator on composite fields belonging to these sectors is also
particularly simple \cite{bks,b1}.

Notice that the $[1,1,1]$ sector at $\D_0=3$ would also belong to this
class, but, as observed in section \ref{d3r64}, it is not realised in
terms of gauge invariant operators.

The representation $\mb{84}$ occurs in (\ref{6x6x6x6d4}) with
multiplicity 2 allowing in principle to construct four independent
operators, two single-trace and two double-trace operators. However
cyclicity of the trace implies that there exist only two independent
gauge-invariant operators. In order to obtain a single-trace operator
in this representation we consider
\ba
\scrO^{(1)[ij][kl]}_{4,\mb{84}} &\!\!=\!\!& \fr{\gy^4N}\, 
\Tr\left\{[\v^i,\v^j][\v^k,\v^l] - \fr4 \left(\d^{ik}[\v^m,\v^j]
[\v^m,\v^l]+\d^{il}[\v^m,\v^j][\v^k,\v^m] \right.\right. \nn \\
&& \left. +\d^{jk}[\v^i,\v^m][\v^m,\v^l]+\d^{jl}
[\v^i,\v^m][\v^k,\v^m]\right) \nn \\
&& \left. +\fr{20}\left(\d^{ik}\d^{jl}-\d^{il}\d^{jk}
\right) [\v^m,\v^n][\v^m,\v^n] \right\} \, ,
\label{O-84-s}
\ea
\ie we need to subtract the singlet, the $\mb{15}$ and the
$\mb{20^\pp}$ from the symmetric part of $\mb{15}\otimes\mb{15}$.

Similarly to project the double trace composite operator in
(\ref{d4scal}) onto the $\mb{84}$ we consider 
\ba
\scrO^{(2)[ij][kl]}_{4,\mb{84}} &\!\!=\!\!& \fr{\gy^4N}\, 
\left\{\rule{0pt}{16pt}\Tr\left(\v^i\v^k\right)\Tr\left(\v^j\v^l\right) 
- \Tr\left(\v^i\v^l\right)\Tr\left(\v^j\v^k\right) \right. \nn \\
&& - \fr4\left[ \d^{ik}\left(\Tr\left(\v^m\v^m\right)
\Tr\left(\v^j\v^l\right)  - \Tr\left(\v^j\v^m\right)
\Tr\left(\v^m\v^l\right) \right) \right. \nn \\
&& + \d^{il} \left( \Tr\left(\v^m\v^k\right)
\Tr\left(\v^j\v^m\right)  - \Tr\left(\v^m\v^m\right)
\Tr\left(\v^j\v^k\right) \right) \label{O-84-d} \\
&& + \d^{jl} \left( \Tr\left(\v^i\v^k\right)
\Tr\left(\v^m\v^m\right)  - \Tr\left(\v^i\v^m\right)
\Tr\left(\v^m\v^k\right) \right) \nn \\
&& \left. + \d^{jk} \left( \Tr\left(\v^i\v^m\right)
\Tr\left(\v^m\v^l\right)  - \Tr\left(\v^i\v^l\right)
\Tr\left(\v^m\v^m\right) \right) \right] \nn \\
&&\left. + \fr{20}\left(\d^{ik}\d^{jl}-\d^{il}\d^{jk} \right)
\left[\rule{0pt}{12pt} \Tr\left(\v^m\v^m\right)
\Tr\left(\v^n\v^n\right)-\Tr\left(\v^m\v^n\right)
\Tr\left(\v^m\v^n\right)\right] \right\} \, . \nn
\ea

Operators in this sector were studied in perturbation theory in
\cite{bkrs1,bkrs2,bks}. The single trace operator (\ref{O-84-s})
belongs to the Konishi multiplet, being a superconformal descendant of
$\scrK_\mb1$ at level $\d^2\bar\d^2$. A linear combination of
(\ref{O-84-d}) and (\ref{O-84-s}) is protected and indeed a
superconformal primary and thus it is the lowest component of a 1/4
BPS multiplet. We should therefore expect to find that the instanton
contributions vanish in this sector.

The computation of instanton contributions to the two-point functions
$\la\scrO^{(r)[ij][kl]}_{4,\mb{84}}(x_1)
\scrO^{(s)[mn][pq]}_{4,\mb{84}}(x_2)\ra$, $r,s=1,2$, at leading order
in $\gy$ requires the expansion of the fields to the lowest order
needed to saturate eight fermion modes at each point, so for the two
above operators we need to compute
\ba
\scrO^{(1)[ij][kl]}_{4,\mb{84}} &\!\!\to\!\!& \fr{\gy^4N}\, \Tr\left(
[\v^{(2)i},\v^{(2)j}][\v^{(2)k},\v^{(2)l}] \right) + \cdots 
\label{O-84-1-satur} \\  
\scrO^{(2)[ij][kl]}_{4,\mb{84}} &\!\!\to\!\!& \fr{\gy^4N}\, \left[
\Tr\left(\v^{(2)i}\v^{(2)k}\right)\Tr\left(\v^{(2)j}\v^{(2)l} \right)
\right. \nn \\ 
&& \left.\hsp{1.3} - \Tr\left(\v^{(2)i}\v^{(2)l}\right)\Tr\left(
\v^{(2)j} \v^{(2)k} \right) + \cdots \right]  \, .
\label{O-84-2-satur}
\ea
Using the expression for $\v^{(2)i}$ in (\ref{phi2-inst-sol}) one can
verify that the expansion of the two operators defined in
(\ref{O-84-s}) and (\ref{O-84-d}) in the instanton background does not
contain the combination (\ref{saturmodes}) needed to produce
non-vanishing two-point functions. The simplest way of showing this is
to exploit the observation of \cite{bks} and choose a component
involving only two complex scalars, \eg
\ba
\scrO^{(1)}_{4,\mb{84}} &\!\!\to\!\!& \fr{\gy^4N} \Tr\left(
[\phi^1,\phi^2][\phi^1,\phi^2]\right) 
\label{O-84-1-com} \\
\scrO^{(2)}_{4,\mb{84}} &\!\!\to\!\!& \fr{\gy^4N} \left[\Tr\left(
\phi^1\phi^2\right)\Tr\left(\phi^1\phi^2\right) - \Tr\left(\phi^1
\phi^1\right)\Tr\left(\phi^2\phi^2\right)\right] \, .
\label{O-84-2-com}
\ea
Notice that with this choice no subtraction of traces is required. 
For both of these operators the instanton profile contains the
combination 
\be
\scrmf^1\scrmf^4\scrmf^2\scrmf^4\scrmf^1\scrmf^4\scrmf^2\scrmf^4 
\label{O-84-zm}
\ee
of zero modes and so all their two-point functions vanish.  Therefore
the two operators in the $\mb{84}$ receive no instanton contribution
to their scaling dimension as expected. 

The calculation just described provides another proof (and the most
direct one) of the fact that the scaling dimension of the Konishi
multiplet is not corrected by instantons. As observed in section
\ref{dim2scal} a direct computation of two-point functions of the
lowest component $\scrK_\mb1$ is rather subtle. It is instead easy to
analyse the superconformal descendant $\scrK^{[ij][kl]}_{\mb{84}}
\equiv \scrO^{(1)[ij][kl]}_{4,\mb{84}}$, which has the same vanishing
anomalous dimension because of supersymmetry. This situation resembles
what is found in perturbation theory \cite{bks}, where the action of
the dilation operator is simpler on the descendant than on the primary
field. This example shows explicitly a general feature, \ie how the
superconformal symmetry of the theory can be used to simplify the
computation of anomalous dimensions.

\subsubsection{$\D_0=4$, $[0,4,0]$}
\label{d4r105}

The sector of $\D_0=4$ operators transforming in the representation
$[0,4,0]\equiv\mb{105}$ is in the same class as the $\mb{20^\pp}$ at
$\D_0=2$ and the $\mb{50}$ at $\D_0=3$ examined in sections
\ref{dim2scal} and \ref{d3r50} respectively. These are operators in
representation with Dynkin labels $[0,\ell,0]$ and bare dimension
$\D_0=\ell$ and when primaries they belong to 1/2 BPS multiplets. The
dimension 4 operators of this type are selected from (\ref{d4scal}) by
taking the completely symmetric and traceless combination of SU(4)
indices. There is a single- and a double-trace operator
\ba
\scrO^{(1)\{ijkl\}}_{4,\mb{105}} &\!\!=\!\!& \fr{\gy^4N}\,
\Tr\left(\v^{\{i}\v^j\v^k\v^{l\}} \right) 
\label{O-105-s} \\
\scrO^{(2)\{ijkl\}}_{4,\mb{105}} &\!\!=\!\!& \fr{\gy^4N}\,
\Tr\left(\v^{\{i}\v^j\right)\Tr\left(\v^k\v^{l\}} \right) \,.
\label{O-105-d} 
\ea
In this case the projection onto the $\mb{105}$ requires the
subtraction of the $\mb{20^\pp}$ and the singlet with coefficients
which are readily determined imposing tracelessness. 

In order to analyse the instanton contributions to these operators we
consider a specific component and the simplest choice is to consider a
component written in terms of a single complex scalar. In this way no
trace subtraction is required and two representatives are for instance 
\be
\scrO^{(1)}_{4,\mb{105}} = \fr{\gy^4N} \, \Tr \left(\phi^1\phi^1
\phi^1\phi^1 \right) 
\label{O-105-s-com}
\ee 
and 
\be
\scrO^{(2)}_{4,\mb{105}} = \fr{\gy^4N} \, \Tr \left(\phi^1\phi^1\right)
\Tr\left(\phi^1\phi^1 \right) \, .
\label{O-105-d-com}
\ee
It is then clear that the corresponding two-point functions are not
corrected at the instanton level since the operators
(\ref{O-105-s-com}) and (\ref{O-105-d-com}) involve fermion modes in
the combination $\left(\scrmf^1\scrmf^4\right)^4$.

\subsubsection{$\D_0=4$, $[1,2,1]$}
\label{d4r175}

The representation $[1,2,1]\equiv\mb{175}$ is contained only in the
decomposition (\ref{6x6x6x6d4}), where it occurs with multiplicity 3. 
We can thus expect single- and double-trace operators made of four
scalars. However cyclicity of the trace implies that actually no
gauge-invariant operator exists in this sector. This can be seen
analysing the origin of the three $\mb{175}$'s in
(\ref{6x6x6x6d4}). They arise from the products
$(\mb{15}\otimes\mb{15})_{\rm a}$,
$(\mb{20^\pp}\otimes\mb{20^\pp})_{\rm a}$ and
$\mb{15}\otimes\mb{20^\pp}$. The first two correspond to operators
of the form 
\be
\Tr\left([\v^{[i}\v^{j]},\v^{[k}\v^{l]}]\right) 
\label{175in15x15}
\ee
and
 \be
\Tr\left([\v^{\{i}\v^{j\}},\v^{\{k}\v^{l\}}]\right) 
\label{175in20x20}
\ee
and thus vanish as traces of commutators. For the third combination
which contains the $\mb{175}$ we should consider 
\be
\Tr\left( \v^{[i}\v^{j]}\v^{\{k}\v^{l\}} \right) \, ,
\label{175in15x20}
\ee
which again vanishes as can be shown by considering~\footnote{I thank
Massimo Bianchi for a discussion on this point.}
\be
\Tr\left( \v^{[i}\v^{j]}\v^{\{k}\v^{l\}} \right) = 
\Tr\left[\left( \v^{[i}\v^{j]}\v^{\{k}\v^{l\}} \right)^\dagger\right]
= - \Tr\left( \v^{[i}\v^{j]}\v^{\{k}\v^{l\}} \right) \, .
\label{175in15x20-zero}
\ee

The representation $\mb{175}$ is thus not realised in terms of
gauge-invariant operators. Notice that if operators in this sector
could be constructed they would be in the same class as those in the
$\mb{84}$ discussed in section \ref{d4r84}, \ie operators transforming
in a representation $[a,b,a]$ with $\D_0=2a+b$: the $\mb{175}$
corresponds to $a=1$, $b=2$.

\subsection{Dimension 5 scalar operators}
\label{dim5scal}

The analysis of gauge-invariant operators becomes increasingly
complicated as their bare dimension grows and it is already very
involved at dimension 5. The number of SU(4) representations which
appear and their multiplicities increase rapidly, making the
construction of a basis of independent operators in each sector rather
non-trivial. The discussion in this section will therefore be less
detailed than in previous sections. Instanton contributions to
anomalous dimensions are expected only for operators in the $\mb6$ of
SU(4)~\footnote{An argument supporting this claim will be presented in
the next section.}. Explicit  calculations of two-point functions in
this sector will be presented in order to display a new feature which
appears, \ie the necessity of including the leading quantum
fluctuations around the classical instanton configuration in order to
compute two-point functions at leading order in the coupling.
Although rather cumbersome, the analysis can however be carried on
along the lines discussed in the previous cases without additional new
conceptual difficulties for all the other sectors.

The combinations of elementary fields which give rise to scalar
composites of bare dimension 5 are
\ba
&& \hsp{-0.8}
\scrO^{(a)i}_{5,\mb6} \sim \Tr\left(F_{\mu\nu}F^{\mu\nu}\v^i\right) 
\,, \qquad \scrO^{(b)i}_{5,\mb6} \sim \Tr\left(F_{\mu\nu}\tilde 
F^{\mu\nu} \v^i\right) \, , \label{d5fieldstr} \\
&& \hsp{-0.8}
\scrO^{ijk}_{5,\mb6\otimes\mb6\otimes\mb6} \sim \Tr\left(
\v^i\scrD_\mu\v^j\scrD^\mu\v^k \right) \, ,\label{d5deriv} \\
&& \hsp{-0.8}
\scrO^{ijAB}_{5,\mb6\otimes\mb6\otimes\mb4\otimes\mb4} \sim
\Tr\left(\v^i\v^j\lambda^{\a A}\lambda^B_\a \right) \, , \qquad 
\scrO^{ij}_{AB;5,\mb6\otimes\mb6\otimes\mbb4\otimes\mbb4} \sim
\Tr\left(\v^i\v^j\bar\lambda_{\adot A}\bar\lambda_B^\adot \right) 
\, , \label{d5fermi} \\
&& \hsp{-0.8}
\scrO^{ijA}_{B;5,\mb6\otimes\mb6\otimes\mb4\otimes\mbb4} \sim 
\Tr\left(\scrD_\mu\v^i\lambda^{\a A}\s^\mu_{\a\adot}
\bar\lambda^\adot_B \right) \, , \label{d5fermi-mix} \\
&& \hsp{-0.8}
\scrO^{AB}_{5,\mb4\otimes\mb4} \sim \Tr \left(F_{\mu\nu}
\lambda^{\a A}\s^{\mu\nu\,\b}_{\:\a}\lambda^B_\b\right) \, , 
\quad \scrO^{AB}_{5,\mb4\otimes\mb4} \sim \Tr \left(\tilde F_{\mu\nu}
\lambda^{\a A}\s^{\mu\nu\,\b}_{\:\a}\lambda^B_\b\right) \,, 
\label{d5fermi-fs} \\
&& \hsp{-0.8}
\scrO_{AB;5,\mbb4\otimes\mbb4} \sim \Tr \left(F_{\mu\nu}
\bar\lambda_{\adot A}\bar\s^{\mu\nu\,\adot}{}_{\bdot}
\bar\lambda^\bdot_B \right) \, , 
\quad \scrO_{AB;5,\mbb4\otimes\mbb4} \sim \Tr \left(\tilde F_{\mu\nu}
\bar\lambda_{\adot A}\bar\s^{\mu\nu\,\adot}{}_{\bdot}
\bar\lambda^\bdot_B \right) \,, 
\label{d5fermib-fs} \\
&& \hsp{-0.8}
\scrO^{AB}_{5,\mb4\otimes\mb4} \sim \Tr\left(\scrD_\mu
\lambda^{\a A}\scrD^\mu\lambda^B_\a \right) \, , \qquad 
\scrO_{AB;5,\mbb4\otimes\mbb4} \sim \Tr\left( \scrD_\mu
\bar\lambda_{\adot A}\scrD^\mu\bar\lambda^\adot_B \right) 
\label{d5fermi-deriv} \\
&& \hsp{-0.8}
\scrO^{ijklm}_{5,\mb6\otimes\mb6\otimes\mb6\otimes\mb6\otimes\mb6} 
\sim \Tr\left( \v^i\v^j\v^k\v^l\v^m \right) \, . \label{d5scal} 
\ea
Moreover for the operators in (\ref{d5fermi}) and in (\ref{d5scal}) we
also need to consider the double-trace operators obtained splitting
the product into two traces in all possible ways. 

Operators involving the field strength only contribute to the $\mb6$
of SU(4). From the combination in (\ref{d5deriv}) we obtain operators
in the sectors appearing in the decomposition
\ba
&& [0,1,0]\otimes[0,1,0]\otimes[0,1,0] = 3[0,1,0]\oplus [2,0,0]\oplus
[0,0,2] \oplus [0,3,0]\oplus 2[1,1,1] \nn \\
&& \Leftrightarrow \mb6\otimes\mb6\otimes\mb6 = 3\cdot\mb6\oplus
\cdot\mb{10}\oplus\mbb{10}\oplus\mb{50}\oplus 2\cdot\mb{64} \, .
\label{6x6x6d5}
\ea
From (\ref{d5fermi}) and the double-trace combinations involving the
same fields we get operators respectively in 
\ba
&&[0,1,0]\otimes[0,1,0]\otimes[1,0,0]\otimes[1,0,0] = 4[0,1,0]
\oplus3[2,0,0]\oplus2[0,0,2] \nn \\
&&{}\!\oplus[0,3,0]\oplus4[1,1,1]\oplus[3,0,1]\oplus[2,2,0] 
\label{6x6x4x4d5} \\
&& \Leftrightarrow \mb6\otimes\mb6\otimes\mb4\otimes\mb4 = 
4\cdot\mb6\oplus3\cdot\mb{10}\oplus2\cdot\mbb{10}\oplus\mb{50}
\oplus4\cdot\mb{64}\oplus\mb{70}\oplus\mb{126} \nn
\ea 
and 
\ba
&&[0,1,0]\otimes[0,1,0]\otimes[0,0,1]\otimes[0,0,1] = 4[0,1,0]
\oplus2[2,0,0]\oplus3[0,0,2] \nn \\
&&{}\!\oplus[0,3,0]\oplus4[1,1,1]\oplus[1,0,3]\oplus[0,2,2] 
\label{6x6xb4xb4d5} \\
&& \Leftrightarrow \mb6\otimes\mb6\otimes\mbb4\otimes\mbb4 = 
4\cdot\mb6\oplus2\cdot\mb{10}\oplus3\cdot\mbb{10}\oplus\mb{50}
\oplus4\cdot\mb{64}\oplus\mbb{70}\oplus\mbb{126} \, . \nn
\ea
The sectors which can contain operators of the form
(\ref{d5fermi-mix}) are 
\ba
&& [0,1,0]\otimes[1,0,0]\otimes[0,0,1] = 2[0,1,0]\oplus[2,0,0]
\oplus[0,0,2]\oplus[1,1,1] \nn \\
&& \Leftrightarrow \mb6\otimes\mb4\otimes\mbb4 = 
2\cdot\mb6\oplus\mb{10}\oplus\mbb{10}\oplus\mb{64} \, .
\label{6x4xb4d5}
\ea
The two types of operators in (\ref{d5fermi-fs}) and the first in
(\ref{d5fermi-deriv}) contribute to
\ba
[1,0,0]\otimes[1,0,0] = [0,1,0]\oplus[2,0,0] \Leftrightarrow 
\mb4\otimes\mb4 = \mb6\oplus\mb{10} 
\label{4x4d5}
\ea
whereas those in (\ref{d5fermi-fs}) and the second in
(\ref{d5fermi-deriv}) contribute to 
\ba
[0,0,1]\otimes[0,0,1] = [0,1,0]\oplus[0,0,2] \Leftrightarrow 
\mbb4\otimes\mbb4 = \mb6\oplus\mbb{10} \, .
\label{b4xb4d5}
\ea
Finally operators of the form (\ref{d5scal}) (and the corresponding
double-trace operators) are found in the decomposition
\ba
&& [0,1,0]\otimes[0,1,0]\otimes[0,1,0]\otimes[0,1,0]\otimes[0,1,0] = 
16[0,1,0]\oplus10\left([2,0,0]\oplus[0,0,2]\right)\nn \\
&&{}\! \oplus 10[0,3,0]\oplus24[1,1,1]\oplus5\left([3,0,1]\oplus[1,0,3]
\right)\oplus\left([2,2,0]\oplus[0,2,2]\right) \nn \\
&&{}\! \oplus[0,5,0]\oplus5[2,1,2]\oplus4[1,3,1] 
\label{d56x6x6x6x6} \\
&& \Leftrightarrow \mb6\otimes\mb6\otimes\mb6\otimes\mb6\otimes\mb6 
= 16\cdot\mb6\oplus10\cdot(\mb{10}\oplus\mbb{10})\oplus10\cdot\mb{50}
\oplus 24\cdot\mb{64}\oplus5\cdot(\mb{70}\oplus\mbb{70})\nn \\
&&{}\!\oplus 6\cdot(\mb{126}\oplus\mbb{126})\oplus\mb{196}\oplus
5\cdot\mb{300}\oplus4\cdot\mb{384} \, . \nn
\ea
From this preliminary analysis it is already apparent that many more
representations appear at $\D_0=5$. Furthermore there are
representations occurring with high multiplicity, which, as already
observed, makes the construction of a basis of operators in the
corresponding sectors rather tedious.

\subsubsection{$\D_0=5$, $[0,1,0]$} 
\label{d5r6}

Operators in the $[0,1,0]\equiv\mb6$ can be obtained from the
combinations (\ref{d5fieldstr}), (\ref{d5fermi}) (which both contain
the $\mb6$ with multiplicity 4), (\ref{d5fermi-mix}) (where it occurs
with multiplicity 2), (\ref{d5fermi-deriv}) and (\ref{d5scal}) (where
it appears with multiplicity 16). Clearly these multiplicities do not
take into account the cyclicity of the trace which reduces the actual
number of independent gauge invariant operators. On the other hand we
must consider single- and double-trace operators of the types in
(\ref{d5fermi}) and (\ref{d5scal}). 

Two independent operators involving field strengths are
\be
\scrO^{(1)i}_{5,\mb6} = \fr{\gy^3N^{1/2}} \, \Tr\left(
F_{\mu\nu}F^{\mu\nu}\v^i\right) \, , \quad 
\scrO^{(2)i}_{5,\mb6} = \fr{\gy^3N^{1/2}} \, \Tr\left(
F_{\mu\nu}\tilde F^{\mu\nu}\v^i\right) \, .
\label{O5-6-fs}
\ee
With three scalars and two covariant derivatives we can construct the
operators 
\be
\scrO^{(3)i}_{5,\mb6} = \fr{\gy^3N^{1/2}}\,\Tr\left( 
\v^i\scrD_\mu\v^j\scrD^\mu\v_j \right) \, , \quad
\scrO^{(4)i}_{5,\mb6} = \fr{\gy^3N^{1/2}}\,\Tr\left( 
\v^j\scrD_\mu\v^i\scrD^\mu\v_j \right) \, . 
\label{O5-6-deriv}
\ee
Single trace operators with two scalars and two fermions are
\ba
\hsp{-1} &&
\scrO^{(5)i}_{5,\mb6} = \fr{\gy^4N} \, \Tr \left(\S^i_{AB} 
\v^j\v^j\lambda^{\a A}\lambda^{B}_\a \right) \,, \quad
\scrO^{(6)i}_{5,\mb6} = \fr{\gy^4N} \, \Tr \left(\S^j_{AB} 
\v^{[i}\v^{j]}\lambda^{\a A}\lambda^{B}_\a \right) \, ,\nn \\
\hsp{-1} &&
\scrO^{(7)i}_{5,\mb6} = \fr{\gy^4N} \, \Tr \left(\S^i_{AB} 
\v^{\{i}\v^{j\}}\lambda^{\a A}\lambda^{B}_\a \right) \,, \quad 
\scrO^{(8)i}_{5,\mb6} = \fr{\gy^4N} \, \Tr \left(
\bar t^{[ijk]}_{(AB)} \v^{[j}\v^{k]}\lambda^{\a A}\lambda^{B}_\a 
\right) \, , \nn \\
\hsp{-1} &&\scrO{(9)i}_{5,\mb6} = \fr{\gy^4N} \, \Tr \left(
\S^j_{AB} \lambda^{\a A}\v^{[i}\lambda^{B}_\a\v^{j]} \right)
\label{O5-6-fer1}
\ea
and similarly with fermions of the opposite chirality
\ba
\hsp{-1} &&
\scrO^{(10)i}_{5,\mb6} = \fr{\gy^4N} \, \Tr \left(\Sbar^{iAB} 
\v^j\v^j\bar\lambda_{\adot A}\bar\lambda^\adot_B \right) \, , \quad
\scrO^{(11)i}_{5,\mb6} = \fr{\gy^4N} \, \Tr \left(\Sbar^{jAB} 
\v^{[i}\v^{j]}\bar\lambda_{\adot A}\bar\lambda^\adot_B \right) \, ,
\nn \\ 
\hsp{-1} &&
\scrO^{(12)i}_{5,\mb6} = \fr{\gy^4N} \, \Tr \left(\Sbar^{iAB} 
\v^{\{i}\v^{j\}}\bar\lambda_{\adot A}\bar\lambda^\adot_B \right)  \,,
\quad \scrO^{(13)i}_{5,\mb6} = \fr{\gy^4N} \, \Tr \left(
t_{[ijk]}^{(AB)} \v^{[j}\v^{k]}\bar\lambda_{\adot A}
\bar\lambda^\adot_B \right)  \, , \nn \\
\hsp{-1} &&
\scrO^{(14)i}_{5,\mb6} = \fr{\gy^4N} \, \Tr \left(\Sbar^{jAB} 
\bar\lambda_{\adot A}\v^{[i}\bar\lambda^\adot_B\v^{j]} \right) \, .
\label{O5-6-fer2}
\ea
Besides these operators there are double-trace operators obtained
splitting the traces in (\ref{O5-6-fer1}) and (\ref{O5-6-fer2}) in all
the ways allowed by the (anti)symmetry properties. The operators in
(\ref{d5fermi-mix}) are 
\ba
\scrO^{(15)i}_{5,\mb6} &\!\!=\!\!& \fr{\gy^3N^{1/2}}\, \Tr\left(
\scrD_\mu\v^i\lambda^{\a A}\s^\mu_{\a\adot}\bar\lambda^\adot_A \right)
\, , \quad \scrO^{(16)i}_{5,\mb6} = \fr{\gy^3N^{1/2}}\, 
\Tr\left(\lambda^{\a A}\scrD_\mu\v^i\s^\mu_{\a\adot}
\bar\lambda^\adot_A \right) \nn \\
\scrO^{(17)i}_{5,\mb6} &\!\!=\!\!& \fr{\gy^3N^{1/2}}\, \Tr\left(
\Sbar^{iC[A} \scrD_\mu\v_i\lambda^{\a B]}\s^\mu_{\a\adot}
\bar\lambda^\adot_C +\fr4 \S^i_{AB} \scrD_\mu \v^{AB} 
\lambda^{\a C}\s^\mu_{\a\adot}\bar\lambda^\adot_C \right) \, .
\label{O5-6-fermix}
\ea
From (\ref{d5fermi-fs}) and (\ref{d5fermib-fs}) we get 
\ba
\scrO^{(18)i}_{5,\mb6} &\!\!=\!\!& \fr{\gy^3N^{1/2}}\, \Tr\left(
\S^i_{AB} F_{\mu\nu} \lambda^{\a A}\sigma^{\mu\nu\,\b}_{\:\a}
\lambda^B_\b \right) \, , \nn \\ 
\scrO^{(19)i}_{5,\mb6} &\!\!=\!\!& \fr{\gy^3N^{1/2}}\, \Tr\left( 
\S^i_{AB} \tilde F_{\mu\nu} \lambda^{\a A}\sigma^{\mu\nu\,\b}_{\:\a} 
\lambda^B_\b \right) \, , \label{O5-6-fermi-fs} \\
\scrO^{(20)i}_{5,\mb6} &\!\!=\!\!& \fr{\gy^3N^{1/2}}\, \Tr\left(
\Sbar^{i\,AB} F_{\mu\nu} \bar\lambda_{\adot A}
\bar\sigma^{\mu\nu\,\adot}{}_{\bdot}\bar\lambda^\bdot_B \right) 
\, , \nn \\
\scrO^{(21)i}_{5,\mb6} &\!\!=\!\!& \fr{\gy^3N^{1/2}}\, 
\Tr\left(\Sbar^{i\,AB} \tilde F_{\mu\nu} \bar\lambda_{\adot A}
\bar\sigma^{\mu\nu\,\adot}{}_{\bdot}\bar\lambda^\bdot_B \right)\,. \nn 
\ea
Operators of the form (\ref{d5fermi-deriv}) actually do not contribute
to the $\mb6$ because the trace of the antisymmetric combination
vanishes. They are only found in the $\mb{10}$ and $\mbb{10}$. 
Finally there are the operators made of only elementary scalars. As a
set of independent gauge-invariant operators we can consider the
single-trace combinations
\ba
&&\hsp{-1} \scrO^{(22)i}_{5,\mb6} = \fr{\gy^5N^{3/2}} \, \Tr\left(
\v^i\v^j\v^j\v^k\v^k \right) \, , \quad \scrO^{(23)i}_{5,\mb6}
= \fr{\gy^5N^{3/2}} \, \Tr\left( \v^i\v^j\v^k\v^j\v^k \right) \, ,
\label{O5-6s-scal} \\
&&\hsp{-1} \scrO^{(24)i}_{5,\mb6} = \fr{\gy^5N^{3/2}} \, \Tr\left(
\v^i\v^j\v^k\v^k\v^j \right) \, , \quad \scrO^{(25)i}_{5,\mb6}
= \fr{\gy^5N^{3/2}} \,\veps^{ijklmn} \,\Tr\left( 
\v^j\v^k\v^l\v^m\v^n \right) \nn 
\ea
and the double-trace combinations
\ba
&&\hsp{-0.5}\scrO^{(26)i}_{5,\mb6} = \fr{\gy^5N^{3/2}} \, 
\Tr\left(\v^i\v^j\right) \Tr\left(\v^j\v^k\v^k\right) \, , 
\quad \scrO^{(27)i}_{5,\mb6} = \fr{\gy^5N^{3/2}} \, 
\Tr\left(\v^i\v^j\v^j\right) \Tr\left(\v^k\v^k\right) \, , \nn \\
&&\hsp{-0.5}\scrO^{(28)i}_{5,\mb6} = \fr{\gy^5N^{3/2}} \, 
\Tr\left(\v^i\v^j\v^k \right)\Tr\left(\v^j\v^k\right)\, .
\label{O5-6d-scal}
\ea

There is therefore a large number of operators in this sector.  Using
the same techniques employed in the previous sections it is easy to
verify that all the above operators can soak up the fermion
superconformal modes in the required combination
(\ref{saturmodes}). The terms we need to consider in the expansion of
all the operators in (\ref{O5-6-fs})-(\ref{O5-6d-scal}) at the
semiclassical level involve 10 fermion modes. Analysing the structure
of the single operators we find that they all contain terms
proportional to
\be
\bar\nu^A\nu^B\left(\zeta^1\right)^2\left(\zeta^2\right)^2
\left(\zeta^3\right)^2\left(\zeta^4\right)^2 \,, 
\label{O5-6-zm}
\ee
so that in the computation of two-point functions,
\be
G^{(r,s)}(x_1,x_2) = \la\scrO^{(r)i}_{5,\mb6}(x_1) 
\scrO^{(s)i}_{5,\mb6}(x_2)\ra \, , \quad r,s=1,\ldots,28 \, ,
\label{2pt-O5-6}
\ee
the integration over the superconformal modes is potentially
non-zero. We should thus expect to find a non vanishing result for all
the two-point functions in this sector, so that the mixing problem in
the one-instanton sector and at leading order in $\gy$ is very
involved.

Since the total number of fermion modes entering the classical
expressions of the operators (\ref{O5-6-fs})-(\ref{O5-6d-scal})
exceeds the minimum required of eight, there is also the possibility
of non-vanishing contributions to two-point functions in which pairs
of fields are contracted.  Ordinary semiclassical contributions to the
two-point functions (\ref{2pt-O5-6}), in which all the fields are
replaced by their background expression in the presence of an
instanton, involve a non trivial five-sphere integration. The profiles
of the operators contain a $\bar\nu\nu$ bilinear each leading to an
integral of the form (\ref{s5int1}). The insertion of two $\bar\nu\nu$
bilinears from the operators induces a factor of $\gy^2$ in the
expectation value.  Similarly a factor of $\gy^2$ is produced by each
propagator in our normalisations, so that the two types of
contributions are of the same order and consistency requires that both
are included.  The counting of zero modes shows that the allowed
contractions are between two $\v^i$'s, between a $\lambda^A$ and a
$\bar\lambda_A$ or between two vectors, $A_\mu$. This is because in
these cases the number of zero modes appearing in the correlation
function is reduced by four, leaving a total of sixteen. The scalar
contraction involves the propagator (\ref{propfin}). The spinor and
vector propagators,  $\la\bar\lambda_A(x_1)\lambda^B(x_2)\ra$ and $\la
A_\mu(x_1)A_\nu(x_2)\ra$, which have not been given explicitly,  can
be deduced from that for the scalar \cite{bccl}~\footnote{It should be
noted that the evaluation of contributions containing  vector
contractions presents subtleties related to infrared divergences. See
\cite{iigt}, chapter 4, and references therein.}.  In the two-point
functions we are considering however we need the sixteen zero modes
distributed in two groups of eight as in (\ref{saturmodes}). This
implies that fermion contractions cannot contribute because after the
contraction the remaining sixteen modes are not evenly
distributed. The same argument applies to vector contractions.

A complete analysis and resolution of the mixing in this sector is
beyond the scope of the present paper, we shall however compute the
two types of contributions for a specific two-point function of
operators in (\ref{O5-6s-scal}) in order to illustrate the features
and difficulties of the calculation.

Consider the correlation function
\ba
G(x_1,x_2) &\!\!=\!\!& \la\scrO^{(23)1}_{5,\mb6}(x_1)
\scrO^{(23)1}_{5,\mb6}(x_2) \ra \nn \\
&\!\!=\!\!& \fr{\gy^{10}N^3} \la \Tr\left[
\left(\v^1\v^i\v^j\v^i\v^j\right)(x_1)\right]  \Tr\left[
\left(\v^1\v^k\v^l\v^k\v^l\right)(x_2)\right] \ra \, .  
\label{spec-2pt-O5-6}
\ea
In order to soak up the sixteen superconformal modes in the moduli
space integration the relevant terms in the expansion of the two
composite operators are $\Tr\left(\v^{(2)1}\v^{(2)i}\v^{(2)j}
\v^{(2)i}\v^{(2)j}\right)$. To compute the expression for the
operators in (\ref{spec-2pt-O5-6}) it is then convenient to rewrite
them in terms of $\v^{AB}$'s as 
\be
\scrO^{(23)i}_{5,\mb6} = \fr{\sqrt{2}\gy^5N^{3/2}}\S^i_{AB}
\veps_{A^\pp B^\pp C^\pp D^\pp} \veps_{A^\dpp B^\dpp C^\dpp D^\dpp}
\Tr\left(\v^{AB}\v^{A^\pp B^\pp}\v^{A^\dpp B^\dpp} \v^{C^\pp D^\pp}
\v^{C^\dpp D^\dpp} \right) \,. 
\label{O5-6-AB}
\ee
and use 
\ba
&& \hsp{-0.6}\Tr\left[\left(\v^{A_1B_1}\v^{A_2B_2}\v^{A_3B_3}
\v^{A_4B_4}\v^{A_5B_5} \right)(x)\right] = 
\frac{2^5\rho^8}{[(x-x_0)^2+\rho^2]^9} \nn \\
&& \hsp{-0.6}\times \left( \left\{ \left[ \left( \rule{0pt}{13pt} 
(\bar\nu^{[A_1}\nu^{A_2]})(\zeta^{B_2}\zeta^{B_3})
(\zeta^{A_3}\zeta^{B_4})(\zeta^{A_4}\zeta^{B_5})
(\zeta^{A_5}\zeta^{B_1}) - (A_4 \leftrightarrow B_4) 
- (A_5 \leftrightarrow B_5) \right. \right. \right. \right. \nn \\
&& \hsp{-0.6} + (A_4 \leftrightarrow B_4,A_5 \leftrightarrow B_5) - 
(A_3 \leftrightarrow B_3) + (A_3 \leftrightarrow B_3,
A_4 \leftrightarrow B_4) + (A_3 \leftrightarrow B_3,
A_5 \leftrightarrow B_5) \nn \\
&& \hsp{-0.6} \left. \left. \! - (A_3 \leftrightarrow B_3, A_4 
\leftrightarrow B_4, A_5 \leftrightarrow B_5) \rule{0pt}{14pt}
\right) - (A_1 \leftrightarrow B_1) - (A_2 \leftrightarrow B_2) 
+ (A_1 \leftrightarrow B_1,A_2 \leftrightarrow B_2) \right] \nn \\
&& \hsp{-0.6} \left. + \mbox{~``cyclic~permutations''}  
\rule{0pt}{13pt} \right\}+\left\{ \rule{0pt}{13pt}
(\bar\nu^{[A_4}\nu^{B_4]})\left[
(\zeta^{A_1}\zeta^{B_2})(\zeta^{A_2}\zeta^{B_3})
(\zeta^{A_3}\zeta^{B_5})(\zeta^{A_5}\zeta^{B_1}) 
\right. \right. \nn \\
&& \hsp{-0.6} - (A_1 \leftrightarrow B_1) - (A_2 \leftrightarrow B_2) 
- (A_3 \leftrightarrow B_3) - (A_5 \leftrightarrow B_5) + 
(A_1 \leftrightarrow B_1,A_2 \leftrightarrow B_2) \nn \\ 
&& \hsp{-0.6} + (A_1 \leftrightarrow B_1,A_3 \leftrightarrow B_3)
+ (A_1 \leftrightarrow B_1,A_5 \leftrightarrow B_5) + 
(A_2 \leftrightarrow B_2,A_3 \leftrightarrow B_3) \nn \\
&& \hsp{-0.6} + (A_2 \leftrightarrow B_2,A_5 \leftrightarrow B_5)
+ (A_3 \leftrightarrow B_3,A_5 \leftrightarrow B_5) - 
(A_1 \leftrightarrow B_1,A_2 \leftrightarrow B_2,A_3 
\leftrightarrow B_3) \nn \\
&& \hsp{-0.6} - (A_1 \leftrightarrow B_1,A_2 \leftrightarrow B_2,A_5 
\leftrightarrow B_5) - (A_1 \leftrightarrow B_1,A_3 \leftrightarrow 
B_3,A_5 \leftrightarrow B_5) \nn \\
&& \left. \hsp{-0.6} - (A_2 \leftrightarrow B_2,A_3 \leftrightarrow 
B_3,A_5 \leftrightarrow B_5) + (A_1 \leftrightarrow B_1,A_2 
\leftrightarrow B_2, A_3 \leftrightarrow B_3,A_5 \leftrightarrow B_5) 
\rule{0pt}{13pt} \right] \nn \\
&& \left. \left. \hsp{-0.6}\rule{0pt}{13pt} 
+ \mbox{~``cyclic~permutations''} \right\} \right)
\label{O5-6-AB-inst}
\ea
which can be obtained after a lengthy but straightforward calculation
utilising the solution (\ref{phi2-inst-sol}) for the scalar field. In
(\ref{O5-6-AB-inst}) only the terms relevant to the computation of
(\ref{spec-2pt-O5-6}) have been kept.

The first contribution to the two-point function (\ref{spec-2pt-O5-6})
in semiclassical approximation, obtained substituting the background
values of the composite operators, is 
\ba
&& G^{(1)}(x_1,x_2) = \int \dr\mu_{\rm phys}\,\er^{-S_{\rm inst}}\, 
\hat\scrO^{(23)1}_{5,\mb6}(x_1;\rho,x_0;\eta,\bar\xi,\nu,\bar\nu) 
\hat\scrO^{(23)1}_{5,\mb6}(x_2;\rho,x_0;\eta,\bar\xi,\nu,\bar\nu) 
\nn \\
&& = \frac{\pi^{-4N}\gy^{4N-10}\er^{2\pi i\tau}}{N^3(N-1)!(N-2)!}  
\int \dr\rho\,\dr^4x_0 \,\prod_{A=1}^4
\dr^2\eta^A\dr^2\bar\xi^A \, \dr^{N-2}\nu^A \dr^{N-2}\bar\nu^A
\,\rho^{4N-13} \nn \\
&& \er^{\frac{\pi^2}{16\gy^2\rho^2}\veps_{ABCD} (\bar\nu^{[A}\nu^{B]})
(\bar\nu^{[C}\nu^{D]})} \frac{\rho^8}{[(x_1-x_0)^2+\rho^2]^9}
\frac{\rho^8}{[(x_2-x_0)^2+\rho^2]^9} \nn \\
&& \left((\bar\nu^{[1}\nu^{4]})+(\bar\nu^{[2}\nu^{3]})\right)^2 
\left[(\zeta^1)^2(\zeta^2)^2(\zeta^3)^2(\zeta^4)^2\right]\!(x_1)
\left[(\zeta^1)^2(\zeta^2)^2(\zeta^3)^2(\zeta^4)^2\right]\!(x_2)
\label{2pt-O5-6-semicl}
\ea
where only the terms giving a non-vanishing contribution have been
included. An overall numerical coefficient has been omitted and will
be reinstated in the final formula. The integrals over the $\nu$ and
$\bar\nu$ modes can be performed with the aid of the generating
function defined in section \ref{sc-corrfunct}, which allows to
replace them with a five-sphere integral of the type
(\ref{s5int1}). The integration over the 16 superconformal modes is
also easily done using (\ref{fermiint}), so that we get
\ba
G^{(1)}(x_1,x_2) &\!\!=\!\!& 
\frac{2^{-2N}(N^2-3N+2)\Gamma(2N-2)\,\er^{2\pi i\tau}}{N^3(N-1)!(N-2)!} 
(x_1-x_2)^8 \nn \\ 
&& \int \frac{\dr^4x_0\dr\rho}{\rho^5}
\frac{\rho^9}{[(x_1-x_0)^2+\rho^2]^9}
\frac{\rho^9}{[(x_2-x_0)^2+\rho^2]^9} \, .
\label{2pt-O5-6-fin-int}
\ea

The second type of contribution involves one scalar contraction
\be
G^{(2)}(x_1,x_2) = \fr{\gy^{10}N^3} \la \Tr\left[
\left(\v^1\v^i\v^j\v^i\v^j\right)(x_1)\right]  \Tr\left[
\left(\v^1\v^k\v^l\v^k\v^l\right)(x_2)\right] 
\raisebox{-10pt}{\hsp{-7.58}
\rule{0.4pt}{4pt}\rule{4.16cm}{0.4pt}\rule{0.4pt}{4pt}}
\hsp{3.2}
\ra + \cdots \, ,  
\label{2pt-O5-6-contr1}
\ee
where the ellipsis stands for the other contractions according to
Wick's theorem. The computation of this correlator is rather involved
because of the complicated form of the scalar propagator in the
instanton background. The insertion the propagator (\ref{propfin})
eventually gives rise to two types of structures. Combining all the
terms which arise from (\ref{2pt-O5-6-contr1}) we obtain (again up to
an overall numerical coefficient)
\ba
&& \hsp{-0.5} G^{(2)}(x_1,x_2) = \frac{2^{-2N}\Gamma(2N-1)\,
\er^{2\pi i\tau}}{N^3(N-1)!(N-2)!}
\int\frac{\dr^4x_0\,\dr\rho}{\rho^5}\,\dr^5\Omega
\prod_{A=1}^4 \dr^2\eta^A\,\dr^2\bar\xi^A \nn \\
&& \hsp{-0.5}\left[ \frac{c_1}{(x_1-x_2)^2}
\frac{\rho^{16}}{[(x_1-x_0)^2+\rho^2]^8[(x_2-x_0)^2+\rho^2]^8}
+\frac{c_2\rho^{18}}{[(x_1-x_0)^2+\rho^2]^9[(x_2-x_0)^2+\rho^2]^9}
\right] \nn \\
&& \hsp{-0.5}\left[(\zeta^1)^2(\zeta^2)^2(\zeta^3)^2(\zeta^4)^2
\right]\!(x_1)\left[(\zeta^1)^2(\zeta^2)^2(\zeta^3)^2(\zeta^4)^2
\right]\!(x_2) \rule{0pt}{18pt} \nn \\
&& \hsp{-0.5} = \frac{2^{-2N}\Gamma(2N-1)\,\er^{2\pi i\tau}}
{N^3(N-1)!(N-2)!} \, (x_1-x_2)^8 \int \frac{\dr^4x_0\,\dr\rho}{\rho^5} 
\, \left[\frac{c_2\rho^{18}}{[(x_1-x_0)^2+\rho^2]^9
[(x_2-x_0)^2+\rho^2]^9} \right. \nn \\
&& \hsp{-0.5} \left. + \frac{c_1}{(x_1-x_2)^2}
\frac{\rho^{16}}{[(x_1-x_0)^2+\rho^2]^8[(x_2-x_0)^2+\rho^2]^8} 
\right] \, . \label{2pt-O5-6-contr2} 
\ea
Summing (\ref{2pt-O5-6-fin-int}) and (\ref{2pt-O5-6-contr2}) we
obtain the complete result for the two-point function
(\ref{spec-2pt-O5-6}) in the one-instanton sector which reads
\ba
G(x_1,x_2) &\!\!=\!\!& \frac{3^2\,5^2\,\pi^{-15}2^{-2N-16}\,
\er^{2\pi i\tau}}{N^3(N-1)!(N-2)!} \, (x_1-x_2)^8 \int \dr^4x_0\,
\dr\rho \nn \\ 
&& \left[ \frac{a_1(N)\rho^{13}}{(y_1^2+\rho^2)^9(y_2^2+\rho^2)^9} 
+ \frac{1}{(x_1-x_2)^2}
\frac{a_2(N)\rho^{11}}{(y_1^2+\rho^2)^8(y_2^2+\rho^2)^8} \right] \, , 
\label{2pt-O5-6-fin}
\ea
where $y_i=x_i-x_0$ and all the numerical factors have been
reinstated. The coefficients $a_1(N)$ and $a_2(N)$ are 
\be
a_1(N) = \frac{2}{3} \,\Gamma(2N-1) \, , \quad a_2(N) = (N^2-3N+2)\,
\Gamma(2N-2)+\frac{2}{3}\,\Gamma(2N-1) \, .
\label{Ndep-d5}
\ee

The final bosonic moduli space integrations in (\ref{2pt-O5-6-fin})
are logarithmically divergent. The second term is exactly of the same
form as that encountered in section \ref{d4r1}. The first term in the
last line of (\ref{2pt-O5-6-fin}) is evaluated in a completely
analogous way and also leads to a simple pole singularity after
dimensional regularisation of the $x_0$ integral. As in the case of
the singlets of dimension 4 we find non-zero entries in the matrix
$K^{rs}$ for operators with $\D_0=5$ transforming in the $\mb6$ of
SU(4). Therefore some of the operators in this sector acquire an
anomalous dimension in the one-instanton sector.

The calculation of the other two-point functions in this sector for
operators made of only elementary scalars can be carried on in a
similar fashion. The result is of the same form as
(\ref{2pt-O5-6-fin}) apart from the numerical coefficient. In
particular the operator $\scrO^{(21)i}_{5,\mb6}$ vanishes in a
one-instanton background, so that two-point functions involving this
operator only receive contribution from terms in which pairs of
scalars are contracted. The double-trace operators in
(\ref{O5-6d-scal}) can be treated in a similar way. The two-point
functions involving the remaining operators in this sector,
(\ref{O5-6-fs})-(\ref{O5-6-fermix}), require higher order terms in the
iterative solution of the field equations for the vector and the
fermions. Without computing these two-point functions it is not
possible to extract the values of the instanton induced anomalous
dimensions of $\D_0=5$ operators in the $\mb6$ of SU(4). The fact that
operators in this sector do receive instanton corrections is however
also confirmed by the analysis of the four-point function in appendix
\ref{ope}. The singularity observed in the OPE studied there
corresponds to the contribution of operators in the sector examined in
this subsection.

\subsubsection{$\D_0=5$, $[2,0,0]$ and $[0,0,2]$} 
\label{d5r10}

In this and the following subsections we briefly discuss the other
SU(4) representations which appear at the level $\D_0=5$. The
discussion will not be detailed. The analysis follows in a
straightforward way what was done in previous sections. What makes the
study of these sectors more involved is the large number of operators
which makes the construction of a basis rather laborious. In
particular we shall only consider single-trace operators. It is
understood that for all operators involving at least four elementary
fields there exist also double-trace operators. From the discussion in
the previous sections it is evident that for a qualitative analysis of
the zero-mode structure the number of traces is not relevant.

Operators in the representation $[2,0,0]\equiv\mb{10}$ are obtained
from (\ref{d5deriv})-(\ref{d5scal}). The combination  (\ref{d5deriv})
can be projected onto the $\mb{10}$ by fully antisymmetrising the
indices on the scalars
\be
\scrO^{(1)[ijk]}_{5,\mb{10}} = \fr{\gy^3N^{1/2}} \, \Tr\left(
\scrD^\mu\v^{[i}\scrD_\mu\v^j\v^{k]} \right) \, .
\label{O5-10-deriv}
\ee
Operators in the $\mb{10}$ containing fermions come from
(\ref{d5fermi})-(\ref{d5fermi-deriv}). We find
\ba
&& \scrO^{(2)(AB)}_{5,\mb{10}} = \fr{\gy^4N}\,\Tr\left(
\v^i\v^i\lambda^{\a(A}\lambda^{B)}_\a\right) \, , \quad 
\scrO^{(3)(AB)}_{5,\mb{10}} = \fr{\gy^4N}\,\Tr\left(
\v^i\lambda^{\a(A}\v_i\lambda^{B)}_\a\right) \, , \nn \\
&& \scrO^{(4)(AB)}_{5,\mb{10}} = \fr{\gy^4N}\,t^{(AB)}_{[ijk]} \,
\Tr\left(\S^i_{CD}\v^j\v^k\lambda^{\a C}\lambda^D_\a \right) 
\, , \nn \\
&& \scrO^{(5)(AB)}_{5,\mb{10}} = \fr{\gy^4N} \, \Tr\left[
\v^i\v^j \left(\S_{ijC}{}^A\lambda^{\a(B}\lambda^{C)}_\a + 
\S_{ijC}{}^B\lambda^{\a(A}\lambda^{C)}_\a \right) \right] \, ,
\nn \\
&& \scrO^{(6)(AB)}_{5,\mb{10}} = \fr{\gy^4N} \, \Tr\left(
\left(\S_{ijC}{}^A\v^i\lambda^{\a(B}\v^j\lambda^{C)}_\a + 
\S_{ijC}{}^B\v^i\lambda^{\a(A}\v^j\lambda^{C)}_\a \right) \right] \, ,
\nn \\ 
&& \scrO^{(7)(AB)}_{5,\mb{10}} = \fr{\gy^2} \, \Tr\left(
\scrD_\mu\lambda^{\a(A}\scrD^\mu\lambda^{B)}_\a \right) \, , \quad
\scrO^{(8)}_{5,\mb{10}} = \fr{\gy^3N^{1/2}} \, t^{(AB)}_{[ijk]} \, 
\Tr\left(\S^{ij\,D}_{\:C}\scrD_\mu \v^k \lambda^{\a C}\s^\mu_{\a\adot}
\bar\lambda^\adot_D \right) \, , \nn \\
&& \scrO^{(9)(AB)}_{5,\mb{10}} = \fr{\gy^3N^{1/2}}\, \Tr\left(
F_{\mu\nu}\lambda^{\a(A}\s^{\mu\nu\,\b}_{\:\a}\lambda^{B)}_\b \right)
\, , \quad \scrO^{(10)(AB)}_{5,\mb{10}} = \fr{\gy^3N^{1/2}}\, \Tr\left(
\tilde F_{\mu\nu}\lambda^{\a(A}\s^{\mu\nu\,\b}_{\:\a}\lambda^{B)}_\b 
\right) \, , \nn \\
&& \scrO^{(11)(AB)}_{5,\mb{10}} = \fr{\gy^4N}\,t^{(AB)}_{[ijk]} \,
\Tr\left(\Sbar^{i\,CD}\v^j\v^k\bar\lambda_{\adot C}\bar\lambda^\adot_D 
\right) \, , \nn \\
&& \scrO^{(12)(AB)}_{5,\mb{10}} = \fr{\gy^4N}\,
t^{(AB)}_{[ijk]} \,\Tr\left(\Sbar^{i\,CD}\v^j\bar\lambda_{\adot C}
\v^k\bar\lambda^\adot_D \right) \, , \nn \\ 
&& \scrO^{(13)(AB)}_{5,\mb{10}} = \fr{\gy^4N} \, 
\Sbar^{C(A}_i\Sbar^{B)D}_j \,\Tr\left(\v^{(i}\v^{j)}
\bar\lambda_{\adot C}\bar\lambda^\adot_D \right) \, , \nn \\
&& \scrO^{(14)(AB)}_{5,\mb{10}} = \fr{\gy^4N} \, 
\Sbar^{C(A}_i\Sbar^{B)D}_j \,\Tr\left(\v^{(i}
\bar\lambda_{\adot C}\v^{j)}\bar\lambda^\adot_D \right) \, .
\label{O5-10-fermi}  
\ea
With only elementary scalars we can construct the following
combinations in the $\mb{10}$
\ba
&& \scrO^{(15)(AB)}_{5,\mb{10}} = \fr{\gy^5N^{3/2}} \, t^{(AB)}_{[ijk]}
\, \Tr\left( \v^l\v^l\v^i\v^j\v^k \right) \, , \nn \\
&& \scrO^{(16)(AB)}_{5,\mb{10}} = \fr{\gy^5N^{3/2}} \, t^{(AB)}_{[ijk]}
\, \Tr\left( \v^l\v^i\v^l\v^j\v^k \right) \, . 
\label{O5-10-scal}
\ea
The operators in the representation $[0,0,2]\equiv\mbb{10}$ are built
in a completely analogous way. Moreover for operators made either of
two scalars and two fermions or of five scalars there are also
double-trace combinations. We should then consider the two-point
functions $\la\scrO^{(r)}_{5,\mb{10}}(x_1)\bar\scrO^{(s)}_{5,\mbb{10}}
(x_2)\ra$ to compute the instanton corrections to the matrix of
anomalous dimensions. By choosing specific components one can verify
that for all the above operators and their conjugates the instanton
contributions to two-point functions vanish. There are always 10
fermion modes involved and depending on the choice of components it
may be possible to soak up the sixteen superconformal modes, but when
this is the case the remaining integral over the $\nu$ and $\bar\nu$
fermion variables vanishes. There is no instanton correction to the
scaling dimension of operators in these sectors.

\subsubsection{$\D_0=5$, $[0,3,0]$}
\label{d5r50}

The construction of operators in the representation
$[0,3,0]\equiv\mb{50}$ is very similar to what was done in the
previous subsection for the $\mb{10}$: in the product of three
$\mb6$'s the $\mb{50}$ is obtained from the totally symmetric and
traceless combination and the $\mb{10}$ arises as totally
antisymmetric combination.

We find the following operators in this sector. From (\ref{d5deriv})
we select the $\mb{50}$ taking
\be
\scrO^{(1)\{ijk\}}_{5,\mb{50}} = \fr{\gy^3N^{1/2}}\, \Tr\left(
\v^{\{i}\scrD_\mu\v^j\scrD^\mu\v^{k\}} \right) \, .
\label{O5-50-deriv}
\ee
With two scalars and two fermions we get
\be
\scrO^{(2)\{ijk\}}_{5,\mb{50}} = \fr{\gy^4N}\, \Tr\left(
\S^{\{i}_{AB}\v^j\v^{k\}}\lambda^{\a A}\lambda_\a^B\right) \, , \quad
\scrO^{(3)\{ijk\}}_{5,\mb{50}} = \fr{\gy^4N}\, \Tr\left(
\Sbar^{AB\{i}\v^j\v^{k\}}\bar\lambda_{\adot A}\bar\lambda^\adot_B
\right) \, .
\label{O5-50-fermi}
\ee
Finally there are the operators made of scalars only
\ba
&& \scrO^{(4)\{ijk\}}_{5,\mb{50}} = \fr{\gy^5N^{3/2}} \, \Tr\left(
\v^l\v_l\v^{\{i}\v^j\v^{k\}} \right) \, , \nn \\ 
&& \scrO^{(5)\{ijk\}}_{5,\mb{50}} = \fr{\gy^5N^{3/2}} \, \Tr\left(
\v^l\v^{\{i}\v_l\v^j\v^{k\}} \right) \, . 
\label{O5-50-scal}
\ea
As in the previous case one can easily verify that none of the above
operators gives rise to non-vanishing two-point functions in the
instanton background. Again depending on the component considered the
vanishing of the two-point functions follows from the five-sphere
integration after re-expressing the dependence on the $\nu$ and
$\bar\nu$ modes in terms of angles $\Omega^{AB}$.

\subsubsection{$\D_0=5$, $[1,1,1]$, $[3,0,1]$, $[1,0,3]$, $[2,2,0]$
and $[0,2,2]$} 
\label{d5r64}

The representations $[1,1,1]\equiv\mb{64}$, $[3,0,1]\equiv\mb{70}$,
$[1,0,3]\equiv\mbb{70}$, $[2,2,0]\equiv\mb{126}$ and
$[0,2,2]\equiv\mbb{126}$ are rather complicated to analyse. They
appear with high multiplicity and to disentangle them one needs
projectors which are not straightforward to construct. We shall only
briefly sketch how one can proceed to define a basis of operators. In
section \ref{discussion} an argument will be given that implies that
there cannot be instanton contributions to two-point functions of
operators in these sectors.

The $\mb{64}$ appears in (\ref{d5deriv}), (\ref{d5fermi}),
(\ref{d5fermi-mix}) and (\ref{d5scal}). The first type of operator
however does not contribute for the same reason discussed in section
\ref{d3r64}. The operators with fermions come from
\ba
&& \hsp{-1} \mb{15}\otimes\mb6 \to \Tr\left(\S^i_{AB}\v^{[j}\v^{k]}
\lambda^{\a A}\lambda^B_\a\right) \! 
\raisebox{-4pt}{$\Big|_{\mb{64}}$} \, , \hsp{0.6} 
\mb{20^\pp}\otimes\mb6 \to \Tr \left(\S^i_{AB}\v^{\{j}\v^{k\}}
\lambda^{\a A}\lambda^B_\a \right) \! 
\raisebox{-4pt}{$\Big|_{\mb{64}}$} \, , \nn \\
&& \hsp{-1} \mb{15}\otimes\mb{10} \to \Tr \left(\v^{[i}\v^{j]}
\lambda^{\a (A}\lambda^{B)}_\a\right) \! 
\raisebox{-4pt}{$\Big|_{\mb{64}}$} \, , \hsp{1}
\mb{20^\pp}\otimes\mb{10} \to \Tr \left(\v^{\{i}\v^{j\}}
\lambda^{\a (A}\lambda^{B)}_\a\right) \! 
\raisebox{-4pt}{$\Big|_{\mb{64}}$} \, , \nn \\
&& \hsp{-1} \mb{15}\otimes\mb6 \to \Tr\left[\scrD_\mu\v^i
\left(\lambda^{\a A}\s^\mu_{\a\adot}\bar\lambda^\adot_B - 
\fr4 \d^A{}_B \lambda^{\a C}\s^\mu_{\a\adot}\bar\lambda^\adot_C
\right)\right] \! \raisebox{-4pt}{$\Big|_{\mb{64}}$} \, ,
\label{O5-64-fermi}
\ea
where the notation indicates that each combination has to be suitably
projected onto the $\mb{64}$, which requires to make it orthogonal to
the operators in the other representations appearing in the same
tensor product. The operators involving spinors in the $\mbb4$ can be
obtained in a similar fashion. The construction of operators in the
$\mb{64}$ made of five scalars is very involved. The $\mb{64}$ appears
in $\mb6\otimes\mb6\otimes\mb6\otimes\mb6\otimes\mb6$ with
multiplicity 24. This does not take into account the cyclicity of the
trace. To incorporate this we can build the operators leaving one
scalar fixed in the first position in the trace and combining the
remaining four scalars respectively into the representations
$\mb{15}$, $\mb{20^\pp}$, $\mb{45}$, $\mbb{45}$, $\mb{84}$ and
$\mb{175}$. The corresponding combinations have been discussed in the
section on operators of bare dimension $\D_0=4$. We then need to
project the tensor products $\mb6\otimes\mb{15}$,
$\mb6\otimes\mb{20^\pp}$, $\mb6\otimes\mb{45}$, $\mb6\otimes\mbb{45}$,
$\mb6\otimes\mb{84}$ and $\mb6\otimes\mb{175}$, which all include the
$\mb{64}$.

The $\mb{70}$ appears in (\ref{d5fermi}) and (\ref{d5scal}). The
operator made of two scalars and two fermions is contained in the
first combination in the second line of (\ref{O5-64-fermi}). The
operators made of only scalars can be extracted with the procedure
outlined in the previous paragraph from $\mb6\otimes\mb{45}$ and
$\mb6\otimes\mb{84}$. For the $\mbb{70}$ one can proceed in a similar
way. 

Operators in the $\mb{126}$ also arise from (\ref{d5fermi}) and
(\ref{d5scal}). One operator of the form (\ref{d5fermi}) is in the
decomposition of the second combination in the second line of
(\ref{O5-64-fermi}). Then there are operators made of five scalars
which can be obtained from $\mb6\otimes\mb{45}$ and
$\mb6\otimes\mb{175}$. Again the conjugate operators in the
$\mbb{126}$ are treated in a completely analogous way.

The same reasoning can be applied to the construction of double-trace
operators in these sectors.

\subsubsection{$\D_0=5$, $[0,5,0]$}
\label{d5r196}

In the representation $[0,5,0]\equiv\mb{196}$ we find one single-trace
and one double-trace operator. These only involve elementary scalars
and belong to the same class as the operators discussed in sections
\ref{dim2scal}, \ref{d3r50} and \ref{d4r105}, \ie the class of scalar
operators of dimension $\D_0=\ell$ transforming in the representation
$[0,\ell,0]$. For such operators it is always possible to choose a
component which can be written in terms of only one complex scalar,
$\phi^I$. For the single-trace operator in the $\mb{196}$ we can take
$\Tr\left(\phi^1\phi^1\phi^1\phi^1\phi^1\right)$, which is immediately
verified to have vanishing two-point functions in the instanton
background. This is in fact a 1/2 BPS operator dual to a third
Kaluza--Klein excited mode of a supergravity scalar in AdS$_5\times
S^5$. A similar argument can be made for the double-trace operator.

\subsubsection{$\D_0=5$, $[2,1,2]$ and $[1,3,1]$}
\label{d5r300-384}

Operators in the representations $[2,1,2]\equiv\mb{300}$ and
$[1,3,1]\equiv\mb{384}$ can only be obtained from the product of five
scalars, \ie from (\ref{d5scal}) and the analogous double-trace
combinations. Taking into account the cyclicity of the trace one finds
that the representation $\mb{384}$ is not realised in terms of
gauge-invariant operators. This is analogous to what was observed for
the $\mb{64}$ at $\D_0=3$ and the $\mb{175}$ at $\D_0=4$. Operators
in the $\mb{300}$ belong to the class \cite{bks} of operators of
dimension $\D_0=2a+b$ transforming in the representation $[a,b,a]$,
for which it is always possible to single out components that only
involve two complex scalars fields, $\phi^I$ and $\phi^J$. From
equations (\ref{phiI-vphiAB}) in appendix \ref{convents} and the
discussion in the previous sections it is then clear that operators of
this type cannot contain the required combination (\ref{saturmodes})
of  superconformal fermion zero-modes and therefore all their
two-point functions vanish in the instanton background. This applies
to both single- and double-trace operators.

\section{Discussion and conclusions}
\label{discussion}

In the previous sections we have analysed instanton contributions to
two-point correlation functions of scalar operators of bare dimension
$\D_0=2,3,4,5$ in the $\scrN$=4 SYM theory in the semiclassical
approximation. In this way it has been possible to identify the
sectors in which operators get instanton corrections to their scaling
dimensions. One of our motivations was to try to understand to what
extent the S-duality of the theory manifests itself in the spectrum of
anomalous dimensions at weak coupling. It is somewhat surprising that
very few operators among those considered are corrected by
instantons. Our results show that there is a large class of operators
that display non-renormalisation properties in topologically
non-trivial sectors. It was already known \cite{bkrs3} that operators
belonging to the Konishi multiplet, which acquire an anomalous
dimension in perturbation theory, are not corrected by instanton
effects. We have now shown that the same is true for the majority of
the scalar operators of bare dimension $\D_0\le5$. As already remarked
these results imply the absence of instanton corrections to a much
larger set of operators. The non-renormalisation properties extend to
all the components of the multiplets for which a representative was
found not to be corrected. Many of the operators that have been
studied in this paper, and for which no instanton corrections arise,
have also been studied in perturbation theory, as well as at strong
coupling via the dual supergravity amplitudes in AdS$_5\times S^5$
\cite{bkrs2,bers,bks,b1,afp,appss}. At the perturbative level the
scaling dimensions of those belonging to long multiplets do get an
anomalous quantum correction. The non-renormalisation properties
observed in this paper are therefore rather surprising, in particular
in view of the S-duality of the theory. Although single operators
might transform in a complicated way under S-duality the full spectrum
of scaling dimensions must be invariant and instantons are expected to
play a crucial r\^ole in implementing the duality as is the case in
the dual type IIB string theory.

A more careful analysis shows however that the observed
non-renormalisation properties are specific to the cases of operators
of relatively small bare dimension considered here. In order to
explain this we need to recall the general discussion of instanton
contributions to two-point functions in section \ref{k12ptfunct}. 
In the one-instanton sector (and up to a non-zero overall numerical
constant) a generic two-point function is given by  
\ba
&&\hsp{-0.8} \la \bar\scrO^r(x_1) \scrO^s(x_2) \ra = c^{rs}\a(m,n;N)\,
\gy^{8+n+m}\,\er^{2\pi i\tau} \int \dr\rho \,\dr^4x_0 \,\dr^5\Omega 
\prod_{A=1}^4 \dr^2\eta^A \, \dr^2 {\bar\xi}^A \rho^{n+m-5} \nn \\ 
&& \hsp{2.35} \hat\scrO^r\!\left(x_1;x_0,\rho,\eta,\bar\xi,
\bar\nu\nu(\Omega)\right)
\hat\scrO^s\!\left(x_2;x_0,\rho,\eta,\bar\xi,\bar\nu\nu(\Omega)\right) 
\, , \label{genk12pt}
\ea
where $n$ an $m$ denote respectively the number of $\nsix$ and $\nten$
bilinears entering the expressions for the operators in the instanton
background. The notation in the second line indicates that these
bilinears have been re-expressed in terms of the angular variables
$\Omega^{AB}$.  The $N$-dependence is contained in the function
$\a(m,n;N)$ defined in equation (\ref{largeN-coeff}). In evaluating
(\ref{genk12pt}) one first computes the integrations over the
superconformal modes which yield a non-vanishing the result if and
only if both operators contain the combination
$(\zeta^1\zeta^1)(\zeta^2\zeta^2)(\zeta^3\zeta^3)(\zeta^4\zeta^4)$.
After performing the integrals over the $\eta$'s and $\bar\xi$'s one
is left with the integration over the five-sphere and over the
original bosonic collective coordinates, $x_0$ and $\rho$. The latter
integrals are in general logarithmically divergent as follows from
dimensional analysis, signalling a contribution to the matrix of
anomalous dimensions.

Now consider the cases of the operators studied in this paper. The
profiles of operators of dimension $\D_0=2$ and $4$ contain  no
dependence on $\Omega^{AB}$, those of operators of dimension $\D_0=3$
and $5$ are linear in $\Omega^{AB}$. Let us assume that there are
operators with $\D_0=2,4$ in two different sectors, \ie transforming
in different (and not conjugate) representations $\mb{r}_1$ and
$\mb{r}_2$, which both receive instanton corrections, so that the
two-point functions
$\la\scrO_{\D_0,\mb{r}_i}(x_1)\scrO_{\D_0,\mb{r}_i}(x_1)\ra_{\rm
inst}$, $i=1,2$, are different from zero. This would lead to a
paradox. Since the five-sphere integral in this case is trivial we
would find that the two-point function
$\la\scrO_{\D_0,\mb{r}_1}(x_1)\scrO_{\D_0,\mb{r}_2}(x_1)\ra_{\rm
inst}$ is also non-zero, but this is forbidden by the SU(4)
symmetry. The same situation would arise if there were instanton
corrections to two-point functions of both an operator of dimension 2
and one of dimension 4. In this case we would have an even worse
situation: there would be a non-vanishing two-point function in which
the two operators have different dimension and this would violate
conformal invariance. A similar argument can be repeated in the cases
$\D_0=3$ and $5$. In these cases the five-sphere integrals are of the
form
\be
\int \dr^5\Omega \, \Omega^{AB}\Omega^{CD} = \fr4\,\veps_{ABCD} \, .
\label{omegaint}
\ee
Again we would find that if there were non-zero two-point functions in
two different sectors, then the mixed two-point functions would also
involve  the same integral (\ref{omegaint}) and so would not vanish,
thus violating the SU(4) symmetry. There is no similar problem
associated with overlapping of different SU(4) representations for
two-point functions of one operator with $\D_0=3,5$ and one with
$\D_0=2,4$. In this case we would get a five-sphere integral of the
form
\be 
\int \dr^5\Omega \, \Omega^{AB}  \, ,
\label{zeroomegaint}
\ee
which vanishes identically. 

This argument shows that there can be instanton corrections to the
anomalous dimensions in one and only one sector among those present at
$\D_0=2,4$ and in one and only one sector among those with
$\D_0=3,5$. This is consistent with our findings in the previous
sections, where by direct inspection we have shown that instantons
only correct operators in the sectors $(\D_0=4,\;[0,0,0])$ and
$(\D_0=5,\;[0,1,0])$. What is special about the cases we have
considered is that they always involve (almost) trivial five-sphere
integrals. The general situation, expected for operators of larger
dimension, is that a more complicated dependence on the angles
$\Omega^{AB}$ should enter in the operator profiles. In this way it is
possible to have non-vanishing contributions in various sectors, but
zero overlap between different sectors since the angular dependence
makes them `orthogonal' with respect to the five-sphere integration
\be 
\int \dr^5\Omega \; \tilde\scrO_{\mb{r}_i}(x_1;x_0,\rho,\Omega)\,
\tilde\scrO_{\mb{r}_j}(x_2;x_0,\rho,\Omega) \neq 0 \quad \mathrm{iff}
\quad \mb{r}_j = \mbb{r}_i \, ,
\label{5sphereselect}
\ee 
where we have indicated with a tilde the expressions of the operators
after the integration over the superconformal zero-modes.  It is
therefore natural to conjecture that the operators analysed here are
special and that generically for larger values of $\D_0$ most of the
operators that receive corrections in perturbation theory are also
corrected by instantons. 

Some of the non-renormalisation properties are not restricted to small
values of the bare dimension. All the operators in the SU(2) subsector
transforming in a representation $[a,b,a]$ with $\D_0=2a+b$, as well
as the other components of multiplets containing such operators, do
not receive instanton corrections, irrespective of the value of
$\D_0$.

The above argument explains why so many operators among those
considered appear to be `protected' against instanton effects, whereas
they get corrections in perturbation theory. On the other hand
comparing the perturbative and non-perturbative sectors we notice that
operator mixing is much more complicated at the instanton level. At
small orders in perturbation theory it is possible to identify subsets
of operators that do not mix with others in the same sector and thus
compute the anomalous dimensions within the subsector. For instance at
one loop two-point functions involving one operator made of only
scalars and one containing fermion bilinears in the same sector
vanish. For this reason in \cite{appss,bks} it was possible to compute
the one-loop anomalous dimension of the SU(4) singlet scalar operators
(\ref{O-1-scal-s})-(\ref{O-1-scal-d}) without having to consider the
mixing with other operators in that sector. The study of instanton
corrections to two-point functions shows that in general whenever a
sector receives correction the mixing occurs among all the operators
in that sector. This phenomenon was to be expected: in general the
mixing begins to be relevant at some order in perturbation theory and
the instanton effects we have considered are subleading with respect
to contributions at any order in the perturbative expansion.

In the present paper we have restricted our attention to the
one-instanton sector. Part of the results remain valid in
multi-instanton sectors. In particular the non-renormalisation results
hold for arbitrary instanton number $K$, since they only rely on the
analysis of the superconformal zero-modes. The calculation in sectors
in which there are non-vanishing contributions is in general much more
complicated for $K>1$. Some cases however can be treated for any $K$
in the large $N$ limit. This is true for sectors in which at leading
order in $\gy$ no non-exact modes enter in the operators and
contractions are also not allowed. This is the case in particular for
the singlet at $\D_0=4$: the two-point functions for arbitrary $K$ in
the large $N$ limit coincide with the one-instanton results of section
\ref{d4r1} up to a calculable $K$-dependent coefficient. This is
because a saddle-point approximation can be used for large $N$. When
there is a non-trivial dependence on the non-exact modes their
evaluation at the saddle-point is complicated and the generalisation
much more difficult.

The extension of the calculations presented here to the case of
orthogonal or symplectic gauge groups is also straightforward. It
should be noted however that the construction of bases of independent
gauge-invariant operators is significantly different in these cases.
We have considered only Lorentz scalars, but the same methods can 
be used to study non-scalar operators. In this case the full set of
PSU(2,2$|$4) quantum numbers, $(\D_0,J_1,J_2;[a,b,c])$ is needed to
identify the various sectors. Therefore the spectrum is richer and the
construction of independent operators in each sector more involved, but
once this is done the computation of anomalous dimensions proceeds on
the same lines as in the cases considered in this paper.

The non-renormalisation properties that we have derived in
semiclassical approximation can be argued to remain valid at higher
orders in  $\gy$. The simplest contributions beyond the semiclassical
result come from additional insertions of $\bar\nu\nu$ bilinears,
which modify the integrations over the angular variables
$\Omega^{AB}$. The resulting five-sphere integrals are only
non-vanishing if an equal number of modes of each flavour is contained
in the combination of $\bar\nu\nu$'s. This implies that including
subleading terms with more fermion modes cannot produce non-zero
results for two-point functions which vanish at leading order. A
similar argument can be made for higher-order corrections involving
contractions because the propagator (\ref{propfin}) is proportional to
$\veps^{ABCD}$.

As previously observed non-protected operators of large dimension are
expected  generically to also receive instanton corrections. It is
therefore natural to ask what the situation is for the operators which
are relevant for the BMN limit. These are operators of large dimension
and  large charge with respect to one of the generators of the
R-symmetry group and the techniques developed here can be applied to
the study of such operators. Instanton effects in the BMN limit and
the comparison with string theory in a plane wave background are
currently under investigation \cite{gks}.

As discussed in the introduction, recently there have been many
interesting developments in connection with the computation of
anomalous dimensions in the $\scrN$=4 SYM theory, leading to new
non-trivial tests of the AdS/CFT correspondence
\cite{bmn,sz,bkpss,cfhm,gkp,r,ft3,bfst,bmsz}. Of particular interest
is the emergence of an integrability structure in the theory
\cite{mz,bks,bs1,b2}, which appears to have a counterpart in the dual
string theory. On the string side this integrability property appears
at the level of the semiclassical analysis of solitonic string
configurations \cite{afrt,bmsz} as well as through the emergence of an
infinite set of non-local classically conserved charges
\cite{bpr,dnw}.  In the gauge theory the integrability arises most
naturally when the problem of computing scaling dimensions for
gauge-invariant operators is reformulated in terms of an eigenvalue
problem for the dilation operator, $\hat D$, \cite{bks,bs1,b2}. $\hat
D$ acts on gauge-invariant composite operators as
\be
\hat D \scrO^{(r)} = \D_r \scrO^{(r)} \, 
\label{dilaction}
\ee
where $\D_r$ is the scaling dimension of $\scrO^{(r)}$. Therefore
knowing the form of the dilation operator allows to compute anomalous
dimensions in a very efficient way expanding (\ref{dilaction}). This
is the approach followed in \cite{bks,b1,b2} at the perturbative level.

The results of the present paper do not seem to be relevant for the
issue of the integrability of $\scrN$=4 SYM. This is because if indeed
an integrable structure survives in the full quantum theory it is
expected to arise only in the planar limit. In this limit instanton
effects are exponentially suppressed, since they produce contributions
of order $\er^{-8\pi^2/\gy^2}\sim\er^{-8\pi^2N/\lambda}$. On the other
hand the possibility of computing instanton induced corrections to the
dilation operator in $\scrN$=4 SYM is very interesting in itself and
our results represent a first step in this direction.  Although we have
not addressed this issue in the previous sections, we can make some
general considerations based on the results obtained for two-point
functions. First of all in the closed SU(2) sector of operators of
dimension $\D_0=2a+b$ transforming in the $[a,b,a]$ of SU(4) the
dilation operator does not receive non-perturbative corrections at
all. The same is true for the even simpler sector of $\D_0=\ell$
operators in the $[0,\ell,0]$.  This has been verified explicitly   in
sections \ref{dim2scal} ($\D_0=2$, $[0,2,0]$), \ref{d3r50} ($\D_0=3$,
$[0,3,0]$), \ref{d4r84} ($\D_0=4$, $[2,0,2]$), \ref{d4r105} ($\D_0=4$,
$[0,4,0]$), \ref{d5r196} ($\D_0=5$, $[0,5,0]$) and \ref{d5r300-384}
($\D_0=5$, $[2,1,2]$ and $[1,3,1]$). Using the fact that operators in
these sectors  have components that can be written in terms of only
two complex scalars it is possible to generalise the result to
arbitrary subsectors of the above type.

In general however the dilation operator contains instanton
corrections as well and it can be expanded as
\be
\hat D = \sum_{n=0}^\infty c_{n}(N)\,\gy^{n}\,\hat D_{n} + 
\sum_{K>0}\sum_{m=0}^\infty c_{(K,m)}(N)\,\gy^{m}\, 
\er^{2\pi iK\tau} \, \hat D_{(K,m)} \, ,
\label{gendilop}
\ee
where the first sum denotes the perturbative contributions and the
second double sum incorporates the instanton corrections including the
perturbative fluctuations around the leading semiclassical term in
each instanton sector. An analogous series of anti-instanton
contributions (proportional to $\er^{2\pi i\bar\tau K}$) has not been
indicated explicitly. 

Focusing on the non-perturbative part, it is natural to construct the
terms $\hat D_{(K,m)}$ not as operators acting on the elementary
fields of the $\scrN$=4 theory, as in perturbation theory, but as
operators acting on the multi-instanton collective coordinates. In
other words one can consider the action of the dilation operator as
realised on the instanton moduli space. In \cite{dkm2} it was shown
that the supersymmetry algebra (in the $\scrN$=2 case) can be realised
in a simple and elegant way on the ADHM collective coordinates (see
appendix \ref{k1adhm}) before imposing the ADHM constraints.  The
construction of the instanton supermultiplet acting with broken
supersymmetries on the bosonic solution, which was outlined in section
\ref{N4instmult} for the SU(2) case, is an example of application of
this idea. As already observed, this approach can be implemented
directly in superspace, at least for $\scrN$=1 supersymmetric
theories. In \cite{bgk} a superspace description of the SU(2)
one-instanton moduli space in $\scrN$=4 SYM was presented and the
corresponding superconformal transformations were used in the
computation of instanton contributions to Wilson loops. The
realisation of symmetries on the instanton moduli space can be
generalised to the whole superconformal group and in particular the
action of dilations can be analysed. The strategy of \cite{bks,b1}
was to write down the most general form of $\hat D$ compatible with
the structure of the perturbative two-point functions and then fix the
unknown coefficients using known values of anomalous dimensions. The
same procedure can be used to determine the form of the dilation
operator on the instanton moduli space. We shall denote the latter by
$\tilde D_{(K,m)}$ to distinguish it from its counterpart acting on
the space of fields. From our analysis of two-point functions it is
clear that $\tilde D_{(K,m)}$  contains eight derivatives with respect
to the variables $\zeta^A$. In the one-instanton sector it also
involves derivatives with respect to $\nu^A$ and $\bar\nu^A$, which
can be replaced by derivatives with respect to the angular variables
$\Omega^{AB}$.  We can argue that $\tilde D_{(1,m)}$ must be of the
form 
\be 
\tilde D_{(1,m)} \sim \gy^8\,c(N)\,\scrd(x_0,\rho)\,
t_N^{A_1B_1\ldots A_mB_m}\,
\frac{\d^8}{\d(\zeta^1)^2\d(\zeta^2)^2\d(\zeta^3)^2\d(\zeta^4)^2}
\frac{\d^m}{\d\Omega^{A_1B_1}\ldots\d\Omega^{A_mB_m}} \, ,
\label{guessdil}
\ee
where the factor of $\gy^8$ comes from the measure and the dependence
on $N$ is contained in the coefficient $c(N)$, which for large $N$
behaves as $\sqrt{N}$, and in the tensor $t_N^{A_1B_1\ldots
A_mB_m}$. The latter is a projector onto SU(4) singlets. It  consists
of various terms corresponding to the different ways of combining
$\bar\nu{[A}\nu^{B]}$ and $\bar\nu^{(A}\nu^{B)}$ bilinears to form a
singlet. The $N$-dependence is determined by the number of SU(4)
indices paired in a $\mb6$ or in a $\mb{10}$. $\tilde D_{(1,m)}$ has
also a dependence on the bosonic collective coordinates which is
encoded in the function $\scrd(x_0,\rho)$. Determining the exact form
of $\tilde D_{(1,m)}$ appears to be feasible, but more data on
instanton induced anomalous dimensions are needed. 

We hope to investigate further this issue in the future. In view of
the previous discussion this requires the study of operators of larger
dimension. The results obtained in this paper mostly show
non-renormalisation properties for a large class of operators, which
however are expected to be specific to small values of $\D_0$. For
greater values of the bare dimension the situation is more involved
because of the larger number of operators appearing. From the point of
view of instanton calculations there is however a simplification
since as $\D_0$ grows fewer terms in the iterative solution for the
elementary fields are required in semiclassical approximation.
Moreover as previously remarked further simplifications arise when
imposing the constraints of PSU(2,2$|$4), which imply that all
operators in a same multiplet have equal anomalous dimension.

\vsp{1} 
\ndt 
\textbf{Acknowledgments}

\vsp{0.5} 
\ndt 
I would like to thank Gleb Arutyunov, Niklas Beisert, Massimo Bianchi,
Michael Green, Jan Plefka, Yassen Stanev and Matthias Staudacher for
useful and stimulating discussions. I am particularly grateful to
Massimo Bianchi and Matthias Staudacher for their comments on the
manuscript. This work was supported in part by the European Commission
RTN programme under contracts HPRN-CT-2000-00122 and HPRN-CT-2000-00131.

\appendix

\section{Conventions and useful relations}
\label{convents}

In this appendix we summarise the notation used in the paper and
collect some useful relations. 

Lower case Latin letters, $i,j,k,\ldots$ are used for the $\mb6$ of
SU(4) and capital letters, $A,B,C,\ldots$ for the $\mb4$. SO(4)
Lorentz spinor and vector indices are indicated by Greek letters
respectively from the  beginning, $\a,\b,\ldots,\adot,\bdot,\ldots$,
and the mid, $\mu,\nu,\ldots$, of the alphabet. SU($N$) colour indices
are denoted by Latin letters from the end of the alphabet,
$r,s,u,v,\ldots$.

The $\scrN$=4 multiplet comprises six real scalars, $\v^i$, four Weyl
fermions, $\lambda^A_\a$, and a vector, $A_\mu$ with field strength
$F_{\mu\nu}$, all transforming in the adjoint representation of the
gauge group. It is often convenient to label the scalars by an
antisymmetric pair of indices in the $\mb4$, $\v^{[AB]}$, subject to
the reality condition
\be
\bar\v_{AB} \equiv \left(\v^{AB}\right)^* = \fr2 \veps_{ABCD}\v^{CD}
\, . \label{realcond}
\ee
In some situations the $\scrN$=1 formulation proves very useful. The
$\scrN$=1 decomposition of the $\scrN$=4 supermultiplet consists of
three chiral multiplets and one vector multiplet and under this
decomposition only a SU(3)$\times$U(1) subgroup of the SU(4)
R-symmetry group is manifest. The six scalars are combined into three
complex fields, $\phi^I$, $I=1,2,3$, according to
\ba
\phi^I &\!\!=\!\!& \fr{\sqrt{2}} \left(\v^I+i\v^{I+3}\right) \nn \\
\phi^\dagger_I &\!\!=\!\!& \fr{\sqrt{2}} \left(\v^I-i\v^{I+3}\right)
\, . \label{complexscal}
\ea
The complex scalars $\phi^I$ and $\phi^\dagger_I$ transform
respectively in the $\mb3_{1}$ and $\mbb3_{-1}$ of SU(3)$\times$U(1). 
The fermions in the chiral multiplets are
\be
\psi^I_\a = \lambda^I_\a \, , \qquad \bar\psi_I^\adot =
\bar\lambda_I^\adot \, , \qquad I=1,2,3 \, ,
\label{n1chifermions}
\ee
transforming in the $\mb3_{3/2}$ and $\mbb3_{-3/2}$. The fourth
fermion and the vector form the $\scrN$=1  vector multiplet,
$\{\lambda_\a = \lambda^4_\a \, , A_\mu\}$, and are SU(3)$\times$U(1)
singlets.  

The two parametrisations of the $\scrN$=4 scalars, $\v^i$ and
$\v^{AB}$, are related by
\be
\v^i = \fr{\sqrt{2}}\,\S^i_{AB} \v^{AB} \, , \qquad 
\v^{AB} = \fr{\sqrt{8}}\,\bar\S^{AB}_i \v^i \, , 
\label{scalAB-scali}
\ee
where $\S^i_{AB}$ ($\bar\S^{AB}_i$) are Clebsch--Gordan coefficients
projecting the product of two $\mb4$'s ($\mbb4$'s) onto the
$\mb6$. These are in other words six-dimensional euclidean sigma
matrices. They are defined as
\ba
&& \S^i_{AB} = (\S^a_{AB},\S^{a+3}_{AB}) 
= (\eta^a_{AB},i\bar\eta^a_{AB}) \nn \\
&& \bar\S_i^{AB} = (\bar\S^a_{AB},\bar\S^{a+3}_{AB}) 
= (-\eta_a^{AB},i\bar\eta^{AB}_a) \, , \rule{0pt}{18pt}
\label{defsig6}
\ea
where $a=1,2,3$ and the 't Hooft symbols $\eta^a_{AB}$ and
$\bar\eta^a_{AB}$ are
\ba
&& \eta^a_{AB} = \bar\eta^a_{AB} = \veps_{aAB} \, , 
\qquad A,B=1,2,3 \, , \nn \\
&& \eta^a_{A4} = \bar\eta^a_{4A} = \d^a_A \, , \nn \\
&& \eta^a_{AB} = - \eta^a_{BA} \, , \qquad
\bar\eta^a_{AB} = - \bar\eta^a_{BA} \, .
\label{etadef}
\ea
Using these definitions we find the following relations among the
scalars in the different formulations
\be
\begin{array}{lll}
\displaystyle \v^1 = \frac{\sqrt{2}}{2} \left( \v^{14}+\v^{23}\right) 
\, , \; & \displaystyle \v^2 = \frac{\sqrt{2}}{2} \left( 
-\v^{13}+\v^{24}\right) \, ,
\; & \displaystyle \v^3 = \frac{\sqrt{2}}{2} \left( 
\v^{12}+\v^{34}\right) \, , \\
\displaystyle \v^4 = \frac{\sqrt{2}}{2} \left( -\v^{14}+\v^{23}\right) 
\, , \; & \displaystyle \v^5 = \frac{\sqrt{2}}{2} \left( 
-\v^{13}-\v^{24}\right) \, , \; & \displaystyle \v^6 = 
\frac{\sqrt{2}}{2} \left( \v^{12}-\v^{34}\right) \rule{0pt}{23pt}
\end{array}
\label{vphii-vphiAB}
\ee
and
\be
\begin{array}{lll}
\displaystyle \phi^1 = 2 \v^{14} \, , \; & \displaystyle 
\phi^2 = 2 \v^{24} \, , \; & \displaystyle \phi^3 = 2 \v^{34} \, , \\
\displaystyle \phi^\dagger_1 = 2 \v^{23} \, , \; & \displaystyle 
\phi^\dagger_2 = -2 \v^{13} \, , \; & \displaystyle \phi^\dagger_3 
= 2 \v^{12} \, . \rule{0pt}{20pt}
\end{array}
\label{phiI-vphiAB}
\ee
The following properties of the six dimensional sigma matrices are of
use 
\be
\veps^{ABCD} \S^i_{CD} = -2 {\bar\S}^{i\,AB} \: , \qquad
\veps_{ABCD} {\bar\S}^{i\,CD} = -2 \S^i_{AB} \, ,
\label{raiseindex}
\ee
\be
\S^i_{AC}\Sbar^{j\,CB} = \delta^{ij} \delta_A^B + \S_A^{ij\,B} \, ,
\label{siga}
\ee
where $\S_A^{ij\,B}=\frac{1}{2}\left(\S^{i\, AC} \Sbar^j_{CB} -
\S^{j\, AC}\Sbar^i_{CB}\right)$, 
\be
\label{sigb}
\Sbar^{i\, AB} \S^i_{CD} = 2(\delta^A_D \delta^B_C - \delta^A_C
\delta^B_D) \, , \quad 
\Sigma^i_{AB}\Sigma^i_{CD} = 2 \veps_{ABCD} \, ,
\label{sigd}
\ee
\be
\S_A^{ij\,B}\S_B^{kl\,A} = -4 \left( \d^{ik}\d^{jl} -
\d^{il}\d^{jk} \right) \, ,
\label{sige}
\ee
\be
\label{multrace}
\tr(\Sbar^i\S^j\Sbar^k\S^l) = 4 \delta^{ij}\delta^{kl} - 4 \delta^{ik}
\delta^{jl} + 4\delta^{il}\delta^{jk} \, .
\ee

\section{One-instanton in $\scrN$=4 SYM}
\label{k1adhm}

In this appendix we briefly review the ADHM description of the
one-instanton sector for the $\scrN$=4 SYM theory. A comprehensive
review, including the treatment of multi-instanton configurations, can
be found in \cite{dhkm}. Here we only recall a few elements useful for
the calculations presented in this paper. 

The classical instanton solution is defined in terms of
a $[N+2]\times [2]$ dimensional matrix $\Delta_{a\a;\adot}$
which is a linear function of the space-time coordinate
$x_{\a\adot}=\s^\mu_{\a\adot}x_\mu$, 
\be
\label{deltadef}
\Delta_{u\a;\adot} = a_{u\a;\adot} +
b_{u\a;}{}^{\b}\, x_{\b\adot} \, ,
\ee
and its conjugate,
\be
\label{deltadefc}
\bar\Delta^{\adot;u\a} = \bar a^{\adot;u\a}
+ x^{\adot \b}\, \bar b_{\b;}{}^{u\a} \, .
\ee
The ADHM gauge field is written in the form
\be
(A_\mu)_{u;}{}^v = {\bar U}_{u;}{}^{r\a} \partial_\mu U_{r\a;}{}^v \,,
\label{adhmgaugef}
\ee
where the complex $[N]\times[N+2]$ matrix $U(x)$ and its hermitian
conjugate ${\bar U}(x)$ satisfy
\ba
&& {\bar U}_{u;}{}^{r\a} U_{r\a;}{}^v = \d_u{}^v \, , \nn \\
&& {\bar \D}^{\adot;r\b} U_{r\b;}{}^v = 0 \, , \qquad
{\bar U}_{u;}{}^{r\a} \D_{r\a;\adot} = 0 \, .
\label{adhmdata1}
\ea
Equation (\ref{adhmgaugef}) gives a gauge configuration with self-dual
field strength provided the matrices $\D$ and ${\bar \D}$ satisfy
\be
{\bar\D}^{\adot;r\b} \D_{r\b;\bdot} =
\d^\adot{}_\bdot f^{-1}(x) \, ,
\label{fintro}
\ee
where $f(x)$ is an arbitrary function.  From this relation and the
definitions (\ref{deltadef}) and (\ref{deltadefc}) it follows that the
coefficients $a$ and $b$ satisfy the bosonic ADHM constraints
\be
\label{consabar}
\bar a^{\adot;u\a}\, a_{u\a;\bdot} =
{\half}\Tr(\bar a a)\, \delta^\adot_\bdot \, ,
\ee
\be
\label{consabarb}
\bar a^{\adot;u\a}\, b_{u\a;}{}^{\b} =
\epsilon^{\b\g}\, \epsilon^{\adot\bdot}\, \bar b_{\g;}{}^{u\a}
\, a_{u\a;\bdot} \, ,
\ee
\be
\label{consbarbb}
\bar b_{\b;}{}^{u\a} \, b_{u\a;}{}^{\g} = \half \Tr (\bar b b)\, 
\delta_\b^\g \, .
\ee

A choice of special frame allows to put the bosonic parameters in the
form
\be
\label{bspec}
b_{u\a;}{}^{\b} = \left(
\begin{array}{c} 0_{u;}{}^{\b} \\
\delta_\a{}^{\b} \end{array} \right) \,,
\qquad \bar b_{\a;}{}^{u\b} =  \left(
\begin{array}{c} 0_{\a;}{}^{u} \\
\delta_\a{}^{\b} \end{array} \right) \, ,
\ee
\be
\label{aspec}
a_{u\a;\adot} =  \left(
\begin{array}{c} w_{u;\adot} \\
\aap_{\a\adot} \end{array} \right) \, ,
\qquad \bar a^{\adot;u\a} =
\left(\bar w^{\adot;u} \, , \bar\aap^{\adot\a} \right) \, ,
\ee
where $\aap_{\a\adot} = \sigma_{\a\adot}^\mu \, \aap_\mu$ and the components
satisfy the matrix constraints
\be
\label{bosconss}
\Tr_2(\tc\,{\bar a}a) = 0 \, , \qquad (\aap_\mu)^* = \aap_\mu \, .
\ee
These ADHM collective coordinates can be easily related to the usual
variables describing the position, size and gauge orientation of the
instanton. The second equation in (\ref{bosconss}) expressing the
reality of $\aap$ allows to identify it with the instanton position,
\be
\aap_\mu = - (x_0)_\mu \, .
\label{aprime-x0}
\ee
Using the first equation in (\ref{bosconss}) the scale size of the
instanton $\rho$ is related to the ADHM variables by
\be
\bar w^{\adot u} w_{u\bdot} = \d^\adot{}_\bdot \, \rho^2 \, .
\label{w-rho}
\ee
The bosonic coordinates $w_{u;\adot}$ and $\bar w^{\adot;u}$
parametrise the embedding of an SU(2) instanton into SU($N$). We start
with the special embedding 
\be
A_\mu = \left(\begin{array}{lr} 
0 & 0 \\ 0 & (A_\mu)^{\rm SU(2)} 
\end{array} \right) \, ,
\label{specemb}
\ee
where $(A_\mu)^{\rm SU(2)}$ is the standard SU(2) instanton solution
\be
(A_\mu)^{\rm SU(2)} = \frac{\rho^2\eta^a_{\mu\nu}(x-x_0)^\nu
(\tau^a)_\a{}^\b}{(x-x_0)^2[(x-x_0)^2+\rho^2]} \, .
\label{su2inst}
\ee
The general SU($N$) configuration is then given by 
\be
A_\mu = \scrH \left(\begin{array}{lr} 
0 & 0 \\ 0 & (A_\mu)^{\rm SU(2)} 
\end{array} \right) \scrH^\dagger \, , 
\label{SUNrotation}
\ee
where 
\be
w_{u\adot} = \rho \, \scrH \left(\begin{array}{c}
0_{[N-2]\times[2]} \\ \one_{[2]\times[2]} \end{array} \right) \, ,
\qquad \scrH \in 
\frac{\mathrm{SU}(N)}{\mathrm{SU}(N-2)\times\mathrm{U}(1)}
\, . \label{wexpl}
\ee

The fermionic collective coordinates enter as $[N+2]\times[1]$
grassmann valued matrices, $\scrM$ and $\bar\scrM$, that satisfy the
ADHM constraints
\be
\label{calmdef}
\bar \scrM^{A\, u\a}\, a_{u\a;\adot} =
- \epsilon_{\adot\bdot} \, \bar a^{\bdot;u\a} \,
\scrM^A_{u\a} \qquad \bar\scrM^{A\,u\a} \,
b_{u\a;}{}^{\b} = \epsilon^{\b\g} \,
\bar b_{\g;}{}^{u\a}\, \scrM^A_{u\a} \, .
\ee
The fermionic matrices can be parametrised by 
\be
\scrM^A_{u\a} =  \left(
\begin{array}{c} \nu^A_u+w_{u;\adot}\,
\mu^{\adot A} \\
\scrM^{\pp\,A}_{\a} \end{array} \right) \equiv 
\left(
\begin{array}{c} \nu^A_u+4w_{u;\adot}\,
\bar\xi^{\adot A} \\
4\eta^A_{\a} \end{array} \right) \, ,
\ee
\be
\label{fermmatb}
\bar \scrM^{A\,u\a} = \left(
\bar\nu^{A\,u} + \bar\mu^A_\adot \,
\bar w^{\adot;u}\, ,\: \bar\scrM^{\pp\,A\,\a} \right)
\equiv \left( -4\bar\xi^A_\adot\bar w^{\adot u} + \bar\nu^{A\,u} 
\, , \: 4\eta^{\a\,A} \right) \, ,
\ee
where the ADHM conditions (\ref{calmdef}) have been used to eliminate
$\bar\mu^A_\adot$ and $\bar\scrM^{\pp\,A\,\a}$ in flavour of the others
and the variables $\nu^A_u$ and $\bar\nu^{A\,u}$ satisfy
\be
\bar w^{\adot;u}\nu^A_u = \bar\nu^{A\,u}w_{u;\adot} = 0 \, .
\label{w-nu-constr}
\ee
The sixteen fermionic collective coordinates $\eta^A_\a$ and
$\bar\xi^A_\adot$ are identified with the zero modes associated
respectively with the Poincar\'e and special supersymmetries broken by
the bosonic instanton solution. The coordinates $\nu^A_u$ and
$\bar\nu^{A\,u}$, whose total number is $8(N-2)$ because of the
constraints (\ref{w-nu-constr}), are the fermionic partners of the
coset variables $w_{u;\adot}$ and $\bar w^{\adot;u}$ parametrising the
gauge orientations. 

We now summarise the expressions for the leading order terms in the
iterative solution of the field equations for the elementary fields in
the $\scrN$=4 SYM multiplet as given in terms of the ADHM variables. All
the fields in the multiplet are in the adjoint representation of the
gauge group SU($N$) and can be represented as $[N]\times[N]$ matrices.

The (self-dual part of the) gauge field strength $F^{(0)-}_{\mu\nu}$
in the instanton background follows from the construction of the gauge
field $A_\mu$ in the previous section. The result is 
\be
\left(F^{(0)\mu\nu}\right)_{u;}{}^{v} = {\bar U}_{u;}{}^{r\a}
b_{r\a;}{}^{\b} \s^{\mu\nu}{}_{\b}{}^{\g} f 
{\bar b}_{\g;}{}^{s\d}U_{s\d;}{}^{v} \, .
\label{fieldstre}
\ee
The Weyl fermions $\lam^{(1)A}_\a$ solve the equation 
\be
\Dscrm^{(0)} \lam^{(1)A}_\a = 0 
\label{direq}
\ee
and in terms of the ADHM variables can be written as
\be
\left(\lam^{(1)A}_\a\right)_{u;}{}^{v} = {\bar U}_{u;}{}^{r\b}
\left(\scrM^A_{r\b} f {\bar b}_{\a;}{}^{s\g}
- \veps_{\a\d} b_{r\b;}{}^{\d} f 
{\bar\scrM}^{As\g}\right) U_{s\g;}{}^{v} \, .
\label{gaugino}
\ee

Analogously the term $\v^{(2)AB}$ in the the solution for the scalar
field is determined by
\be
\scrD^2 \v^{(2)AB} = \frac{i}{\sqrt{2}} \{\lam^{(1)A},\lam^{(1)B}\} 
\, . \label{scaleq}
\ee
The solution was constructed in \cite{dkm1}. In the $\scrN$=4 SYM
theory in the superconformal phase of interest here 
it can be written in the form
\ba
\v^{(2)AB} &=& \half {\bar U}_{u;}{}^{r\a}
\left(\scrM^B_{r\a} f {\bar\scrM}^{As\b} - \scrM^A_{r\a} f 
{\bar\scrM}^{Bs\b} \right) U_{s\g;}{}^{v}  \nn \\
&+& \half{\bar U}_{u;}^{r\a} \left(
\begin{array}{cc} 0_{r;}{}^{s} & 0_{r;}{}^{\b} \\
0_{\a;}{}^{s} & \scrA^{AB} \d_\a^\b
\end{array} \right) U_{s\b;}{}^{v} \, ,
\label{scalar}
\ea
where $\scrA^{AB}$ is defined by
\be
\scrA^{AB} = \fr{2\sqrt{2}\rho^2} \left( 
{\bar\scrM}^{Ar\a} \scrM^B_{r\a} - {\bar\scrM}^{Br\a} \scrM^A_{r\a}
\right) \, .
\label{eqcalA}
\ee

The gauge invariant composite operators we are interested in are
traces over colour indices of products of elementary fields. In
evaluating correlation functions of such operators in the
semiclassical approximation in the instanton background one must then
compute expressions of the form
\be
\scrO = \Tr_N \left({\bar U}{\tilde F}U
\ldots {\bar U}{\tilde\lam} U \ldots {\bar U}{\tilde\v} U
\ldots\right) \, .
\ee
It is convenient to rewrite such expressions as traces over
$[N+2]\times[N+2]$ matrices in the following way
\ba
\scrO &\!\!=\!\!& \Tr_N \left({\bar U}{\tilde F}U \ldots
{\bar U}{\tilde\lam} U \ldots {\bar U}{\tilde\v} U \ldots\right)  \nn \\
&\!\!=\!\!& \Tr_{N+2}\left[ (\scrP {\tilde F}) \ldots 
(\scrP{\tilde\lam}) \ldots (\scrP{\tilde\v}) \ldots \right] \, ,
\label{projtrace}
\ea
where we have defined the projection operator
\be
\scrP_{u\a;}{}^{v\b} = U_{u\a;}{}^{r} {\bar U}_{r;}{}^{v\b} =
\d_{u;}{}^{v}\d_{\a}^{\b} - \D_{u\a;}{}^{\g} f 
\Db_{\g;}{}^{v\b} \, ,
\label{proj}
\ee
where $\D$ and $\Db$ are the matrices of bosonic ADHM variables
defined in (\ref{deltadef}) and (\ref{deltadefc}).

In the one-instanton sector all the previous formulae can be made very
explicit. The function $f(x)$ entering the ADHM construction is
\be
f = f(x;x_0,\rho) =
\frac{1}{(x-x_0)^2 + \rho^2} = \frac{1}{y^2 + \rho^2} \, .
\label{foneinst}
\ee
and the projector $\scrP$ becomes
\be
\scrP_{u\a;}{}^{v\b} = \d_{u;}{}^{v}\d_\a^\b -
\frac{1}{y^2+\rho^2} \left(
\begin{array}{cc} w_{u;\adot} {\bar w}^{\adot;v}
& \quad w_{u;\adot} y^{\adot \b} \\
y_{\a\adot}{\bar w}^{\adot;v} & \quad
y^2 \d_\a^\b \end{array} \right) \, .
\label{projoneinst}
\ee
Notice in particular that it satisfies
\be
\Tr_{N+2}\left[\scrP(x)\right] = N \, .
\label{trproj}
\ee
As observed above the gauge invariant composite operators can be
expressed as traces over $[N+2]\times[N+2]$ matrices. In the following
we give the formulae for the `projected' $[N+2]$-dimensional matrices 
corresponding to the solutions (\ref{fieldstre}), (\ref{gaugino}) and
(\ref{scalar}) for the elementary fields. For this purpose we
introduce the notation for a generic Yang--Mills field $\Phi$
\ba
&& \Phi_{u;}{}^{v} = {\bar U}_{u;}{}^{r\b} \tilde{\Phi}_{r\b;}{}^{s\g}
U_{s\g;}{}^{v} \nn \\
&& {\hat \Phi}_{u\a;}{}^{v\b} = \scrP_{u\a;}{}^{r\g}
\tilde{\Phi}_{r\g;}{}^{v\b} \, .
\label{hatfield}
\ea
For the field strength $F_{\mu\nu}$ we get~\footnote{In the remainder
of this appendix we omit the superscript denoting the number of
fermion modes on the fields. Superscripts in parenthesis refer here to
matrix elements. }
\ba
({\hat F}_{\mu\nu})_{u\a;}{}^{v\b} = \left(
\begin{array}{cc} 0 & \quad
\left({\hat F}^{(1)}_{\mu\nu}\right)_{u;}{}^{\b} \\
0 & \quad \left({\hat F}^{(2)}_{\mu\nu}\right)_{\a;}{}^{\b}
\end{array} \right) \, ,
\label{hatF}
\ea
where
\be
\left({\hat F}^{(1)}_{mn}\right)_{u;}{}^{\b} = -
\frac{4}{(y^2+\rho^2)^2} w_{u;\adot} y^{\adot\g} \s_{mn\g}{}^{\b} \: ,
\qquad \left({\hat F}^{(2)}_{mn}\right)_{\a;}{}^{\b} =
\frac{4}{(y^2+\rho^2)^2} \rho^2 \s_{mn\a}{}^{\b} \: .
\label{hatFcomp}
\ee
The solution for the fermions $\lam^A_\a$ is
\be
({\hat\lam}^A_{\a})_{u\b;}{}^{v\g} = \left( \begin{array}{cc} \left(
{\hat\lam}^{(1)A}_\a \right)_{u;}{}^{v} & \quad
\left({\hat\lam}^{(2)A}_{\a}\right)_{u;}{}^{\g} \\
\left({\hat\lam}^{(3)A}_\a \right)_{\b;}{}^{v} & \quad
\left({\hat\lam}^{(4)A}_{\a}\right)_{\b;}{}^{\g} \end{array} \right) \, ,
\label{hatlambda}
\ee
where
\ba
\left({\hat\lam}^{(1)A}_\a \right)_{u;}{}^{v} &=&
\frac{1}{(y^2+\rho^2)^2} \veps_{\a\d} w_{u;\adot} y^{\adot \d}
(-4{\bar\xi}^A_\bdot {\bar w}^{\bdot;v}+{\bar\nu}^{A\,v}) \nn \\
\left({\hat\lam}^{(2)A}_\a \right)_{u;}{}^{\g} &=&
\frac{1}{(y^2+\rho^2)^2} \left[4\left(y^2
w_{u;\adot}{\bar\xi}^{\adot A} \d_\a^\g -
w_{u;\adot}y^{\adot\d}\eta^A_\d \d_\a^\g +
\veps_{\a\d}w_{u;\adot}y^{\adot\d}
\eta^{\g A} \right) \right. \nn \\
&+& \left. (y^2+\rho^2)\nu^A_u\d_\a^\g \right] \nn \\
\left({\hat\lam}^{(3)A}_\a \right)_{\b;}{}^{v} &=&
\frac{\rho^2}{(y^2+\rho^2)^2} \veps_{\a\b}
(4{\bar\xi}^A_\adot {\bar w}^{\adot;v} -
{\bar\nu}^{A\,v}) \nn \\
\left({\hat\lam}^{(4)A}_\a \right)_{\b;}{}^{\g} &=&
\frac{4\rho^2}{(y^2+\rho^2)^2} \left(-y_{\b\adot} {\bar\xi}^{\adot A}
\d_\a^\g + \eta^A_\b \d_\a^\g - \veps_{\a\b}\eta^{\g A} \right) \, .
\label{hatlambcomp}
\ea
Finally the scalar field $\v^{AB}$ reads
\be
({\hat\v}^{AB})_{u\b;}{}^{v\g} = \left( \begin{array}{cc} \left(
{\hat\v}^{(1)AB} \right)_{u;}{}^{v} & \quad
\left({\hat\v}^{(2)AB}\right)_{u;}{}^{\g} \\
\left({\hat\v}^{(3)AB}\right)_{\b;}{}^{v} & \quad
\left({\hat\v}^{(4)AB}\right)_{\b;}{}^{\g} \end{array} \right) \, ,
\label{hatscalar}
\ee
where
\ba
\left({\hat\v}^{(1)AB}\right)_{u;}{}^{v} &\!\!=\!\!&
\frac{1}{4(y^2+\rho^2)^2} \left\{ y^2 \left[ -16
({\bar\xi}^{\adot B}{\bar\xi}^A_\bdot - {\bar\xi}^{\adot A}
{\bar\xi}^B_\bdot)w_{u;\adot}{\bar w}^{\bdot;v}
\right. \right. \nn \\
&\!\!+\!\!& \left. 4 w_{u;\adot}
({\bar\xi}^{\adot B} {\bar\nu}^{A\,v}-{\bar\xi}^{\adot A}
{\bar\nu}^{B\,v}) \right] \nn \\
&\!\!+\!\!&(y^2+\rho^2) \left[ -4({\bar\xi}^B_\adot
\nu^A_u-{\bar\xi}^A_\adot \nu^B_u){\bar w}^{\adot;v} +
(\nu^B_u{\bar\nu}^{A\,v} - \nu^A_u{\bar\nu}^{B\,v})\right]  \nn \\
&\!\!+\!\!& \left. y^{\adot\d}\left[ 16(\eta^B_\d{\bar\xi}^A_\bdot -
\eta^A_\d{\bar\xi}^B_\bdot)w_{u;\adot}{\bar w}^{\bdot;v} -
4w_{u;\adot} (\eta^B_\d {\bar\nu}^{Av} - \eta^A_\d {\bar\nu}^{Bv})
\right]\right\} \nn \\
\left({\hat\v}^{(2)AB}\right)_{u;}{}^{\g} &\!\!=\!\!&
\frac{1}{4(y^2+\rho^2)^2} \left\{ 16y^2
w_{u;\adot}({\bar\xi}^{\adot B}\eta^{\g A} - {\bar\xi}^{\adot A}
\eta^{\g B}) + 4 (y^2+\rho^2)(\nu^B_u \eta^{\g A} - \nu^A_u
\eta^{\g B})  \rule{0pt}{18pt} \right. \nn \\
&\!\!-\!\!& \left. w_{u;\adot} \left[
16y^{\adot\d} (\eta^B_\d \eta^{\g A} - \eta^A_\d \eta^{\g B}) +
\frac{1}{2} \frac{y^2+\rho^2}{\rho^2} y^{\adot\g} ({\bar\nu}^{Au}
\nu^B_u - {\bar\nu}^{Br} \nu^A_r) \right] \right\} \nn \\
\left({\hat\v}^{(3)AB}\right)_{\b;}{}^{v} &=&
\frac{1}{4(y^2+\rho^2)^2}\left\{ \rho^2
\left[ 16y_{\b\adot}({\bar\xi}^{\adot B}{\bar\xi}^A_\bdot
- {\bar\xi}^{\adot A}{\bar\xi}^B_\bdot) {\bar w}^{\bdot;v}
- 4y_{\b\adot} ({\bar\xi}^{\adot B} {\bar\nu}^{Av} -
{\bar\xi}^{\adot A} {\bar\nu}^{Bv} )  \right. \right. \nn \\
&\!\!-\!\!& \left.\left. 16 (\eta^B_\b{\bar\xi}^A_\adot -
\eta^A_\b{\bar\xi}^B_\adot) {\bar w}^{\adot;v} + 4(\eta^B_\b
{\bar\nu}^{Av} - \eta^A_\b {\bar\nu}^{Bv}) \right]\right\} \nn \\
\left({\hat\v}^{(4)AB}\right)_{\b;}{}^{\g} &=&
\frac{\rho^2}{4(y^2+\rho^2)^2} \left[
-16y_{\b\adot}({\bar\xi}^{\adot B}\eta^{\g A} - {\bar\xi}^{\adot B}
\eta^{\g A}) +16 (\eta^B_\b \eta^{\g A} - \eta^B_\b \eta^{\g A})
\right.  \nn \\
&\!\!+\!\!& \left. \frac{1}{2}\frac{y^2+\rho^2}{\rho^2}
\delta_\b^\g ({\bar\nu}^{Ar} \nu^B_r - {\bar\nu}^{Br} \nu^A_r)
\right] \, .
\label{phi2-inst-sol}
\ea

In the calculation of correlation functions at leading order in $\gy$
in non-trivial topological sectors we need the expression for the
propagators in the instanton background. The propagator for the
adjoint scalars in the one-instanton background takes the form
\ba
&& \la \tilde\v_{\:r\a}^{AB\:u\g}(x) \tilde\v_{\:s\b}^{CD\:v\d}(y)\ra 
\equiv \tilde{G}^{ABCD}[_{r\a,s\b}\hspace*{-21pt}^{u\g,v\d}](x,y) \nn \\
&& = \frac{\gy^2 \veps^{ABCD}}{2\pi^2 (x-y)^2}\left\{ \rule{0pt}{16pt}
\!\left[\scrP(x) \scrP(y)\right]_{r\a}{}^{\!v\d} \left[\scrP(y)
\scrP(x)\right]_{s\b}{}^{\!u\g} \right. \nn \\
&& -\frac{1}{N} \left(
\left[\scrP(x)\right]_{r\a}{}^{\!u\g}
\left[\scrP(y)\scrP(x)\scrP(y)\right]_{s\b}{}^{\!v\d} +
\left[\scrP(y)\right]_{s\b}{}^{\!v\d}
\left[\scrP(x)\scrP(y)\scrP(x)\right]_{r\a}{}^{\!u\g}
\right) \nn \\
&& \left.+\frac{1}{N^2} \left[\scrP(x)\right]_{r\a}{}^{\!u\g}
\left[\scrP(y)\right]_{s\b}{}^{\!v\d}
\, \Tr_{N+2} \left[\scrP(x)\scrP(y)\right] \right\} \nn \\
&& +\frac{\gy^2\veps^{ABCD}}{4\pi^2 \, \rho^2} \left\{ \rule{0pt}{16pt}
\!\left[ \scrP(x)b\bar b \scrP(x)\right]_{r\a}{}^{\!u\g} \, \left[
\scrP(y)b\bar b \scrP(y)\right]_{s\b}{}^{\!v\d} \right. \nn \\
&& - \frac{1}{N}\left(
\left[\scrP(x)\right]_{r\a}{}^{\!u\g}\left[\scrP(y)b\bar b\scrP(y)
\right]_{s\b}{}^{\!v\d}\, \Tr_2\left[\bar b\scrP(x)b\right]
\right. \nn \\
&& \left. + \left[\scrP(y)\right]_{s\b}{}^{\!v\d}\left[\scrP(x)b\bar
b\scrP(x)\right]_{r\a}{}^{\!u\g} \, \Tr_2\left[\bar b\scrP(y)b
\right] \right) \nn \\
&& \left. + \frac{1}{N^2} \left[\scrP(x)\right]_{r\a}{}^{\!u\g}
\left[\scrP(y)\right]_{s\b}{}^{\!v\d} \,
\Tr_2\left[ \bar b\scrP(x)b\right]\Tr_2\left[\bar b\scrP(y)b
\right] \right\} \, . 
\label{propfin}
\ea
The propagators for the fermions, $\la\bar\lambda_{A\adot}
\lambda^B_\a\ra$, and for the vector, $\la A_\mu A_\nu\ra$, can be
deduced from the scalar Green function \cite{bccl}.

\section{OPE analysis of a four-point function}
\label{ope}

In this appendix we present the calculation of the four-point function
(\ref{4ptope}),
\be
G(x_1,x_2,x_3,x_4) = \la \scrQ^{i_1j_1k_1}(x_1)\scrQ^{i_2j_2}(x_2)
\scrQ^{i_3j_3k_3}(x_3)\scrQ^{i_4j_4}(x_4) \ra \, .
\label{4ptope1}
\ee 
As discussed in section \ref{d3r6rev} the double pinching limit,
$x_{12}\to 0$, $x_{34}\to 0$, allows to extract the anomalous dimension
of the $\D_0=3$ operator in the $\mb6$ of SU(4) from its contribution
to the OPE. The computation of the four-point function can be
drastically simplified by a suitable choice of components in
(\ref{4ptope1}). It is convenient to work with complex fields in the
$\scrN$=1 formulation. We use the SU(4)$\to$SU(3)$\times$U(1)
branching rules
\ba
&& \mb{20^\pp} \to \mb6_2 \oplus \mbb6_{-2} \oplus \mb8_0
\label{decomp50} \\
&& \mb{50} \to \mb{10}_3 \oplus \mbb{10}_{-3} \oplus \mb{15}_1 
\oplus \mbb{15}_{-1} \, ,
\label{decomp20}
\ea
where the subscript denotes the U(1) charge. Under this decomposition
the operators in the $\mb{20^\pp}$ and $\mb{50}$, $\scrQ^{ij}$ and
$\scrQ^{ijk}$, decompose respectively into 
\ba
\scrC^{IJ} &\!\!=\!\!& \fr{\gy^2} \, \Tr\left(\phi^I\phi^J\right) 
\; \in \mb6_2  \label{cpo-6+2} \\
\bar\scrC_{IJ} &\!\!=\!\!&
\fr{\gy^2} \, \Tr\left(\phi^\dagger_I\phi^\dagger_J\right) 
\; \in \mbb6_{-2} \label{cpo-6b-2} \\
\bar\scrV_I^J &\!\!=\!\!& \fr{\gy^2}
\Tr\left(\phi^\dagger_I\phi^J\right) - 
\fr{3\gy^2} \d_I{}^J \Tr\left(\phi^\dagger_K\phi^K\right)  
\; \in \mb8_0 \label{cpo-8_0}
\ea
and 
\ba
\scrC^{IJK} &\!\!=\!\!& \fr{\gy^3N^{1/2}}\,\Tr\left(
\phi^I\phi^J\phi^K+\phi^I\phi^K\phi^J
\right) \; \in \mb{10}_3 \label{cpo-10+3} \\
\bar\scrC_{IJK} &\!\!=\!\!& \fr{\gy^3N^{1/2}} \,
\Tr\left(\phi^\dagger_I\phi^\dagger_J\phi^\dagger_K
+\phi^\dagger_I\phi^\dagger_K\phi^\dagger_J\right) \; \in 
\mbb{10}_{-3} \label{cpo-10b-3} \\
\scrV_I{}^{JK} &\!\!=\!\!& \fr{\gy^3N^{1/2}}\,\Tr\left(
\phi^\dagger_I\phi^J\phi^K+\phi^\dagger_I \phi^K\phi^J\right) 
- \fr{4\gy^3N^{1/2}} \left[ \d_I{}^J 
\Tr\left(\phi^\dagger_L \phi^L\phi^K + \phi^\dagger_L\phi^K\phi^L
\right) \right. \nn \\
&& \left. + \d_I{}^K\Tr\left(\phi^\dagger_L\phi^L\phi^J + 
\phi^\dagger_L \phi^J\phi^L \right)\right] \; \in \mb{15}_1 
\label{cpo-15+1} \\
\bar\scrV^I{}_{JK} &\!\!=\!\!& \fr{\gy^3N^{1/2}}\, \Tr\left(
\phi^I\phi^\dagger_J\phi^\dagger_K+\phi^I \phi^\dagger_K
\phi^\dagger_J\right) - \fr{4\gy^3N^{1/2}} \left[ \d^I{}_J 
\Tr\left(\phi^L \phi^\dagger_L\phi^\dagger_K 
+ \phi^L\phi^\dagger_K\phi^\dagger_L \right) \right. \nn \\
&& \left. + \d^I{}_K\Tr\left(\phi^L\phi^\dagger_L\phi^\dagger_J + 
\phi^L \phi^\dagger_J\phi^\dagger_L \right)\right] 
\; \in \mbb{15}_{-1} \, .
\label{cpo-15b-1}
\ea
A simple choice of components, which leads to a contribution of the
$\mb6$ in the $x_{12}\to 0$ $x_{34}\to 0$ channel, is then
\be
G(x_1,x_2,x_3,x_4) = \la \scrC^{113}(x_1)\bar\scrC_{11}(x_2)
\scrV_3{}^{11}(x_3)\bar\scrC_{11}(x_4)\ra \, .
\label{4ptope2}
\ee
Recalling the relation between the complex scalars $\phi^I$ and the
$\v^{AB}$'s we find
\be
\scrC^{113} \sim \Tr\left(\v^{14}\v^{14}\v^{34} \right) \, ,
\quad \bar\scrC_{11} \sim \Tr\left( \v^{23}\v^{23} \right) \, ,
\quad \scrV_3{}^{11} \sim \Tr\left( \v^{12}\v^{14}\v^{14} \right)
\, , \label{4ptflavour}
\ee
which shows that (\ref{4ptope2}) can saturate the sixteen
superconformal modes.

In the one-instanton sector this correlation function receives  two
types of contributions. The operators in (\ref{4ptope2}) contain at
least twenty fermionic modes. Therefore one can either replace all the
fields by their classical instanton solutions, in which case four of
the fermion modes have to be of the $\nu$ and $\bar\nu$ type, or
contract one pair of scalars.

For the first type of contribution substituting the classical
expressions for the composite fields and after some simple Fierz
rearrangements we get
\ba
&& G^{(1)}(x_1,x_2,x_3,x_4) = \int \dr\mu_{\rm phys}\,
\er^{-S_{\rm inst}}\,\hat\scrC^{113}(x_1;x_0,\rho;\zeta,\nu,\bar\nu)
\,\:\:\hat{\!\!\!\bar{\scrC}}_{11}(x_2;x_0,\rho;\zeta,\nu,\bar\nu) 
\nn \\
&& \hsp{3.8} 
\hat\scrV_3{}^{11}(x_3;x_0,\rho;\zeta,\nu,\bar\nu) 
\,\:\:\hat{\!\!\!\bar{\scrC}}_{11}(x_4;x_0,\rho;\zeta,\nu,\bar\nu)
\nn \\
&& = \frac{\pi^{-4N}\gy^{4N-10}\er^{2\pi i\tau}}{N(N-1)!(N-2)!}
\int\dr^4x_0\,\dr\rho \prod_{A=1}^4 \dr^2\eta^A\,\dr^2\bar\xi^A\,
\dr^{N-2}\nu^A\,\dr^{N-2}\bar\nu^A \, \er^{-S_{4F}}\nn \\
&& \frac{\rho^4[\zeta^1(x_1)\zeta^1(x_1)][\zeta^4(x_1)\zeta^4(x_1)]
(\bar\nu^{[3}\nu^{4]})}{(y_1^2+\rho^2)^5} \, 
\frac{\rho^4[\zeta^2(x_2)\zeta^2(x_2)][\zeta^3(x_2)\zeta^3(x_2)]}
{(y_2^2+\rho^2)^4} \nn \\
&& \frac{\rho^4[\zeta^1(x_3)\zeta^1(x_3)][\zeta^4(x_3)\zeta^4(x_3)]
(\bar\nu^{[1}\nu^{2]})}{(y_3^2+\rho^2)^5} \, 
\frac{\rho^4[\zeta^2(x_4)\zeta^2(x_4)][\zeta^3(x_4)\zeta^3(x_4)]}
{(y_4^2+\rho^2)^4} \, ,
\label{comp4pt1}
\ea
where the spinor indices on pairs of $\zeta$'s in square brackets are
understood to be contracted. In (\ref{comp4pt1}) as usual an  overall
numerical constant has been omitted and is to be restored in the final
result. The integrations over the superconformal fermion modes can
then be performed using (\ref{fermiint}). The integrations over the
$\nu$ and $\bar\nu$ modes can be treated similarly to what was done in
previous cases and after simple manipulations one is left with
five-sphere integrals of the form (\ref{s5int1}). After computing all
the fermionic integrals we obtain
\ba 
&& \hsp{-0.5} G^{(1)}(x_1,x_2,x_3,x_4) =
\frac{3^4\pi^{-15}2^{-2N-15}(N^2-3N+2)\,\Gamma(2N-2)\,
\er^{2\pi i\tau}}{N(N-1)!(N-2)!}\,x_{13}^4x_{24}^4 
\label{compt4pt2} \\ 
&& \hsp{-0.5} \int\dr^4x_0\,\dr\rho \,
\frac{\rho^{13}}{[(x_1-x_0)^2+\rho^2]^5
[(x_2-x_0)^2+\rho^2]^4[(x_3-x_0)^2+\rho^2]^5[(x_4-x_0)^2+\rho^2]^4}
\nn \, ,
\ea 
where $x_{pq}=(x_p-x_q)$. The final integrations over $\rho$ and
$x_0$ can be rewritten as
\ba
&& \int\dr^4x_0\,\dr\rho\,\frac{\rho^{13}}{[(x_1-x_0)^2+\rho^2]^5
[(x_2-x_0)^2+\rho^2]^4[(x_3-x_0)^2+\rho^2]^5[(x_4-x_0)^2+\rho^2]^4} 
\nn \\
&& = c\,\frac{\del}{\del x^2_{13}}\prod_{i<j}
\frac{\del}{\del x_{ij}^2} \, B(x_1,x_2,x_3,x_4) \, , 
\label{boxderiv}
\ea
where $c$ is a numerical constant and we have introduced the box
integral, $B(x_1,x_2,x_3,x_4)$, defined as
\be 
B(x_1,x_2,x_3,x_4) = \int \dr^4x \,
\fr{(x-x_1)^2(x-x_2)^2(x-x_3)^2(x-x_4)^2} \, .
\label{boxdef}
\ee
The four-point function (\ref{compt4pt2}) is finite and in the double
limit $x_{12}\to 0$, $x_{34}\to 0$ its leading singularity is
\be 
G^{(1)}(x_1,x_2,x_3,x_4) \hsp{0.3} \longrightarrow 
\raisebox{-12pt}{\hsp{-0.86}${\scriptstyle x_{12}\to 0}$\hsp{0.31}} 
\raisebox{-22pt}{\hsp{-1.2}${\scriptstyle x_{34}\to 0}$\hsp{0.3}}
\frac{1}{x_{13}^5x_{24}^5} \log\left(
\frac{x_{12}^2x_{34}^2}{x_{13}^2x_{24}^2}\right) \, .
\label{opelim1}
\ee

The second type of contribution involves scalar propagators. Recalling
that $\la\v^{AB}\v^{CD}\ra\sim\veps^{ABCD}$, from
(\ref{4ptope2})-(\ref{4ptflavour}) it follows that the allowed
contractions are
\ba
&& G^{(2)}(x_1,x_2,x_3,x_4) \sim \la \Tr\left[\left(\v^{14}
\v^{14}\v^{34} \right)(x_1)\right] \Tr\left[\left( \v^{23}\v^{23} 
\right)(x_2)\right] \raisebox{-10pt}{\hsp{-6.35}
\rule{0.4pt}{4pt}\rule{3.86cm}{0.4pt}\rule{0.4pt}{4pt}}
\hsp{2.3} \nn \\
&& \Tr\left[\left( \v^{12}\v^{14}\v^{14} \right)(x_3)\right] 
\Tr\left[\left( \v^{23}\v^{23} \right)(x_4)\right] \ra + \cdots 
\label{4ptwick} \\
&& + \la\Tr\left[\left(\v^{14}\v^{14}\v^{34} \right)(x_1)\right] 
\Tr\left[\left( \v^{23}\v^{23} \right)(x_2)\right]
\Tr\left[\left( \v^{12}\v^{14}\v^{14} \right)(x_3)\right] 
\raisebox{-10pt}{\hsp{-9.05}
\rule{0.4pt}{4pt}\rule{5.98cm}{0.4pt}\rule{0.4pt}{4pt}}
\hsp{2.9}
\Tr\left[\left( \v^{23}\v^{23} \right)(x_4)\right] \ra \, , \nn
\ea
where the dots in (\ref{4ptwick}) stand for the other contractions of
the same type as those indicated as required by Wick's theorem.  The
first type of contraction, between a $\v^{14}$ and a $\v^{23}$, leads
however to a vanishing contribution since after replacing the
remaining fields with their instanton solution one is forced to put
three $\zeta^1$ modes at the same point, $x_3$. We are thus left with
only one possible contraction, the one on the last line of
(\ref{4ptwick}).   To evaluate the corresponding contribution to the
four-point function we use the propagator (\ref{propfin}). The
calculation is rather involved and gives
\ba
&& G^{(2)}(x_1,x_2,x_3,x_4) = \frac{3^4\pi^{-15}2^{-2N-13}
\Gamma(2N-1)\,\er^{2\pi i\tau}}{N(N-1)!(N-2)!} 
\int\dr^4x_0\,\dr\rho\,\dr^5\Omega
\prod_{A=1}^4 \dr^2\eta^A\,\dr^2\bar\xi^A \nn \\
&& \left[ \frac{4}{(x_1-x_3)^2}
\frac{\rho^8}{(y_1^2+\rho^2)^4(y_3^2+\rho^2)^4} 
-\frac{(3N^2+10N+12)}{N^2} \frac{\rho^{10}}
{(y_1^2+\rho^2)^5(y_3^2+\rho^2)^5} \right] \nn \\
&& \frac{\rho^8}{(y_2^2+\rho^2)^4(y_4^2+\rho^2)^4}
\left[(\zeta^1\zeta^1)(\zeta^4\zeta^4)(x_1)\right]
\left[(\zeta^1\zeta^1)(\zeta^4\zeta^4)(x_3)\right] \nn \\
&& \left[(\zeta^2\zeta^2)(\zeta^3\zeta^3)(x_2)\right]
\left[(\zeta^2\zeta^2)(\zeta^3\zeta^3)(x_4)\right]
\rule{0pt}{18pt} \, ,
\label{4ptcontr1}
\ea
where the $\nu$ and $\bar\nu$ integrals have been replaced by a
five-sphere integral since in this contribution there is no explicit
dependence on these variables in the integrand. The integrations over
the superconformal modes can now be performed and we get
\ba
&& G^{(2)}(x_1,x_2,x_3,x_4) = \frac{3^4\pi^{-15}2^{-2N-13}\Gamma(2N-1)
\,\er^{2\pi i\tau}}{N(N-1)!(N-2)!} (x_1-x_3)^4(x_2-x_4)^4 \nn \\
&& \int\dr^4x_0\,\dr\rho \, \frac{\rho^{11}}{[(x_1-x_0)^2+\rho^2]^4
[(x_2-x_0)^2+\rho^2]^4[(x_3-x_0)^2+\rho^2]^4[(x_4-x_0)^2+\rho^2]^4}
\nn \\
&& \left[ \frac{4}{(x_1-x_3)^2} -\frac{(3N^2+10N+12)}{N^2} 
\frac{\rho^2} {(y_1^2+\rho^2)(y_3^2+\rho^2)} \right] \, ,
\label{4ptcontr2}
\ea
which can be rewritten in terms of the box integral (\ref{boxdef})
similarly to the case of $G^{(1)}(x_1,x_2,x_3,x_4)$. In the limit
$x_{12}\to 0$ $x_{34}\to 0$ the singularity is again of the type
\be 
G^{(2)}(x_1,x_2,x_3,x_4) \hsp{0.3} \longrightarrow 
\raisebox{-12pt}{\hsp{-0.86}${\scriptstyle x_{12}\to 0}$\hsp{0.31}} 
\raisebox{-22pt}{\hsp{-1.2}${\scriptstyle x_{34}\to 0}$\hsp{0.3}}
\frac{1}{x_{13}^5x_{24}^5} \log\left(
\frac{x_{12}^2x_{34}^2}{x_{13}^2x_{24}^2}\right) \, .
\label{opelim2}
\ee
In conclusion the OPE of the complete four-point function
(\ref{4ptope2}) does not present a singularity corresponding to the
contribution of a scalar operator of bare dimension $\D_0=3$ in the
$\mb6$ of SU(4) in the $x_{12}\to 0$, $x_{34}\to 0$ channel. Therefore 
the operator 
\be
\scrO^i_{3,\mb6} = \frac{1}{\gy^3N^{1/2}} 
\Tr \left( \v^i\v^j\v^j \right) \, .
\label{dim3-6a}
\ee
which according to the analysis of section \ref{d3r6} could have an
instanton induced anomalous dimension appears instead to be protected
at the instanton level. 

The singularities observed in (\ref{opelim1}) and (\ref{opelim2}) on
the other hand correspond to the contribution of operators of bare
dimension 5. Since in section \ref{dim5scal} it was argued that at the
level of dimension 5 operators only the ones in the $\mb6$ can have an
instanton contribution, the result of the above OPE analysis confirms
that indeed at least one of the $\D_0=5$ operators in the $\mb6$ has
scaling dimension corrected by instantons. The study of a single
four-point function however does not allow to identify which
operators in this sector receive corrections.


\begin{thebibliography}{123}

\addcontentsline{toc}{section}{References}


\bibitem{m}{J.M.~Maldacena, ``The large $N$ limit of superconformal
field theories and supergravity'', \atmp{2}{1998}{231} 
[\ijtp{38}{1999}{1113}] [\hepth{9711200}].}

\bibitem{gkp}{S.S.~Gubser, I.R.~Klebanov and A.M.~Polyakov,
``Gauge theory correlators from non-critical string theory'',
\plb{428}{1998}{105} [\hepth{9802109}].}

\bibitem{w}{E.~Witten, ``Anti-de Sitter space and holography'', 
\atmp{2}{1998}{253} [\hepth{9802150}].}

\bibitem{bkrs1}{M.~Bianchi, S.~Kovacs, G.C.~Rossi and Ya.S.~Stanev, 
``On the logarithmic behavior in $\scrN$=4 SYM theory'', 
\jhep{08}{1999}{020} [\hepth{9906188}].}

\bibitem{adslog}{H.~Liu, ``Scattering in anti-de Sitter space and
operator product expansion'', \prd{60}{1999}{106005}
[\hepth{9811152}]; \\ 
J.H.~Brodie and M.~Gutperle, ``String corrections to four point
functions in the AdS/CFT  correspondence'', \plb{445}{1999}{296}
[\hepth{9809067}]; \\
E.~D'Hoker, D.Z.~Freedman, S.D.~Mathur, A.~Matusis and L.~Rastelli,
``Graviton exchange and complete 4-point functions in the AdS/CFT
correspondence'', \npb{562}{1999}{353} [\hepth{9903196}].}

\bibitem{cftlog}{B.~Eden, P.S.~Howe, C.~Schubert, E.~Sokatchev and 
P.C.~West, ``Four-point functions in $\scrN$=4 supersymmetric 
Yang-Mills theory at two loops'', \npb{557}{1999}{355} 
[\hepth{9811172}].}

\bibitem{bmn}{D.~Berenstein, J.M.~Maldacena and H.~Nastase,
``Strings in flat space and pp waves from $\scrN$=4 super Yang
Mills'', \jhep{04}{2002}{013} [\hepth{0202021}].}

\bibitem{bfhp}{M.~Blau, J.~Figueroa-O'Farrill, C.~Hull and
G.~Papadopoulos, ``A new maximally supersymmetric background of IIB
superstring theory'', \jhep{01}{2002}{047} [\hepth{0110242}].}

\bibitem{mt}{R.R.~Metsaev and A.A.~Tseytlin, ``Exactly solvable model
of superstring in plane wave Ramond-Ramond  background'',
\prd{65}{2002}{126004} [\hepth{0202109}].}

\bibitem{sz}{A.~Santambrogio and D.~Zanon, ``Exact anomalous
dimensions of $\scrN$=4 Yang-Mills operators with large R charge'', 
\plb{545}{2002}{425} [\hepth{0206079}].}

\bibitem{bkpss}{N.~Beisert, C.~Kristjansen, J.~Plefka, G.W.~Semenoff
and M.~Staudacher, ``BMN correlators and operator mixing in $\scrN$=4 
super Yang-Mills theory'', \npb{650}{2003}{125} [\hepth{0208178}].}

\bibitem{cfhm}{N.R.~Constable, D.Z.~Freedman, M.~Headrick and
S.~Minwalla, ``Operator mixing and the BMN correspondence'',
\jhep{10}{2002}{068} [\hepth{0209002}].}

\bibitem{kpss}{C.~Kristjansen, J.~Plefka, G.~W.~Semenoff and 
M.~Staudacher, ``A new double-scaling limit of $\scrN$=4 super 
Yang-Mills theory and PP-wave  strings'', \npb{643}{2002}{3} 
[\hepth{0205033}].}

\bibitem{cfhmmps}{N.R.~Constable, D.Z.~Freedman, M.~Headrick, 
S.~Minwalla, L.~Motl, A.~Postnikov and W.~Skiba, ``PP-wave string 
interactions from perturbative Yang-Mills theory'', 
\jhep{07}{2002}{017} [\hepth{0205089}].}

\bibitem{gkp2}{S.S.~Gubser, I.R.~Klebanov and A.M.~Polyakov,
``A semi-classical limit of the gauge/string correspondence'', 
\npb{636}{2002}{99} [\hepth{0204051}].}

\bibitem{p}{A.M.~Polyakov, ``Gauge fields and space-time'', 
\ijmpa{17S1}{2002}{119} [\hepth{0110196}].}

\bibitem{ft0}{S.~Frolov and A.A.~Tseytlin, ``Semiclassical
quantization of rotating superstring in AdS$_5\times S^5$'', 
\jhep{06}{2002}{007} [\hepth{0204226}].}

\bibitem{r}{J.G.~Russo, ``Anomalous dimensions in gauge theories from
rotating strings in  AdS$_5\times S^5$'', \jhep{06}{2002}{038}
[\hepth{0205244}].}

\bibitem{bmsz}{N.~Beisert, J.A.~Minahan, M.~Staudacher and K.~Zarembo,
``Stringing spins and spinning strings'', \jhep{09}{2003}{010}
[\hepth{0306139}].} 

\bibitem{t1}{A.A.~Tseytlin, ``On semiclassical approximation and
spinning string vertex operators in  AdS$_5\times S^5$'', 
\npb{664}{2003}{247} [\hepth{0304139}].}

\bibitem{ft1}{S.~Frolov and A.A.~Tseytlin, ``Multi-spin string
solutions in AdS$_5\times S^5$'', \npb{668}{2003}{77}
[\hepth{0304255}].}

\bibitem{ft2}{S.~Frolov and A.A.~Tseytlin, ``Quantizing three-spin
string solution in AdS$_5\times S^5$'', \jhep{07}{2003}{016}
[\hepth{0306130}].} 

\bibitem{ft3}{S.~Frolov and A.A.~Tseytlin, ``Rotating string
solutions: AdS/CFT duality in non-supersymmetric  sectors'',
\plb{570}{2003}{96} [\hepth{0306143}].}

\bibitem{afrt}{G.~Arutyunov, S.~Frolov, J.~Russo and A.A.~Tseytlin,
``Spinning strings in AdS$_5\times S^5$ and integrable systems'',
\hepth{0307191}.} 

\bibitem{bfst}{N.~Beisert, S.~Frolov, M.~Staudacher and A.A.~Tseytlin,
``Precision spectroscopy of AdS/CFT'', \hepth{0308117}.}

\bibitem{j}{R.A.~Janik, ``BMN operators and string field theory'',
\plb{549}{2002}{237} [\hepth{0209263}].}

\bibitem{mz}{J.A.~Minahan and K.~Zarembo, ``The Bethe-ansatz for
$\scrN$=4 super Yang-Mills'', \jhep{03}{2003}{013} [\hepth{0212208}].}

\bibitem{bkps}{N.~Beisert, C.~Kristjansen, J.~Plefka and M.~Staudacher,
``BMN gauge theory as a quantum mechanical system'',
\plb{558}{2003}{229} [\hepth{0212269}].}

\bibitem{bks}{N.~Beisert, C.~Kristjansen and M.~Staudacher,
``The dilatation operator of $\scrN$=4 super Yang-Mills theory'', 
\npb{664}{2003}{131} [\hepth{0303060}].}

\bibitem{bs1}{N.~Beisert and M.~Staudacher, ``The $\scrN$=4 SYM 
integrable super spin chain'', \npb{670}{2003}{439} [\hepth{0307042}].} 

\bibitem{msw}{G.~Mandal, N.V.~Suryanarayana and S.R.~Wadia,
``Aspects of semiclassical strings in AdS$_5$'', 
\plb{543}{2002}{81} [\hepth{0206103}].}

\bibitem{bpr}{I.~Bena, J.~Polchinski and R.~Roiban, ``Hidden
symmetries of the AdS$_5\times S^5$ superstring'', \hepth{0305116}.}

\bibitem{dnw}{L.~Dolan, C.R.~Nappi and E.~Witten, ``A relation between
approaches to integrability in superconformal Yang-Mills theory'', 
\hepth{0308089}.}

\bibitem{bgkr}{M.~Bianchi, M.B.~Green, S.~Kovacs and G.C.~Rossi,
``Instantons in supersymmetric Yang-Mills and D-instantons in IIB  
superstring theory'', \jhep{08}{1998}{013} [\hepth{9807033}].}

\bibitem{dhkmv}{N.~Dorey, T.J.~Hollowood, V.V.~Khoze, M.P.~Mattis and
S.~Vandoren, ``Multi-Instanton Calculus and the AdS/CFT
Correspondence in $\scrN$=4 Superconformal Field Theory'',
\npb{552}{1999}{88} [\hepth{9901128}].}

\bibitem{dkmv}{N.~Dorey, V.V.~Khoze, M.P.~Mattis and S.~Vandoren,
``Yang-Mills instantons in the large-$N$ limit and the AdS/CFT  
correspondence'', \plb{442}{1998}{145} [\hepth{9808157}].}

\bibitem{gk}{M.B.~Green and S.~Kovacs, ``Instanton-induced Yang--Mills
correlation functions at large $N$ and their  AdS$_5 \times S^5$
duals'', \jhep{04}{2003}{058} [\hepth{0212332}].}

\bibitem{bkrs3}{M.~Bianchi, S.~Kovacs, G.C.~Rossi, Ya.S.~Stanev,
``Properties of the Konishi multiplet in $\scrN$=4 SYM theory'',
\jhep{05}{2001}{042} [\hepth{0104016}].}

\bibitem{bms}{M.~Bianchi, J.F.~Morales and H.~Samtleben,
``On stringy AdS$_5\times S^5$ and higher spin holography'',
\jhep{07}{2003}{062} [\hepth{0305052}].}

\bibitem{b1}{N.~Beisert, ``The complete one-loop dilatation operator
of $\scrN$=4 super Yang-Mills theory'', \hepth{0307015}.}

\bibitem{b2}{N.~Beisert, ``Higher loops, integrability and the near
BMN limit'', \jhep{09}{2003}{062} [\hepth{0308074}].}

\bibitem{adhm}{M.F.~Atiyah, N.J.~Hitchin, V.G.~Drinfeld and
Yu.I.~Manin, ``Construction of instantons'', \pla{65}{1978}{185}.}

\bibitem{akmrv}{D.~Amati, K.~Konishi, Y.~Meurice, G.C.~Rossi and 
G.~Veneziano, ``Nonperturbative Aspects in Supersymmetric Gauge
Theories'', \prep{162}{1988}{169}.}

\bibitem{dhkm}{N.~Dorey, T.J.~Hollowood, V.V.~Khoze and
M.P.~Mattis, ``The Calculus of Many Instantons'', 
\prep{371}{2002}{231} [\hepth{0206063}].}

\bibitem{dkm1}{N.~Dorey, V.V.~Khoze and M.P.~Mattis, ``On
mass-deformed $\scrN$=4 supersymmetric Yang-Mills theory'',
\plb{396}{1997}{141} [\hepth{9612231}].}

\bibitem{cgt}{E.F.~Corrigan, P.~Goddard and S.~Templeton, ``Instanton
Green's functions and tensor products'', \npb{151}{1979}{93}.}

\bibitem{bccl}{L.S.~Brown, R.D.~Carlitz, D.B.~Creamer and C.~Lee, 
``Propagation functions in pseudoparticle fields'',
\prd{17}{1978}{1583}.} 

\bibitem{bvv}{A.V.~Belitsky, S.~Vandoren and P.~van Nieuwenhuizen,
``Yang-Mills and D-instantons'', \cqg{17}{2000}{3521}
[\hepth{0004186}].} 

\bibitem{sv}{M.A.~Shifman and A.I.~Vainshtein,
``Instantons versus supersymmetry: Fifteen years later'',
\hepth{9902018}.} 

\bibitem{agj}{D.~Anselmi, M.~Grisaru and A.~Johansen, ``A critical
behaviour of anomalous currents, electric-magnetic universality and
CFT$_4$'', \npb{491}{1997}{221} [\hepth{9601023}].}

\bibitem{bkrs2}{M.~Bianchi, S.~Kovacs, G.C.~Rossi and Ya.S.~Stanev,
``Anomalous dimensions in $\scrN$=4 SYM theory at order $g^4$'',
\npb{584}{2000}{216} [\hepth{0003203}].}

\bibitem{aeps}{G.~Arutyunov, S.~Frolov and A.C.~Petkou, ``Operator
product expansion of the lowest weight CPOs in $\scrN$ = 4 SYM$_4$ at
strong coupling'',  \npb{586}{2000}{547} [\hepth{0005182}], Erratum
\ibid{609}{2001}{539}.}

\bibitem{dp}{V.K.~Dobrev and V.B.~Petkova, ``All Positive Energy
Unitary Irreducible Representations of Extended Conformal
Supersymmetry'', \plb{162}{1985}{127}.}

\bibitem{fz}{S.~Ferrara and A.~Zaffaroni, ``Superconformal field
theories, multiplet shortening, and the  AdS$_5$/SCFT$_4$
correspondence'', \hepth{9908163}.}

\bibitem{dfs}{E.~D'Hoker, D.Z.~Freedman and W.~Skiba, ``Field Theory
Tests for Correlators in the AdS/CFT Correspondence'',
\prd{59}{1999}{045008} [\hepth{9807098}].}

\bibitem{afp}{G. Arutyunov, S. Frolov and A.C. Petkou, ``Perturbative 
and instanton corrections to the OPE of CPO's in $\scrN$=4 SYM$_4$'',
\npb{602}{2001}{238} [{\tt hep-th/0010137}], Erratum
\ibid{609}{2001}{540}.} 

\bibitem{dhhr}{E.~D'Hoker, P.~Heslop, P.~Howe and A.V.~Ryzhov,
``Systematics of quarter BPS operators in $\scrN$=4 SYM'',
\jhep{04}{2003}{038} [\hepth{0301104}].}

\bibitem{appss}{G.~Arutyunov, S.~Penati, A.C.~Petkou, A.~Santambrogio 
and E.~Sokatchev, ``Non-protected operators in $\scrN$=4 SYM and
multiparticle states of AdS$_5$ SUGRA'', \npb{ 643}{2002}{49} 
[\hepth{0206020}].}

\bibitem{bers}{M.~Bianchi, B.~Eden, G.C.~Rossi and Ya.S.~Stanev,
``On operator mixing in $\scrN$=4 SYM'', 
\npb{646}{2002}{69} [\hepth{0205321}].}

\bibitem{iigt}{{\it Instantons in Gauge Theories}, ed. M.A.~Shifman, 
World Scientific (1994)} 

\bibitem{gks}{M.B.~Green, S.~Kovacs and A.~Sinha, work in progress.}

\bibitem{dkm2}{N.~Dorey, V.V.~Khoze and M.P.~Mattis, ``Multi-instanton
calculus in $\scrN$=2 supersymmetric gauge theory. II: Coupling to
matter'', \prd{54}{1996}{7832} [\hepth{9607202}].}

\bibitem{bgk}{M.~Bianchi, M.B.~Green and S.~Kovacs,
``Instanton corrections to circular Wilson loops in $\scrN$=4
supersymmetric Yang--Mills'', \jhep{04}{2002}{040} [\hepth{0202003}];
``Instantons and BPS Wilson loops'', \hepth{0107028}.}


\end{thebibliography}
\end{document}